\documentclass[
 aip,
 jcp,
 amsmath,amssymb,
  reprint
]{revtex4-2}

\usepackage{graphicx}
\usepackage{graphics,amssymb,amsmath,epsfig,color,textgreek}
\usepackage{dcolumn}
\usepackage{bm}

\usepackage[dvipsnames]{xcolor}
\usepackage[utf8]{inputenc}
\usepackage[T1]{fontenc}
\usepackage{array}
\usepackage{booktabs}
\usepackage{graphicx}
\usepackage{mathptmx}
\usepackage{etoolbox}
\usepackage{float}
\usepackage{tabularx}
\usepackage{braket}
\usepackage{booktabs}
\usepackage{tikz}
\usepackage{bbm}
\usepackage{hyperref}
\hypersetup{colorlinks,linkcolor=blue,anchorcolor=blue,citecolor=blue,urlcolor=blue}
\hypersetup{colorlinks,linkcolor=blue,anchorcolor=blue,citecolor=blue,urlcolor=blue}
\usetikzlibrary{decorations.pathreplacing}

\newcommand{\nocontentsline}[3]{}
\let\origcontentsline\addcontentsline
\newcommand\stoptoc{\let\addcontentsline\nocontentsline}
\newcommand\resumetoc{\let\addcontentsline\origcontentsline}

\newcommand{\bracketmatrix}[4]{
\begin{tikzpicture}[baseline=(m.center), scale=1]
\node (m) at (0,0) {$\displaystyle
\begin{bmatrix}
#1 & \cdots & \textbf{0} \\
\vdots & \ddots \\
\textbf{0} & & #2
\end{bmatrix}$};

\draw[decorate, decoration={brace, amplitude=2pt, raise=1pt}]
 (-0.55,0.75) -- (-0.3,0.75)
node[midway, above=2pt] {\scriptsize $#3$};

\draw[decorate, decoration={brace, amplitude=5pt, raise=4pt}]
(-0.65, 0.3) -- (-0.65,0.6)
node[midway, left=8pt, rotate=0] {\scriptsize $#4$};
\end{tikzpicture}
}

\makeatletter
\def\@email#1#2{
 \endgroup
 \patchcmd{\titleblock@produce}
  {\frontmatter@RRAPformat}
  {\frontmatter@RRAPformat{\produce@RRAP{*#1\href{mailto:#2}{#2}}}\frontmatter@RRAPformat}
  {}{}
}
\makeatother

\usepackage{orcidlink}

\begin{document}

\preprint{AIP/123-QED}

\title{Scalable Quantum Computational Science: \\ A Perspective from Block-Encodings and Polynomial Transformations}

\author{Kevin J. Joven\,\orcidlink{0000-0003-4730-7053}}
\affiliation{Department of Electrical and Computer Engineering, North Carolina State University, Raleigh, North Carolina 27695, USA}

\author{Elin Ranjan Das\,\orcidlink{0009-0009-8690-1523}}
\affiliation{Department of Electrical and Computer Engineering, North Carolina State University, Raleigh, North Carolina 27695, USA}

\author{Joel Bierman}
\affiliation{Department of Electrical and Computer Engineering, North Carolina State University, Raleigh, North Carolina 27695, USA}

\author{Aishwarya Majumdar \orcidlink{0009-0008-2800-0455}}
\affiliation{Department of Electrical and Computer Engineering, North Carolina State University, Raleigh, North Carolina 27695, USA}

\author{Masoud Hakimi Heris}
\affiliation{Department of Electrical and Computer Engineering, North Carolina State University, Raleigh, North Carolina 27695, USA}

\author{Yuan Liu\,\orcidlink{0000-0003-1468-942X}}
\email{q\_yuanliu@ncsu.edu}
\affiliation{Department of Electrical and Computer Engineering, North Carolina State University, Raleigh, North Carolina 27695, USA}
\affiliation{Department of Computer Science, North Carolina State University, Raleigh, North Carolina 27695, USA}
\affiliation{Department of Physics, North Carolina State University, Raleigh, North Carolina 27695, USA}

\begin{abstract}
    Significant developments made in quantum hardware and error correction recently have been driving quantum computing towards practical utility. However, gaps remain between abstract quantum algorithmic development and practical applications in computational sciences. In this Perspective article, we propose several properties that scalable quantum computational science methods should possess. We further discuss how block-encodings and polynomial transformations can potentially serve as a unified framework with the desired properties. Recent advancements on these topics are presented including construction and assembly of block-encodings, and various generalizations of quantum signal processing (QSP) algorithms to perform polynomial transformations. The scalability of QSP methods on parallel and distributed quantum architectures is also highlighted. Promising applications in simulation and observable estimation in chemistry, physics, and optimization problems are presented. We hope this Perspective serves as a gentle introduction of state-of-the-art quantum algorithms to the computational science community, and inspires future development on scalable quantum computational science methodologies that bridge theory and practice. 
\end{abstract}

\maketitle

\tableofcontents

\section{Introduction}

Progress in quantum hardware \cite{acharya2024quantum,ni2023beating,sivak2023real,gupta2024encoding,brock2025quantum,putterman2025hardware,egan2021fault,bluvstein2024logical,altman2021quantum}, fault-tolerance \cite{knill1997theory,michael2016new,gidney2024magic,gottesman2022ft}, quantum control \cite{vandersypen2004nmr,koch2019quantum,liu2024hybrid}, and quantum algorithms \cite{cao2019quantum,bharti2022noisy,cerezo2021variational,chan2024quantum,PRXQuantum.2.040203} in the past decades has made large-scale quantum computers closer than ever. Just as how classical computers accelerated discoveries in the past century \cite{thijssen2007computational,mcweeny1989method,waterman2018introduction}, quantum computers, especially fault-tolerant (FT) ones, can be a powerful addition to the toolbox of computational scientists to accelerate understanding and discoveries in science and engineering. Recent advancements have indeed started to push quantum computing from small- to utility-scale applications, together with the help of classical high-performance computing (HPC) systems \cite{alexeev2024quantum}.

No doubt that classical HPC plays important roles in quantum, but scaling quantum computational methods themselves faces many challenges. For one, there remains a huge gap between quantum algorithm development and practical applications in computational science. Quantum computers work in a fundamentally different way than classical ones. Historically, most fault-tolerant quantum algorithms were discovered by mathematicians or computer scientists, where the original analysis is very different from what computational scientists use on a daily basis. From a practical perspective, the limited availability of real quantum hardware resources makes it challenging for computational scientists to perform extensive experimental execution of quantum algorithms on actual hardware for problems of interest to iterate and improve algorithm construction.

Bridging this gap is of critical importance for sustained development of quantum computational science. 
This is not only because computational scientists often know the most demanding practical applications that can test the ultimate limit of any computing machines but also because they have the best classical computational methods that can help benchmark performance of quantum computers and pin down applications for practical quantum advantage. Advocating quantum computational methods in domain applications is particularly important in the coming early fault-tolerant era (with a logical qubit error rate of $10^{-5} \sim 10^{-8}$). This is a regime where the complexity of quantum computing methods will increase enough such that they are likely capable of tackling problems beyond classical computers. While at the same time, simple reasoning that we can perform at the current stage on the performance of quantum algorithms may become very limited.
To overcome this gap, beyond continued improvement on quantum hardware, demystifying fault-tolerant quantum algorithms for computational scientists is essential for bridging theory and practice.

In this Perspective, we define properties that scalable quantum computational science methods should possess. Searching over the current algorithmic landscape, we identify a class of key quantum algorithmic primitives, the quantum signal processing (QSP) algorithm \cite{low2016methodology,low2014optimal,PhysRevLett.118.010501} and its generalizations \cite{10.1145/3313276.3316366,PRXQuantum.2.040203,rossi2021mqsp,martyn2025parallel,rossi2023quantum,laneve2024quantumsignalprocessingsun,lu2024quantumsignalprocessingquantum,singh2025nonabelianquantumsignalprocessing,SinananSingh2024singleshotquantum,tan2023perturbative,motlagh2023generalized}, which satisfies these properties and can serve as a core primitive for scalable quantum computational science methodology development. We provide a pedagogical overview of the two building blocks of QSP algorithms, i.e., the block-encoding and polynomial transforms, and review the most recent development of this family of algorithms. We discuss error tradeoffs between block-encoding and polynomial transforms, as well as basic tools to construct them and assemble them in a modular fashion into complex quantum computing methods. Whenever possible, we provide explicit quantum circuits and focus on connections from abstract algorithms to practical applications in chemistry, physics, and other domain applications. We hope our work helps to demystify QSP algorithms for computational scientists and inspire future community efforts toward methodology development in scalable quantum computational science. 

The rest of the article is organized as follows. We define notions of scalable quantum computational science methods and introduce basic notations in Sec. \ref{sec:quest}. Secs. \ref{sec:block_encoding} and \ref{sec:polynomial-transformation} each describes the art in the two pillars of QSP algorithms, i.e., block-encoding and polynomial transforms. Sec. \ref{sec:scale} briefly discusses ways to scale QSP algorithms to parallel and distributed quantum architectures. Sec. \ref{sec:app} provide illustrative examples on common applications in physical science and beyond, followed by conclusion in Sec. \ref{sec:conclusion}.

\section{Quest for Scalable Quantum Computational Science}
\label{sec:quest}

Computational science aims to efficiently simulate physical systems and processes with computers. This idea, while simple as it sounds, presents significant challenges for classical computers. As a result, significant sacrifice on accuracy is needed to accommodate computation of practical relevance into classical computers of reasonable size. This unfortunately leads to the loss of predictive power for a large class of problems \cite{}.

The power of quantum computers can potentially allow us to escape this doom by making a much better tradeoff between accuracy and efficiency, providing that a collection of scalable quantum computational methods can be developed to finally achieve \emph{predictive} power for computational problems.
While it is challenging to establish all quantum computational methods all at once, we can nonetheless outline a set of properties for such methods. 
The properties we will present are somewhat different from the current status of the field; the goal is to open the opportunity of constructing a quantum computational science paradigm that significantly differs from the classical ones.
We present this set of properties below:

\textbf{Property 1. Exhibit well-characterized and quantifiable resource cost.} The methods should possess well-defined bounds that quantify key quantities such as error rates, time complexity, qubit count, energy consumption, and gate count. These bounds must at least enable their estimation based on other measurable parameters. This is in sharp contrast with current NISQ algorithms that are heuristic or variational in nature. We note that opportunities still exist for heuristic algorithms on quantum computers.

\textbf{Property 2. Be resource-efficient and possess a flexible trade-off between various resource types.} For a given application problem, the computing method should be able to be easily adapted to various versions that can be implemented in an efficient manner on available resources and hardware (serial, parallel, and distributed). The notion of \emph{efficiency} should be precisely defined, but may vary from application to applications and may depend on the best known classical algorithm runtime to solve that problem. The method should maintain an optimal trade-off between accuracy and resource cost, space (classical and quantum), and time (gate depth and sample complexity) cost.

\textbf{Property 3. Adaptability, programmability, and modularity.} The method should be sufficiently adaptable to accommodate diverse types of applications that can take different inputs and may require different levels of outputs, including hybrid implementations that integrate classical and quantum data. It should also exhibit scalability, enabling the incremental incorporation of additional functionalities when required by the application. The method should be easy enough to program, allowing a tunable level of abstraction and modular assembly from gates to arithmetic to modules.

These properties can then serve as a reference for the development of methods. 

\begin{figure*}[htbp!]
    \centering
    \includegraphics[width=0.8\textwidth]{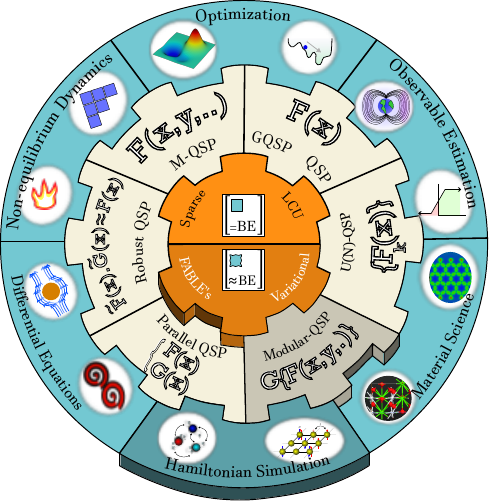} 
    \caption{ Figure illustrates an overview of block-encoding and polynomial transformation techniques as building blocks of scalable quantum computational science for various applications. Each algorithmic primitive in the middle (light grey) can be used to tackle different applications in the outer part (light blue). For example, the dark color aims to perform Hamiltonian simulation by combining modular-QSP with approximate block-encoding.}
   \label{fig1}
\end{figure*}

\subsection{Block-encoding and Polynomial Transformation as Promising Candidates}

Searching through the landscape of existing quantum computing algorithms, we identify primitives that best satisfy these properties: quantum signal processing (QSP) algorithms and their variants. In the following, we briefly explain how the two building blocks of QSP algorithms, block-encoding (BE) and polynomial transformations, make QSP-type algorithms strong candidates for future quantum computational science methods.

It is well established that block-encoding and polynomial transformations possess well-characterized resource costs for given error thresholds \cite{PRXQuantum.2.040203}. These bounds enable the estimation of fundamental resource requirements such as error rates, gate counts, qubit counts, time complexity, and query complexity for a given quantum algorithm, as we will present in Sec. \ref{sec:block_encoding} and \ref{sec:polynomial-transformation}. In contrast to variational or heuristic that depend on classical training parameters, block-encoding and polynomial transformations offer a clear and systematic framework for analyzing the trade-off between computational efficiency and accuracy.
Additionally, block-encoding and polynomial transformations provide modular implementation schemes and different-level of abstractions that can be adapted to different hardware platforms. These methods also offer a straightforward means of accommodating arbitrary input data and support the possibility of incremental implementations. Given the versatility of function approximation achievable within the QSP framework, these approaches can be applied to a broad range of practical problems.

Nonetheless, the potential advantages of block-encoding techniques and polynomial methods are currently constrained by the limited circuit sizes achievable on current quantum devices. These methods are, however, anticipated to become practical within Fault-Tolerant Application-Scale Quantum (FASQ) systems \cite{eisert2025mindgapsfraughtroad} and Fault-Tolerant Quantum Computing (FTQC)\cite{shor1997faulttolerantquantumcomputation}, thereby positioning block-encoding and polynomial transformations as strong candidate techniques in practical applications.

With these three properties being potentially addressed through ongoing research on block-encoding and polynomial transformations, further advances are likely to emerge from the development of new methods that advance the current state of the art. Figure \ref{fig1} illustrates the central concept discussed in this work, emphasizing various versions of polynomial transformations and block-encoding as central techniques for addressing different applications. 

Additionally, Figure \ref{fig1} presents a general workflow for both techniques. In the block-encoding framework, the system of interest is defined and subsequently encoded based on the available computational resources (e.g., number of qubits). For polynomial transformations, one specifies a target observable (e.g., energy) and the corresponding function that represents the computation (e.g., time evolution), which is then adapted using a suitable set of approximation methods. We note that the early version of block-encoding is also called ``qubitization"~\cite{Low_2019}, because after block-encoding each singular value or eigenvalue of the block-encoded matrix lives in a $2\times 2$ block. However, later generalizations of QSP methods can achieve a much broader class of computations. Therefore, in this work, we will use the combination of block-encoding and polynomial transform to refer to the broader class of related quantum algorithms. All of these methods are discussed in detail in Sections \ref{sec:block_encoding} and \ref{sec:polynomial-transformation}. 

\subsection{A Primer and Notations}

\begin{table*}[htbp!]
\centering
\small
\setlength{\tabcolsep}{4pt}
\renewcommand{\arraystretch}{1.2} % a bit more spacing for readability

\resizebox{\textwidth}{!}{%
\begin{tabular}{|c|c|c|c|}
\hline
\textbf{Quantum Circuit Representation} & \textbf{Equation} & \textbf{Description} & \textbf{Matrix} \\
\hline

% Row 1
\raisebox{-1.4\height}{\includegraphics[width=1.2cm]{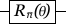}} &
\raisebox{-2.4\height}{$R(\theta,\phi,\lambda)$} &
\parbox[c]{7cm}{
General $SU(2)$ rotation on the Bloch sphere.\\
Any single-qubit gate can be synthesized using this rotation.\\
For $\hat{n}=(\sin(\theta)\cos(\phi),\sin(\theta)\sin(\phi),\cos(\theta))$.
} &
\raisebox{-1.2\height}{$\begin{bmatrix}
e^{i(\lambda+\phi)}\cos(\theta) & e^{i\phi}\sin(\theta) \\
e^{i\lambda}\sin(\theta) & -\cos(\theta)
\end{bmatrix}$} \\
\hline

% Row 2
\raisebox{-0.7\height}{\includegraphics[width=0.6cm]{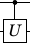}} &
\raisebox{-0.9\height}{$C_1 U = |0\rangle\langle0|\otimes I + |1\rangle\langle 1| U$} &
\parbox[c]{7cm}{
General control-1 $U$ gate.\\
This is a $4\times4$ matrix controlled by the first qubit when it is in state $|1\rangle$.
} &
\raisebox{-0.7\height}{$\begin{bmatrix} I & \mathbf{0} \\ \mathbf{0} & U \end{bmatrix}$} \\
\hline

% Row 3
\raisebox{-0.9\height}{\includegraphics[width=2.8cm]{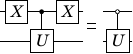}} &
\raisebox{-1.8\height}{$C_0 U = |1\rangle\langle 1|\otimes I + |0\rangle\langle 0| \otimes U$} &
\parbox[c]{7cm}{
General control-0 $U$ gate controlled by state $|0\rangle$.\\
The $X$ gates serve as an easy implementation trick based on a controlled-$U$ gate.
} &
\raisebox{-1.2\height}{$\begin{bmatrix} U & \mathbf{0} \\ \mathbf{0} & I \end{bmatrix}$} \\
\hline

% Row 4
\raisebox{-0.7\height}{\includegraphics[width=2.0cm]{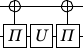}} &
\raisebox{-1.7\height}{$C_{\Pi}U = \Pi \otimes U + (I-\Pi) \otimes I$} &
\parbox[c]{7cm}{
General projector operator.\\
Apply $U$ onto the subspace $\Pi$.\\
For QSP subroutines this is usually the $|0\rangle \langle 0|$ projector.
} &
\raisebox{-0.5\height}{\bracketmatrix{U}{I}{\Pi}{\Pi}} \\
\hline

% Row 5
\raisebox{-1.0\height}{\includegraphics[width=3.8cm]{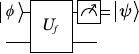}} &
\raisebox{-2.8\height}{$\langle\psi| U_f |\phi\rangle$} &
\parbox[c]{7cm}{
Measurement projection within a unitary implementation.\\
Extracts information in the subspace where block-encoding lives.
} &
\raisebox{-1.4\height}{$\begin{bmatrix}
\colorbox{lightgray}{$\langle \psi| U_f | \phi \rangle$} & \langle \psi| U_f | \phi_{\bot} \rangle \\
\langle \psi_{\bot}| U_f | \phi \rangle & \langle \psi_{\bot}| U_f | \phi_{\bot} \rangle
\end{bmatrix}$} \\
\hline

% Row 6
\raisebox{-0.8\height}{\includegraphics[width=2.6cm]{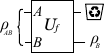}} &
\raisebox{-1.8\height}{\ensuremath{\rho_B = \text{Tr}_{A}(\rho_{AB})}} &
\parbox[c]{7cm}{
Partial trace over qubit register $A$.\\
Discard $A$ to obtain a density matrix on register $B$ rather than performing a projective measurement.
} &
\raisebox{-0.9\height}{$\rho_{AB} = \begin{bmatrix} \colorbox{lightgray}{$\rho_{00}$} & \rho_{01} \\ \rho_{10} & \colorbox{lightgray}{$\rho_{11}$} \end{bmatrix} \Rightarrow \rho_B = \rho_{00} + \rho_{11}$} \\
\hline

\end{tabular}%
}

\caption{Quantum gates and useful circuit implementations used in this work for block-encodings and polynomial transformations.}
\label{table1}
\end{table*}

We begin this section by introducing the foundational primitives necessary to understand both block-encoding and polynomial transformations necessary for the rest of the paper, summarized as a representative circuit in Table \ref{table1}. While many introductory references \cite{nielsen2010quantum, Djordjevic2021} offer more extensive and pedagogically detailed discussions, the focus here is limited to the key concepts most relevant to the analyses presented in subsequent sections of this work.

\textbf{Qubits:} The smallest unit of information in quantum computing is a qubit: a quantum analog of the classical bit. A qubit is a two-level system, whose state can be in a quantum superposition of $\ket{0}$ and $\ket{1}$:
\begin{equation}
    |\psi\rangle = \alpha |0\rangle + \beta|1\rangle,
    \alpha, \beta \in \mathbb{C},
\end{equation}
where $\mathbb{C}$ is the set of complex numbers, and $\ket{\cdot}$ is the Dirac bra-ket notation to represent a quantum state. A collection of many qubits can also exhibit other quantum properties such as entanglement. For an $n$-qubit state, the vector representation is a $2^n$-dimensional column vector, with the dimension doubling for each additional qubit,
$
    |\psi\rangle = \sum_{n=0}^{2^n-1} \alpha_n |n\rangle, \alpha_n \in \mathbb{C}.
$
The complex coefficients $\alpha_n$, known as probability amplitudes, satisfy the Born rule $\sum |\alpha_n|^2 = 1$.

\textbf{One-qubit Rotation:} Any single qubit gate $U$ can be represented as a general rotation over a given axis $\hat{n}$,
\begin{align}
    R_{\hat{n}}(\theta) = \exp(-i\frac{\theta}{2}(\hat{n} \cdot \vec{\sigma})),
\end{align}
where $\hat{n}=(n_x,n_y,n_z)$ is a unit vector and $\vec{\sigma} =(\sigma_x, \sigma_y, \sigma_z) = (X,Y,Z)$ denotes the Pauli matrices. The matrix representation is presented in Table \ref{table1}. Through the work we will use the $X,Y,Z$ convention. Only a small set of discrete gates are required to achieve universal quantum computation, as established by the Solovay-Kitaev theorem \cite{dawson2005solovaykitaevalgorithm}. A common universal set is $\{S, H, \mathrm{CNOT}\} + T$, where $S = \begin{bmatrix}
    1 & 0 \\ 0 & i
\end{bmatrix}$, $H = \frac{1}{\sqrt{2}}\begin{bmatrix}
    1 & 1 \\ 1 & -1
\end{bmatrix}$, $T = \begin{bmatrix}
    1 & 0 \\ 0 & e^{i \frac{\pi}{4}}
\end{bmatrix}$ 
are single-qubit gates, while $\mathrm{CNOT}$ is a two-qubit controlled-NOT gate, analogous to its classical counterpart.

\textbf{Multiqubit controlled-$U$.} A controlled-$U$ is illustrated in Table \ref{table1}, in which one qubit serves as the control and the other as the target. The operation gate works as follows: if the control qubit is in the $|1\rangle$, the unitary $U$ is applied to the target qubit, and the identity operation $I$ is applied if the qubit is in $\ket{0}$. In the context of polynomial transformations, as exemplified by Generalized-QSP in Section \ref{sec:single-poly-single-var}, it is often useful to define a controlled-$U$ operation conditioned on the $\ket{0}$ state instead of $\ket{1}$. This operation can be implemented by adding conjugated $X$ gates on the control qubit, as illustrated in Table \ref{table1}. In addition to two-qubit gates, multi-control multi-target quantum gates can be defined whose implementation is based on using as few two-qubit gates as possible, potentially supported by ancilla qubits \cite{PhysRevA.52.3457, 10293178}.

\textbf{Projector and projector-controlled operation.} The previous controlled operations can be generalized through the use of projection operators (also projectors). A projector $\Pi$ is defined as an operator where $\Pi^2 = \Pi$. Then a projector $\Pi$-controlled $U$ operation, i.e., $C_{\Pi}U$ can be defined that selectively apply $U$ to the subspace flagged or projected by $\Pi$. For instance, the controlled-$U$ in the previous paragraph is a special case where the projector $\Pi =|0\rangle\langle0|$(or $|1\rangle\langle1|$). This projection is essential for polynomial transformations, as it enables the application of a given operation to a specific subspace where the block-encoded matrix lives and the polynomial transformation happens. 

\textbf{Measurement and post-selection.} Measurement allows extraction of classical information from quantum computation. Measurement is particularly important for QSP methods, because the desired computation results are often only block-encoded, or exist in a subspace of the entire unitary. Therefore a measurement is necessary to extract or post-select the computational outcome. Sec. \ref{sec:block_encoding} will provide an example of this.

\section{Block-Encodings and Construction Techniques}
\label{sec:block_encoding}

Building on the motivation and notations in previous section, we now turn to the theory and practice for constructing block-encodings. Sec. \ref{ssec:dilation-be} introduces block-encoding (BE) from a general perspective of matrix dilation theory, presents a few explicit mathematical constructions of BEs, and highlight the importance of approximate BEs as computationally more efficient alternative to exact BEs. Sec. \ref{ssec:assemble-be} builds on these dilation theory and present methods to assemble multiple BEs together, including realizing addition, subtraction, multiplication of BEs. Sec. \ref{ssec:circ-be} give several explicit circuits for constructing and assembling BEs, most of which are known from previous work. Sec. \ref{ssec:soft-be} compiles existing software tools for generating explicit circuits for BEs and benchmarks their performance using simple molecular systems. 

\subsection{Block-Encoding and Unitary Matrix Dilation}
\label{ssec:dilation-be}

The idea of block-encoding is simple -- any amplitude or matrix where computation needs to be performed upon using quantum computers first has to be encoded inside a unitary matrix. Block-encoding refers to the process of encoding the amplitude or matrix as a block of a larger unitary matrix. This process necessarily requires introducing ancillary qubits due to the enlarged Hilbert space. 
Interestingly, this process of embedding a matrix into a larger matrix with some properties has been studied in applied math under a different name, \textit{matrix dilation} \cite{unitary-dilation}, independent of quantum computing development.

Specifically, for a general (can be non-normal) matrix $A \in \mathbb{C}^{M\times N}$, a \textit{unitary dilation} of $A$ is
\begin{align}
    U = \begin{bmatrix}
        A & * \\
        * & *
    \end{bmatrix},
\end{align}
such that $U$ is a unitary matrix. Ref. \cite{Low_2019} shows that $A$ has a unitary dilation if and only if $\lVert A \rVert \le 1$ (i.e., $A$ is a contraction). This means any matrix can be encoded as a sub-matrix of a unitary that can be later encoded onto a quantum computer using a quantum circuit. There are several different ways of producing a unitary dilation. More formal definitions deal with algebraic contractions of the matrix $A$ on a separable Hilbert space \cite{unitary-dilation}. In the following, we discuss some unitary dilations of the matrix $A \in \mathbb{C}^{N\times N}$ and the applicability of each of them.  

\textbf{Unitary Dilation of Hermitian Matrices.}
The most widely known unitary dilation is for Hermitian matrix. Given a Hermitian matrix $H$, one can construct a unitary matrix $U$ such that
\begin{equation}
    U = \begin{pmatrix}
        H/\alpha & * \\
        * & *
    \end{pmatrix}.
    \label{eq:u-be}
\end{equation}
The parameter $\alpha \geq \lVert H\rVert$ is a rescaling constant that is required for the unitarity of the whole matrix $U$, and $\lVert \cdot \rVert$ is the matrix norm. The choice of location as upper left block in $U$ of Eq. \eqref{eq:u-be} is only by convention. $H$ can be encoded in any other location of $U$ as well. 
Using the measurement projection from Table \ref{table1}, we can show that
\begin{equation}
    H/\alpha = (\langle0|\otimes I) U (|0\rangle \otimes I).
    \label{eq:lcu}
\end{equation}

\textbf{Unitary Dilation via Polar Decomposition.} For general square matrix $A$ that are not necessarily Hermitian, its polar decomposition $A = PV$ always exists, where $V\in \mathbb{C}^{N\times N}$ is unitary and $P$ is a positive semi-definite Hermitian matrix defined as $P = (A A^\dagger)^{1/2}$, also known as the left polar decomposition. Then we can prove that

\begin{align}
    U = \begin{bmatrix}
        A & (I - P^2)^{\frac{1}{2}} V \\
        -(I - P^2)^{\frac{1}{2}} V & A^{\dagger}
    \end{bmatrix} \in \mathbb{C}^{2N\times 2N}
\end{align}
is a unitary dilation of $A$ by simply verifying $U U^\dagger = U^\dagger U = I$.

As a special case, let $V=e^{i\theta}$ with $\theta \in [0,2\pi]$ as a global phase, the unitary dilation can be written as
\begin{align}
    U = \begin{bmatrix}
        A & e^{-i\theta}\sqrt{I - A A^{\dagger}} \\
        -e^{i\theta}\sqrt{I - A A^{\dagger}} & A^{\dagger}
    \end{bmatrix}
\end{align}
These different form for representing a matrix can be important in quantum circuit realization of (approximate) block-encodings. 

\textbf{Unitary dilation via Hermitian Dilation in $\mathbb{C}^{4N\times 4N}$.} Now we consider a slightly different way of dilating a general non-square (non-normal) matrix $A\in \mathbb{C}^{N\times N}$. We first construct a Hermitian dilation $H \in \mathbb{C}^{2N \times 2N}$. Next, a dilation of $H$ in $\mathbb{C}^{4N\times 4N}$ can be constructed using the previously developed method. This construction via an intermediate step of a Hermitian matrix enables us to use properties of Hermitian matrices even when the matrix $A$ is not Hermitian. This can sometimes be helpful to simplify our block-encodings.

One of the simple way to recast any matrix $A \in \mathbb{C}^{N\times N}$ (including non-normal) into a Hermitian matrix $H \in \mathbb{C}^{2N\times 2N}$ is by defining
\begin{align}
    H = 
    \begin{bmatrix}
        0 & A \\
        A^\dagger & 0
    \end{bmatrix}.
    \label{eq:hermitian}
\end{align}
In this way, we can use two additional ancilla to block-encode a non-normal matrix $A$. Moreover, the eigenvalues of $H$ always appears in pairs with opposite signs, and it can be shown that the eigenvalues of $H$ are actually $\pm \sigma(A)_j$ for $j=1,2,...,N$. This is a key relationship in connecting results for eigenvalues of Hermitian matrices to results for singular values of general matrices.

Building on our previous results on the block-encoding of a rescaled Hamiltonian $H / \alpha $, and by incorporating the auxiliary operator $\sqrt{I - H^2/\alpha^2}$, we can construct a unitary operator $U$ that serves as a block-encoding of an arbitrary matrix $A$ using Eq. \eqref{eq:hermitian}
\begin{align}
    U = 
    \begin{bmatrix}
    0 &A/\alpha  & P & 0 \\
    A^\dagger/\alpha  & 0 & 0 & Q \\
    P & 0 & 0 & -A/\alpha \\
    0 & Q & -A^\dagger /\alpha & 0
    \end{bmatrix},
\end{align}
where
\begin{align}
    P = \sqrt{I - A A^\dagger  / \alpha^2 }, ~~~
    Q = \sqrt{I -  A^\dagger A /\alpha^2}.
\end{align}
Here, $A$ is block-encoded in the $\ket{00}\bra{01}$ block. It can be moved to the diagonal block by simply multiply $I \otimes X \otimes I$ 
\begin{align}
    U' = U (I \otimes X \otimes I) = 
    \begin{bmatrix}
    A /\alpha & 0 & 0 & P\\
     0 & A^\dagger /\alpha  & Q & 0 \\
     0 & P & -A /\alpha  &  0 \\
    Q & 0 & 0 & - A^\dagger /\alpha 
    \end{bmatrix}.
\end{align}
This particular dilation of $A$ is important when one needs to have a BE $U$ that is also Hermitian. As an application, this particular approach can be useful for chiral symmetry in non-Hermitian systems \cite{PhysRevB.103.014111}. 

\textbf{Unitary Dilation with Minimum Dimension.} Let $\delta = {\rm rank} (I - A^\dagger A)$, so $0\le \delta \le n$, and $\delta = 0$ if and only if $A$ is unitary. It follows that ${\rm rank}(I - A A^\dagger) = \delta $. Since $A$ is a contraction, so both $I - A A^\dagger$ and $I - A^\dagger A$ are positive definite. This means there exist non-singular matrices $X, Y \in \mathbb{C}^{N\times N}$ such that
\begin{align}
    I - A A^\dagger = X \begin{bmatrix}
        I_\delta & 0 \\
        0 & 0
    \end{bmatrix} X^\dagger,
    \text{ and }
    I - A^\dagger A = Y^\dagger \begin{bmatrix}
        0 & 0 \\
        0 & I_\delta
    \end{bmatrix} Y,
\end{align}
and $I_\delta \in \mathbb{C}^{\delta \times \delta}$. Define
\begin{align}
    B &= X \begin{bmatrix}
        I_\delta \\ 0
    \end{bmatrix} \in \mathbb{C}^{N \times \delta},
    ~C = -\begin{bmatrix} 0 & I_\delta \end{bmatrix} Y \in \mathbb{C}^{\delta \times N}, \text{ and }\\
    D &= \begin{bmatrix} 0 & I_\delta \end{bmatrix} Y A^\dagger (X^\dagger)^{-1} \begin{bmatrix} I_\delta \\ 0\end{bmatrix} \in \mathbb{C}^{\delta \times \delta},
\end{align}
then
\begin{align}
    U = \begin{bmatrix}
        A & B \\
        C & D
    \end{bmatrix} \in \mathbb{C}^{N+\delta \times N+\delta}
\end{align}
is a unitary dilation of $A$. In fact, this is the unitary dilation of $A$ with the smallest dimension. 

\textbf{Unitary Dilation in General $\mathbb{C}^{(k+1)N \times (k+1)N}$ Space.} At the other end of the spectrum, it can be useful to construct a very large unitary dilation with a much enlarged large Hilbert space, as this will proven to be useful for assemble block-encodings.

Any unitary dilation in $\mathbb{C}^{2N \times 2N}$
\begin{align}
    U = \begin{bmatrix}
        A & Z_{12} \\
        Z_{21} & Z_{22}
    \end{bmatrix} \in \mathbb{C}^{2N \times 2N}
\end{align} 
developed above admits a simple generalization to the space of $\mathbb{C}^{(k+1)N \times (k+1)N}$ in the following way. Consider the block matrix $V = [V_{ij}]_{i,j=1}^{k+1}$ where each block is defined as $V_{11} = A, V_{12} = Z_{12}, V_{k+1, 1} = Z_{21}, V_{k+1, 2} = Z_{22}, V_{2,3} = V_{3,4} = ...=V_{k,k+1} = I$, and all the other blocks are zero:
\begin{align}
    V = \begin{bmatrix}
        A & Z_{12} & 0 & \cdots & 0 \\
        0 & 0 & I &   &\vdots \\
        \vdots & \vdots & & \ddots  & \vdots \\
        0 & 0 & 0 &  &I \\
        Z_{21} & Z_{22} & 0 & \cdots  & 0
    \end{bmatrix}.
    \label{highD-dilation}
\end{align}
It can be shown that $V$ is indeed a unitary dilation of $A$ for arbitrary $k\ge 2$.

\textbf{Approximate BEs.}
Previous methods implement an exact dilation of a matrix $A$. Nevertheless, depending on the size of the matrix and the sparsity the exact implementation can be resource-intensive to implement \cite{kuklinski2025efficientblockencodingsrequirestructure}. This inspires the study of approximate BEs. 

In general, for a given matrix $U_A$ that block-encode $A$, we said that $\tilde{U}_{A}$ is an approximation of $U_A$ with error $\epsilon$, if
\begin{equation}
    ||U_A-\tilde{U}_A|| < \epsilon.
\end{equation}
Such approximations have been studied using the Fast Approximate BLock-encoding (FABLE) algorithm that block-encodes an arbitrary matrix in $\mathbb{C}^{N\times N}$ using $O(N^2)$ one- and two-qubit gates \cite{9951292}. Modifications of the same algorithm have been implemented that perform better in sparse matrices and lose efficiency when there are symmetries in it, called S-FABLE and LS-FABLE \cite{sfablelsfable}. Recursive implementations have also been proposed to construct approximations that use smaller block-encodings, easy to implement, and utilize ancilla qubits for the efficient synthesis of block-encodings \cite{PhysRevA.102.052411}.

Recently, quantum-classical variational algorithms have also been studied for block encoding for applications in non-Hermitian dynamics and open quantum systems \cite{li2025variationalquantumalgorithmunitary}, suggesting the importance of efficient block-encoding methods for quantum simulations. Ideas from perturbation theory and similarity transform known in many-body physics and quantum chemistry can also be used to construct approximate block-encodings \cite{}. Introducing approximation in block-encodings has important consequences for error analysis of QSP algorithms, as we will discuss in Sec. \ref{ssec:tradeoff-poly}.

\subsection{Methods To Assemble Block-Encodings}
\label{ssec:assemble-be}

Once block-encodings are constructed for individual matrices, ways to assemble multiple block-encodings together will be important for scalable computation. In particular, elementary arithmetic such as addition, subtraction, and multiplication between block-encodings will be desired.

\textbf{Addition and Subtraction.} Given two block-encoding matrices of $A,B \in \mathbb{C}^{N\times N}$ as $U_A,U_B \in \mathbb{C}^{2N\times 2N}$, respectively, using the unitary dilation form, we can implement the addition or subtraction using the dilation of the form
\begin{equation}
    C_{U_A} = \begin{pmatrix}
        U_A & 0 \\ 0 & I
    \end{pmatrix}, ~~~~~
    C_{U_B} = \begin{pmatrix}
        I & 0 \\ 0 & U_B
    \end{pmatrix}
\end{equation}
where $C_{U_A}, C_{U_B}$ are unitary. By multiplication of these matrices and conjugation using $(H \otimes I)$ we have
\begin{equation}
    (H\otimes I ) C_{U_A} C_{U_B} (H\otimes I ) = \begin{bmatrix}
        (U_A+U_B)/2 & (U_A-U_B)/2 \\
        (U_A-U_B)/2 & (U_A+U_B)/2
    \end{bmatrix},
\end{equation}
which allows us to implement addition on the upper left corner and subtraction on the upper right.

This process can be implemented iteratively to sum many matrices, in similar spirit as the linear combination of unitaries method \cite{childs2012hamiltonian}. If there is a constant scaling factor $\alpha$ for each matrix as in Eq. \eqref{eq:u-be}, the result will also apply but only change the success probability to get the proper result. This success probability can be boosted using recent techniques of amplitude amplification \cite{martyn2023efficient}.

\textbf{Multiplication and Power.} Using the results from the previous dilations, we can combine two different block-encoding to form the multiplication of two matrices, given $A, B \in \mathbb{C}^{N\times N}$ in the general dilation on the $ \mathbb{C}^{(k+1)N \times (k+1)N}$ space for $k=2$
\begin{align}
    V_A = \begin{bmatrix}
        A & Z_{12} & 0 \\
        0 & 0 & I \\
        Z_{21} & Z_{22} & 0
    \end{bmatrix}, \text{ and }
    V_B = \begin{bmatrix}
        B & Y_{12} & 0 \\
        0 & 0 & I \\
        Y_{21} & Y_{22} & 0
    \end{bmatrix}.
\end{align}
It is easy to verify that
\begin{align}
    V_A V_B = \begin{bmatrix}
        AB & AY_{12} & Z_{12} \\
        Y_{21} & Y_{22} & 0 \\
        Z_{21} B & Z_{21} Y_{12} & Z_{22}
    \end{bmatrix},
\end{align}
which means that we can form the multiplication of $AB$ by simple multiplying their block-encodings together. The particular block-encoding can be implemented using an architecture based on qutrits. 

In addition, this method can be generalized to multiplication of $p$ matrices. Given $A_1, A_2, ..., A_p \in  \mathbb{C}^{N\times N}$, and given their dilation as $V_1, V_2, ..., V_p \in \mathbb{C}^{(p+1)N \times (p+1)N}$. Therefore, what we have
\begin{align}
    V_1 V_2 \cdots V_p = \begin{bmatrix}
        A_1 A_2 \cdots A_p & * \\
        * & *
    \end{bmatrix} \in \mathbb{C}^{(p+1)N \times (p+1)N}
\end{align}
is a unitary dilation of $A_1 A_2 \cdots A_p$. Note that the total resource requirement for encoding each $V_j$ is $\log\big[ (p+1)n \big]$, where the number of ancilla qubits is $\log(p+1)$. Observe that the number of ancilla qubits needed scales only in a logarithmic form as the number of matrices which are being multiplied, as also pointed out by a recent work \cite{vasconcelos2025methodsreducingancillaoverheadblock}. This is in analogous to the case of the LCU approach (where unitaries are summed up to produce the block-encoding of sums of matrices instead of products), where the number of ancilla qubits is also equal to $\log(p)$ where $p$ is the total number of unitaries that is summed up.

\subsection{Quantum Circuit Realization}
\label{ssec:circ-be}

There are many ways to realize a block-encoding circuits based on the properties of the matrix being block-encoded. Designing efficient implementation of BE circuits itself is an active area of research \cite{vasconcelos2025methodsreducingancillaoverheadblock,yuan2025cobblecompilingblockencodings,kuklinski2025efficientblockencodingsrequirestructure}. The problem is challenging as the structure of the matrix being block-encoded has to be properly exploited by quantum circuits to achieve efficient block-encoding \cite{kuklinski2025efficientblockencodingsrequirestructure}. The structure of a matrix is often basis-dependent, which makes it even more challenging. Some algorithms leverage the structure of the sparsity matrix to perform efficient block-encoding \cite{doi:10.1137/22M1484298, 9951292}. We review some existing methods for constructing BE circuits.

\begin{figure}[H]
    \centering
    \includegraphics[width=0.3\textwidth]{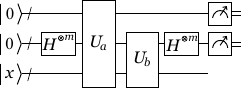}
    \caption{Circuit to block-encode a sparse matrix $A$.}
    \label{fig:sp-be}
\end{figure}

\textbf{Block-encoding of Sparse Matrices.}\cite{doi:10.1137/22M1484298} For sparse and well-structured matrices, we can block-encode a matrix $A \in \mathrm{C}^{N \times N}$, where $N = 2^n$ and $n$ the number of qubits, given a sequence of operators $(I\otimes H^m \otimes I) U_a (I \otimes U_b) (I\otimes H^m \otimes I)$ in the following way, depicted as a quantum circuit in Fig. \ref{fig:sp-be}.

The $U_H$ operation applies a set of Hadamard gates $H^{\otimes m}$ to the second register, where $m$ is the register size. The second register implements the encoding of the $(x,y)$ matrix element of $A$, $A_{x,y}$ as the following
\begin{equation}
    U_a ( |0\rangle |x\rangle |y\rangle ) = (A_{x,y}|0\rangle + \sqrt{1-|A_{x,y}|^2}|1\rangle )|x\rangle |y\rangle 
\end{equation}

The third register implements the $\text{mod}$ operation to the two registers $U_b(|x\rangle|y\rangle) = |x\rangle | \text{ mod }((x+y),2) \rangle $. While this implementation provides an exact, step-by-step circuit construction for building a block encoding, the following method can be considered easier to construct and more intuitive. However, for sparse matrices, the current implementation remains more efficient.

\begin{figure}[H]
    \centering
    \includegraphics[width=0.2\textwidth]{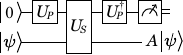}
    \caption{Circuit to assemble the linear combination of unitaries.}
    \label{fig:LCU}
\end{figure}

\textbf{Linear Combination of Unitaries. }One naive way to construct BEs is the Linear Combination of Unitaries (LCU) method \cite{childs2012hamiltonian}. For this technique, each operator $A$ can be decomposed as a sum of unitary operators $U_i$, such as
\begin{equation}
    A = \sum_{i}^{N} \alpha_i U_i
    \label{eq:LCU}
\end{equation}
where $\alpha_i$ are the coefficients. This circuit can be implemented as depicted in Fig. \ref{fig:LCU}, using two operators, called select and prepare, denoted as $U_S$ and $U_P$, respectively.

The non-unitary operator $A$ is now block-encoded, for most cases, in the upper-left corner. Then after a post-selection of the $|0\rangle^{m}$ state, refer to Table \ref{table1}, where $m$ is the extra Hilbert space necessary to block-encode the system ($m=\log_2(N)$ for $A$ given in Eq. \eqref{eq:LCU}), we can implement the non-unitary operator for a given initial state $|\psi_i \rangle$. Now we will explain each implementation.

\textbf{Prepare circuit.} The purpose of the prepare state is to implement a quantum state that encodes all the decomposition coefficients of the matrix $A$, defined as:
\begin{equation}
    U_P|0\rangle = \sum_{i=0}^{N}\sqrt{\frac{|\alpha_i|}{\lambda}} |i\rangle \label{prep-be}
\end{equation}
Here, $\lambda = \sum_i |\alpha_i|$ denotes the normalization factor. The main idea is to encode each coefficient and select the corresponding unitary operator based on its value. 

\begin{figure}[H]
    \centering
    \includegraphics[width=0.45\textwidth]{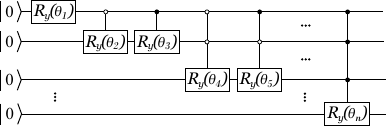}
    \caption{Circuit to implement the prepare operation $U_P$. Here the angles $\theta_i$ can be efficiently calculated for a given state.}
    \label{fig:state_preparation}
\end{figure}

Various approaches exist for implementing the prepare operation, which refers to the method of state preparation employed in quantum circuit applications such as Quantum Phase Estimation (QPE). In this work, for illustrative purposes, we employ the \textit{"Divide-and-Conquer"} method \cite{araujo2021divide} that allows us to implement the LCU. The prepare circuit is depicted in Fig. \ref{fig:state_preparation}.

The inverse prepare circuit can be calculated by flipping the initial circuit state and changing the phase of the $R_y(\theta)$ gates, following the inverse operation in Table \ref{table1}.

In Sec. \ref{ssec:soft-be} we present a compendium list of software to implement block-encodings. 

\textbf{Select circuit.} The select operation $U_S$ apply a given unitary $U_i$ to the state $|\psi\rangle$, that means:
\begin{equation}
    U_S|k\rangle |\psi\rangle = |k\rangle U_k|\psi\rangle \label{select-be}
\end{equation}
The following quantum circuit depicts the implementation of the select operation using multi-controlled $U_k$ gates. 

\begin{figure}[H]
    \centering
    \includegraphics[width=0.25\textwidth]{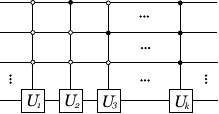}
    \caption{Circuit to assemble the select operation using multi-control gates.}
    \label{fig:select_operation}
\end{figure}

The previous implementation of encoding each qubit position in a qubit gate is known as a multiplex gate with $M$ controls, which will have $2^M$ possible controlled-unitary gates for a different $U$ implementation. In this case each $U$ represent a Pauli string that represents a gate. More efficient methods for decomposing these quantum gates have been proposed in the literature \cite{10293178} that requires a decomposition for a $n$-qubit multi-controlled $SU(2)$ gate proportional $20n$. 

\textbf{Projecting to the correct sector.} The end of the implementation will have the matrix $A$ by projecting the state $|0\rangle$ to the  block-encoding of $A$:
\begin{equation}
    A = \alpha (\langle 0| \otimes I )(U_P^{\dagger}  U_S U_P) (|0\rangle \otimes I)
    \label{lcu-proj}
\end{equation}

While the LCU method can be used to encode a Hamiltonian into a larger Hilbert space, its effectiveness strongly depends on the success probability of measuring the $|0\rangle \langle0|$ subspace. This probability decreases as the norm of the encoded matrix increases. To address this limitation, techniques that amplify the success probability have been developed, such as oblivious amplitude amplification \cite{10.1145/2591796.2591854}. This technique requires that the matrix $A^{-1}=A^{\dagger}$ be antisymmetric for the amplification  to be possible. Usually this condition is possible for the majority of the systems.

LCU is known for its conceptual and implementation simplicity. Moreover, the LCU method is not just reserved for block-encoding methods; it can also serve as the main tool for the simulation of the time evolution operator by expressing the exponential term as a Taylor series and implementing it usually on an LCU structure. This approach will be explained in more detail in Sec. \ref{ss:RTE}, also in the context of Hamiltonian simulation. 

% One of the most important operators for physical and chemical systems is the time evolution of a Hamiltonian $H$. Such evolution can be carried out with the time evolution operator for a time-independent case as $U = \exp(-i Ht)$, where we assume that Planck's constant is the unit $(\hbar = 1)$ and that we do not have any dissipation or injection of energy, i.e., we have a closed system. This operator, together with any other non-unitary operators, can be hard using Taylor expansion and LCU methods as explained in Sec \ref{sec:app}. 

\subsection{Software to construct block-encodings and Benchmarking}
\label{ssec:soft-be}

In this section, we present the quantum software frameworks that support BE methods. 
\emph{Approximate block-encoding} techniques such as FABLE, S-FABLE, and LS-FABLE, have been implemented by the Quantum Computing Lab~\cite{fable_repo} and are accessible through \emph{PennyLane}~\cite{pennylane_fable}, providing a dedicated \emph{FABLE} operator and tutorials for constructing block-encoded unitaries. 

\emph{Variational (quantum--classical) block-encoding} approaches are supported by \emph{Classiq}'s hybrid variational quantum linear solver (\emph{VQLS}) modules~\cite{classiq_vqls}, using classical optimization together with \emph{LCU} quantum circuits. 
This provides a flexible variational route to approximate block-encoded representations without requiring full ancilla-based unitaries. 

For methods based on \emph{LCU}, many platforms offer explicit implementations. 
\emph{OpenFermion}~\cite{openfermion_lcu} implements \emph{lcu\_util} for Hamiltonian synthesis; 
\emph{Riverlane}'s \emph{pauli\_lcu}~\cite{riverlane_lcu} provides Pauli-term combinations; 
\emph{Google Qualtran}~\cite{qualtran_docs} enables chemistry resource estimation through block-encoded \emph{LCU}s; 
and both \emph{PennyLane}~\cite{pennylane_lcu} and \emph{Qrisp}~\cite{qrisp_lcu} supply \emph{LCU} primitives. 
\emph{Classiq}~\cite{classiq_vqls} additionally supports hybrid extensions. 

\emph{Sparse-matrix} block-encodings have been demonstrated in the \textit{Explicit-Block-Encodings} repository~\cite{explicit_be}, while supporting frameworks such as \emph{Sequential-Quantum-Gate-Decomposer}~\cite{sqgd_repo}, 
\emph{QGOpt}~\cite{qgopt_docs}, and \emph{GateDecompositions.jl}~\cite{gate_decomp_jl} focus on efficient gate synthesis and multiplexor optimization. 
Additional primitives for \emph{divide-and-conquer state preparation}~\cite{dcsp_repo}, \emph{unary iteration}~\cite{unary_iteration}, and \emph{amplitude-amplification} appear in \emph{PennyLane}~\cite{pennylane_amp_amp_demo}, \emph{Classiq}~\cite{oaa_docs}, and \emph{Qrisp}~\cite{qrisp_amp_amp_docs}. 
For \emph{Taylor-series LCUs} and time-evolution applications, \emph{Quantinuum/CQCL}'s \textit{qtnm-tts}~\cite{qtnm_tts} and \textit{QITE}~\cite{qite_repo} repositories offer open implementations. All referenced toolkits are summarized in Table~\ref{tab:software_BE}. 

\begin{table}[h!]
\centering
\small
\setlength{\tabcolsep}{4pt}
\renewcommand{\arraystretch}{1.1}
\caption{Summary of publicly available SDKs and repositories implementing block-encoding techniques.}
\label{tab:software_BE}
\resizebox{\columnwidth}{!}{
\begin{tabular}{|c|c|}%{p{5 cm}|p{3cm}}

  \hline \hline
\textbf{Method} & \textbf{Associated SDKs / Frameworks} \\
  \hline

Approximate block-encoding &  \\ (FABLE / S-FABLE / LS-FABLE) & FABLE~\cite{fable_repo}, PennyLane~\cite{pennylane_fable} \\[4pt]
  \hline
Variational (quantum--classical) BE & Classiq VQLS module~\cite{classiq_vqls} \\[4pt]
  \hline
Sparse-matrix BE & Explicit-Block-Encodings~\cite{explicit_be} \\[4pt]
  \hline
LCU (Linear Combination of Unitaries) & OpenFermion~\cite{openfermion_lcu}, Riverlane~\cite{riverlane_lcu}, Qualtran~\cite{qualtran_docs}, \\ &PennyLane~\cite{pennylane_lcu}, Qrisp~\cite{qrisp_lcu}, Classiq~\cite{classiq_vqls} \\[4pt]
  \hline
Divide-and-Conquer state-prep & DCSP~\cite{dcsp_repo} \\[4pt]
  \hline
Unary iteration for multi-controls & PennyLane~\cite{unary_iteration} \\[4pt]
  \hline
Efficient gate decomposition  & Sequential-Quantum-Gate-Decomposer~\cite{sqgd_repo}, \\ / multiplex optimization & QGOpt~\cite{qgopt_docs}, GateDecompositions.jl~\cite{gate_decomp_jl} \\[4pt]
  \hline
Amplitude / Oblivious amplitude amplification & PennyLane~\cite{pennylane_amp_amp_demo}, Classiq~\cite{oaa_docs}, Qrisp~\cite{qrisp_amp_amp_docs} \\[4pt]
  \hline
Taylor-series LCU (time evolution) & qtnm-tts~\cite{qtnm_tts}, QITE~\cite{qite_repo} \\
  \hline \hline

\end{tabular}
}
\end{table}

In order to compare the performance of different quantum software frameworks, without any integrated circuit optimization, in implementing block-encoded Hamiltonians, we examined the gate statistics obtained for the H$_2$ molecular system. 
Block-encoded unitaries representing the H$_2$ Hamiltonian were constructed and decomposed using three major platforms: \emph{Qiskit}, \emph{PyTKET}, and \emph{Cirq}. 
The interatomic distance was fixed at $r = 0.5~\text{\AA}$, and one- and two-electron integrals were calculated with the \texttt{STO--3G} basis using \texttt{PySCF}. We perform BE for both first- and second-quantization representation.
For this minimal basis, there are only two molecular orbitals ($\sigma_g$ and $\sigma_u$), giving two-electrons each with four spin-orbitals. Under first-quantization, this is mapped to 4 qubits. 
For the second-quantized case, the Hamiltonian operators were mapped to qubits with both the Jordan--Wigner and Bravyi--Kitaev transformations and normalized by their spectral norm $\alpha = \lVert H \rVert$ to ensure unitary block-encodings. 

Each normalized Hamiltonian was simulated as $\tilde{U}_H = e^{-iHt}$ with $t = 1.0$~a.u. using the prepare $U_P$ and select $U_S$ construction described in Sec.\ref{ssec:circ-be}. 
Gate counts were extracted automatically after decomposition to the $\{R_Z, R_X, R_Y, \mathrm{CNOT}\}$ basis and before any circuit optimization passes. 
Since \emph{PyTKET} and Cirq do not directly support large ($32\times32$) unitaries, the full block-encoding was recursively divided into sixteen $8\times8$ sub-unitaries using the cosine--sine decomposition (\emph{CSD}), and the total counts were summed. 
All simulations were performed in \emph{Python~3.12} on \emph{Ubuntu~22.04}, and all scripts, integral data, and notebooks are provided in our GitHub repository for reproducibility~\cite{ncsuQCSrepo}.

\begin{table}[h!]
\centering
\small
\caption{Gate counts for block-encoded unitaries of the H$_2$ molecular Hamiltonian under first- and second-quantization (Jordan--Wigner and Bravyi--Kitaev) mappings. 
Note that the asterisk ($^*$) denotes frameworks that do not directly support decomposition of the full $5$-qubit block-encoded unitary. 
Therefore, the unitary was recursively decomposed into sixteen $8\times8$ ($3$-qubit) subunitaries using the cosine--sine decomposition (CSD), and total gate counts were obtained by summing over all subcircuits. 
}
\begin{tabular}{lcccc}
\toprule
\textbf{Framework} & \textbf{R$_Z$} & \textbf{R$_X$} & \textbf{R$_Y$} & \textbf{CNOT} \\
\midrule
\multicolumn{5}{l}{\textbf{First Quantization}} \\

Qiskit\cite{aleksandrowicz2019qiskit} & 896 & 528 & 20 & 235 \\
PyTKET$^*$\cite{sivarajah2020tket} & 1297 & 1570 & 2552 & 342 \\
Cirq$^*$\cite{cirq_developers2023cirq} & 903 & 392 & 258 & 282 \\
\midrule
\multicolumn{5}{l}{\textbf{Second Quantization (Jordan--Wigner)}} \\

Qiskit & 2217 & 1478 & -- & 423 \\
PyTKET$^*$ & 1300 & 1587 & 2519 & 343 \\
Cirq$^*$ & 812 & 364 & 253 & 297 \\
\midrule
\multicolumn{5}{l}{\textbf{Second Quantization (Bravyi--Kitaev)}} \\

Qiskit & 915 & 532 & 20 & 236 \\
PyTKET$^*$ & 1313 & 1554 & 2524 & 346 \\
Cirq$^*$ & 843 & 373 & 252 & 298 \\
\bottomrule
\end{tabular}
\vspace{5pt}

\end{table}

\section{Polynomial Transforms and Angle Finding}
\label{sec:polynomial-transformation}

\begin{figure*}[htbp!]
    \centering
    \includegraphics[width=1.0\textwidth]{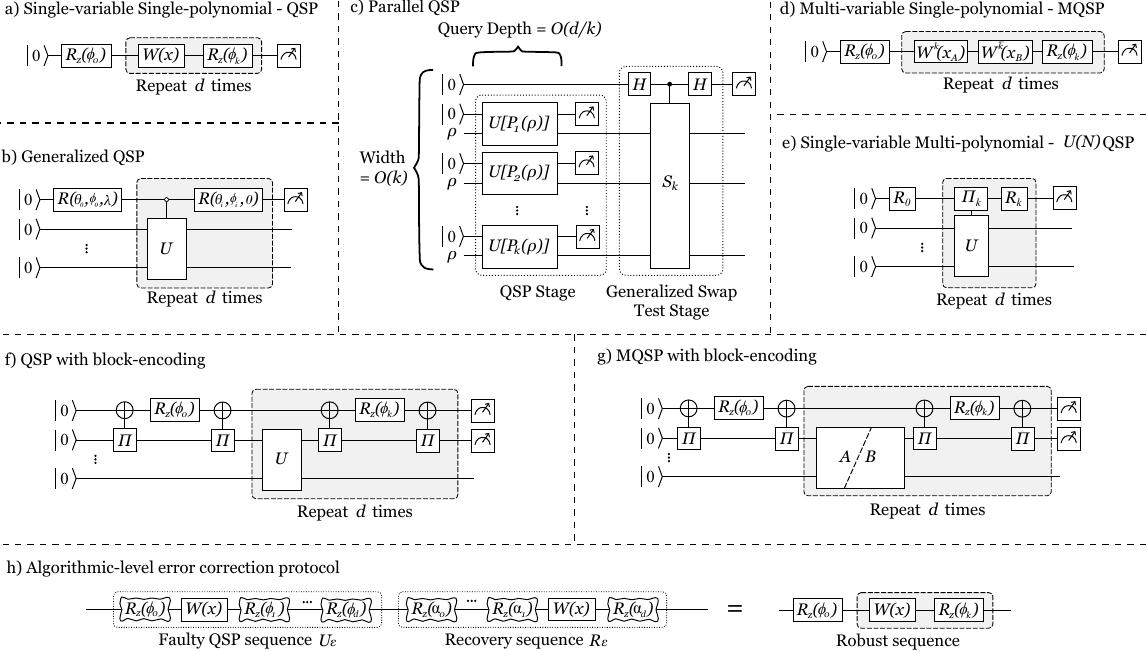}
    \caption{Zoo of polynomial transformations realized by QSP algorithms. The original formulation of single-variable single-polynomial transform \textbf{(a)} and the generalized QSP \textbf{(b)} that remove the parity constraint on the polynomials. More recent development of parallel QSP that is suitable on parallel quantum architectures \textbf{(c)}. Generalization to the multi-variate polynomial as in M-QSP \textbf{(d)} and multiple polynomials of a single-variable as in $U(N)$ QSP \textbf{(e)}. Circuit diagrams of normal QSP \textbf{(f)} and M-QSP \textbf{(g)} explicitly with projectors and additional ancilla to impart the signal processing rotations. In M-QSP the quantum gate has a probability to be the operator $A$ or $B$. For structured noise, a recovery QSP sequence can be concatenated after a faulty QSP to obtain a robust QSP sequence that suppresses overall error at the algorithmic level \textbf{(h)}. The curved boundary boxes represent faulty signal processing gates. The corresponding QSVT circuit can be constructed similarly by using two different projectors in the circuit instead of one. 
    }
    \label{fig2}
\end{figure*}

One way to represent any computation on the BEs is via a polynomial. Over the past decade, QSP algorithms have been developed to effectively realize desired polynomial transformations on the BEs \cite{PhysRevLett.118.010501,PRXQuantum.5.020368,10.1145/3313276.3316366, PRXQuantum.2.040203,dong2024feedforward,chakraborty2025quantumsingularvaluetransformation,laneve2024quantumsignalprocessingsun,lu2024quantumsignalprocessingquantum,Martyn_2025,dong2024feedforwardquantumsingularvalue}. In this section, we review existing art to perform polynomial transforms. We start from QSP and GQSP that can perform a single polynomial transform of one variable in Sec. \ref{sec:single-poly-single-var}. This is followed by the $U(N)$-QSP algorithm that can perform multiple polynomial transforms simultaneously of a single variable in Sec. \ref{ssec:un-qsp}. Sec. \ref{ssec:mqsp} discusses algorithms that can perform a polynomial transform of multiple commuting or non-commuting variables. Sec. \ref{ssec:poly-assemble-alec} presents ways to assemble polynomials together as well as considerations of building robustness into QSP algorithms, i.e., the idea of algorithmic-level error correction (ALEC). We continue the topic of robust QSP with a discussion of error tradeoffs between BEs and polynomial transforms in Sec. \ref{ssec:tradeoff-poly}. Finally, Sec. \ref{ssec:soft-poly} reviews existing software tools for QSP angle finding. Fig. \ref{fig2} presents an overview of all the quantum circuits for polynomial transformations discussed in this work.

\subsection{Single polynomial transformation of a single variable}
\label{sec:single-poly-single-var}

The standard form of QSP interleaves two orthogonal single-qubit rotations, the signal operator $W(x)$ and the signal processing operator $S(\phi)$. The signal operator is fixed and rotates the qubit by the same angle in every iteration, while the signal processing operator are varied to control the shape of the polynomial transformation applied as illustrated in Fig. \ref{fig2}a.

If $W(x) = e^{i\frac{\theta}{2}X}$ is a $x$-rotation operator with rotation angle $\theta = -2 \cos^{-1}({x})$, then the signal operator can be expressed as
\begin{equation}
    W(x) 
    = \begin{pmatrix}
        x & i\sqrt{1-x^2} \\
        i\sqrt{1-x^2} & x
    \end{pmatrix}.
\end{equation}
The signal processing operator $S(\phi)$ can be a $z$-rotation given by
\begin{equation}
    S(\phi) = e^{i \phi Z}
\end{equation}
which rotates the qubit by angle $-2\phi$.
For a sequence of tuple $\vec{\phi} = \{\phi_0, \phi_1, \ldots, \phi_d\} \in \mathbb{R}^{d+1}$, the overall unitary operation on the qubit $U_{\vec{\phi}}(x)$ can be expressed as 
\begin{equation}
    U_{\vec{\phi}}(x) = S(\phi_0) \prod_{k=1}^{d} W(x) S(\phi_k)
\end{equation}
The block-form realized by $U_{\vec{\phi}}(x)$ is of the form,
\begin{equation}
    U_{\vec{\phi}}(x) = \begin{pmatrix}
        P(x) & i Q(x)\sqrt{1-x^2} \\
        iQ^*(x)\sqrt{1-x^2} & P(x)
    \end{pmatrix}
    \label{eq:qsp}
\end{equation}
where, $P(x)$, and $Q(x)$ are polynomials of $x$ that satisfy the following conditions:
\begin{enumerate}
    \item (expressivity) deg($P$) $\leq$ $d$, deg($Q$) $\leq$ $d-1$,
    \item (parity) $P(x)$ has parity $d$ mod $2$,
    \item (unitarity) $|P(x)|^2 + (1-x^2)|Q(x)|^2 = 1$, $\forall$ $x \in [-1,1]$.
\end{enumerate}

The current approach implements a polynomial transformation for a single variable $x$. Additionally, connections have been found with space-time dual quantum circuits and Lorentz transformations when considering a complex parameter \cite{bastidas2024complexification}. While useful, most practical applications require applying it to a matrix. In order to do that, it is necessary to implement a block-encoding of the desired matrix and project over the subspace where the matrix lies in the expanded Hilbert space \cite{Low_2019}. Fig. \ref{fig2}f depicts the circuit implementation.

Extending the work of QSP, Generalized Quantum Signal Processing (GQSP) \cite{PRXQuantum.5.020368} proposed to replace the fixed-axis qubit signal processing operator with an arbitrary $SU(2)$ rotation illustrated in Fig \ref{fig2}b. 
The signal operator in the standard GQSP is given by $A = \begin{bmatrix}
    U & 0
    \\
    0 & \mathbb{I} 
\end{bmatrix}$ which is a $\ket{0}\bra{0}$-controlled application of the desired unitary operator $U$ of the form $U =e^{ix},~x\in \mathbb{R}$, as shown in Table \ref{table1}.
The signal processing operator given by
\begin{align}
     R(\theta, \phi, \lambda) = 
\begin{bmatrix}
    e^{i(\lambda+\phi)}\cos{(\theta)} & e^{i\phi}\sin{(\theta)}
    \\
    e^{i\lambda}\sin{(\theta)} & -\cos{(\theta)}
\end{bmatrix} \otimes \mathbb{I}
\end{align}
is an arbitrary $SU(2)$ rotation tensored with identity operator $\mathbb{I}$.
Interleaving $A$ and $R(\theta_k, \phi_k, \lambda_k)$. It is just necessary to define one $\lambda$ parameter for the sequence that realizes the following block form,
\begin{equation}
    \left( \prod_{k=1}^{d} R(\theta_k, \phi_k, 0) A \right) R(\theta_0, \phi_0, \lambda) = \begin{bmatrix}
        P(U) & * \\
        Q(U) & * 
    \end{bmatrix}
    \label{eq:gqsp}
\end{equation}

The constraints on the polynomials thus realized are 
\begin{enumerate}
    \item $P, Q \in \mathbb{C}[x]$, 
    \item $\text{deg}(P), \text{deg}(Q) \leq d$
    \item $\forall x \in \mathbb{R}, |P(e^{ix})|^2 + |Q(e^{ix})|^2 = 1$.
\end{enumerate}

This generalization lifts the second parity constraints imposed by QSP and can realize Laurent polynomials with complex coefficients as well, unlike QSP, which can only realize Laurent polynomials with real coefficients. GQSP has also been proven to have connections with other areas that could potentially be applied to different fields, such as the nonlinear Fourier transform \cite{laneve2025generalizedquantumsignalprocessing}. Moreover, in applications like Hamiltonian simulation, it has been shown that GQSP can double the efficiency of the implementation over traditional methods \cite{PhysRevA.110.012612}.

\subsection{Multiple polynomial transformations of a single variable}
\label{ssec:un-qsp}

In QSP \eqref{eq:qsp} and GQSP \eqref{eq:gqsp}, a single control qubit is used to implement polynomial transformations on the eigenvalues of a target unitary $U \in U(2)$, yielding a unitary of the form  
\begin{align}
    \begin{bmatrix}
P_{0,0}(U) & P_{0,1}(U) \\
P_{1,0}(U) & P_{1,1}(U)
\end{bmatrix},
\end{align}
where each entry $P_{jk}(U)$ is a (real for QSP, complex for GQSP) polynomial in $U$. These techniques realize two polynomials $P$ and $Q$ (under some constraints) that transform a given unitary $U$ (the signal). This is useful, but the polynomials realized are still only two. 

By introducing an ancilla with $n$-qubits (instead of 1), the QSP algorithm can be generalized to simultaneously realize $N=2^n$ polynomial transformations of a single variable $U$ in one circuit. This has recently been accomplished by the so-called $U(N)$-QSP \cite{lu2024quantumsignalprocessingquantum,  laneve2024quantumsignalprocessingsun} algorithm as depicted in Fig. \ref{fig2}e. More specifically, given any unitary $U$ as a signal that one would like to transform over, it is possible to construct a quantum circuit that uses $L$ calls a projector controlled-$U$ operation $C_\Pi U$ (block-encoding of $U$)
\begin{align}
    \textbf{P}(U) = \left[ \prod_{l=1}^L R_l C_{\Pi}U \right] V_0 
\end{align}
that realizes $N$ degree-$L$ polynomial transformations over $U$ simultaneously
\begin{align}
    \textbf{P}(U) = \begin{bmatrix}
    P_{0,0}(U) & P_{0,1}(U) & \cdots & P_{0, N-1}(U)\\
    P_{1,0}(U) & P_{1,d1}(U) & \cdots & P_{1, N-1}(U) \\
    \vdots & \vdots & \ddots & \vdots \\
    P_{N-1,0}(U) & P_{1,1}(U) & \cdots & P_{N-1, N-1}(U)
    \end{bmatrix},
\end{align}
for a given polynomial \emph{matrix} $ \textbf{P}(z) = \{ P_{jk}(z) \}$ .
The shape of the polynomials are controlled by the unitary operations on the $n$-qubit ancilla $R_l$. The degree of each such polynomial $\{P_{jk}(U)\}$ is no more than $L$\cite{lu2024quantumsignalprocessingquantum}.

Similarly, if given a unitary $U$ and complex polynomial matrix $\textbf{P}(z)$ with constraints on its singular values, it is possible to construct a quantum circuit with $L$ calls to controlled-$U$ to realize a polynomial matrix $\textbf{P}(U)$. The $U(N)$ QSP and QSVT are useful in implementing a more general polynomial transformation framework with applications in quantum amplitude estimation and encoding multivariate functions.

\subsection{One polynomial multi-variables}
\label{ssec:mqsp}

Many important problems involve functions of multiple variables. Multivariate quantum signal processing provides a framework to address such problems. The operators involved in such protocols can be either commuting operators corresponding to the abelian case \cite{Rossi2022multivariable}, or non-commuting operators corresponding to the non-abelian case \cite{singh2025nonabelianquantumsignalprocessing,11129874}.

The bivariate form of the abelian M-QSP can be useful for interrogating joint properties of two commuting signals $x_A$ and $x_B$ encoded in operators $A(x_A) = e^{i\cos^{-1}(x_A)X}$ and $B(x_B) = e^{i\cos^{-1}(x_B)X}$ respectively. Such a protocol of length $d$ can be defined by a length-$d$ binary string $s \in \{ 0, 1\}^d$ and a set of phases $\Phi = \{\phi_0, \phi_1 \cdots, \phi_d\} \in \mathbb{R}^{d+1}$ such that the desired unitary $U_{(s, \Phi)}(x_A, x_B)$ can be realized by a sequence of operators given by\cite{rossi2021mqsp}
\begin{equation}
    U_{(s, \Phi)}(x_A, x_B) = e^{i\phi_0 Z} \prod_{k=1}^{d} A^{s_k}(x_A)B^{1-s_k}(x_B)e^{i\phi_k Z}
    \label{eq:m-qsp}
\end{equation}
where $ x_A, x_B \in [-1, 1]^2$ as illustrated in Fig. \ref{fig2}d and the generalization for matrices in Fig. \ref{fig2}g. The M-QSP ansatz in Eq. \eqref{eq:m-qsp} can generate complicated multivariable transformations of eigenvalues of commuting variables, finding applications in multi-channel discrimination problems \cite{rossi2021mqsp} and providing a pathway for coherent control of the dynamics of multiple commuting subsystems efficiently.

One important application of the non-abelian QSP is realizing polynomial transformations on systems involving the position $\hat{x}$ and momentum $\hat{p}$ quadratures of a qumode (oscillators) coupled to a qubit \cite{11129874, singh2025nonabelianquantumsignalprocessing, liu2024hybrid}. Given access to $X$-rotations parameterized by a set of phases $\{ \phi_j^{(\kappa)}, \phi_j^{(\lambda)}\}$ on the qubit, a desirable unitary $U_d (\hat{w}, \hat{v})$ of non-commuting operators $\hat{w}$ and $\hat{v}$ on the joint oscillator-qubit system can be obtained via the following sequence,
\begin{align}
    U_d(\hat{w}, \hat{v}) &= e^{i\phi_o X} \prod_{j=1}^{d} W_z^{(\kappa)}e^{i\phi_j^{(\kappa)}X}
     W_z^{(\lambda)}e^{i\phi_j^{(\lambda)}X}
     \\
     &=
     \begin{bmatrix}
         F_d(\hat{w}, \hat{v}) & iG_d(\hat{w}, \hat{v})
         \\
         iG_d(\hat{v}^{-1}, \hat{w}^{-1}) & F_d(\hat{v}^{-1}, \hat{w}^{-1})
     \end{bmatrix}
\end{align}
where $W_z^{(\kappa)} = e^{-i\frac{\kappa}{2}\hat{x}Z}$ is a block-encoding of the operator $\hat{w}=e^{-i\frac{\kappa}{2}\hat{x}}$ and $W_z^{(\lambda)} = e^{-i\frac{\lambda}{2}\hat{p}Z}$ is a block-encoding of the other quadrature operator $\hat{v} = e^{-i\frac{\lambda}{2}\hat{p}}$. This sequence implements a bivariate Laurent polynomial transformation on the non-commuting operators $\hat{w}$ and $\hat{v}$ of the form 
\begin{align}
    F_d(\hat{w}, \hat{v}) = \sum_{r,s = -d}^{d}  f_{r,s} \hat{w}^{r}\hat{v}^{s}, \quad 
    G_d(\hat{w}, \hat{v}) = \sum_{r,s = -d}^{d}  g_{r,s} \hat{w}^{r}\hat{v}^{s}
\end{align}
where the complex coefficients $f_{r,s} $ and $ g_{r,s}$ are determined from the phases $\{ \phi_j^{(\kappa)}, \phi_j^{(\lambda)}\}$.

\subsection{Polynomial Assembly and Algorithmic-level Error Correction}
\label{ssec:poly-assemble-alec}

Having explored the diverse forms of Quantum Signal Processing (QSP) in the previous sections, spanning cases for multiple variables (both commuting and non-commuting), multi-polynomials, and higher dimensions, assembling them in a modular way becomes important for scalability. The modular QSP approach as proposed by Ref. \cite{rossi2023modular} builds on combining LEGO-like blocks of QSP-based operators called \emph{gadgets}, which block-encode multi-variable functions. Such \emph{gadgets} can be constructed from the Abelian M-QSP approach discussed in the previous section. Basic arithmetic operations like inversion, negation, multiplication, etc., can be defined on \emph{gadgets} thus obtained, such that arranging various \emph{gadgets} can realize the desirable super operator consisting of multiple multivariate polynomials.

In the early fault-tolerant era, small gate errors in a long quantum computation can still accumulate to a degree that is non-negligible. Analyzing the performance of fault-tolerant quantum algorithms with structured or small logical errors is therefore an important problem. The iterative nature of QSP algorithms provides algorithmic structures for gate error to propagate in a structured fashion into algorithms. This provides a pathway to analyze and correct errors in QSP at the algorithmic level. 

Ref. \cite{tan2023perturbative} proposed a way to analyze error propagation in QSP via perturbation theory. In particular, a coherent error of $\phi \rightarrow (1+\epsilon) \phi$ is assumed to happen for all signal processing rotation angles, for some small $\epsilon$. A procedure was proposed to efficiently propagate these local gate errors to the entire QSP algorithm level. Moreover, once these algorithm-level errors are known, it becomes possible to correct these errors by appending a recovery QSP sequence after the original one, such that the overall errors from the two QSP sequences cancel each other out. This is the basic idea of algorithmic-level error correction protocol \cite{tan2023errorcorrectionquantumalgorithms}, as shown in Fig. \ref{fig2}h. Future generalizations of this protocol to stochastic error and combination with standard error-correcting code could serve as more powerful ways to curb errors at the algorithmic level.

\subsection{Error Tradeoff Between Block-Encodings and Polynomial Transforms}
\label{ssec:tradeoff-poly}

Taking block-encoding and polynomial transforms together as QSP algorithms, there are important algorithmic-level error tradeoffs that we discuss in this section.

\textbf{Deterministic case.} Suppose an approximate block-encoding for a matrix has error $\epsilon_b$. Query these approximate block-encoding $d$ times in a circuit to realize a degree-$d$ polynomial transform; the overall error in the worst case will simply add up to produce $d \epsilon_b$. 
On the other hand, assuming the degree-$d$ polynomial approximation to the target function $f(x)$ has an error of $\epsilon_f(d)$ (we know the precise form of this from function approximation theory; for example, Ref. \cite{PhysRevLett.118.010501} derived this for Hamiltonian simulation), then it immediately follows that the maximally allowed error on block-encoding has to satisfy $d \epsilon_b \sim \epsilon_f(d) $ such that the block-encoding error does not ruin the long circuits that QSP performs for polynomial transform. Solving for $\epsilon_b$, we have
\begin{align}
    \epsilon_b \sim \epsilon_f(d) / d.
    \label{eq:eps_be}
\end{align}
This puts an error bound on thinking about any approximate block-encoding strategies in QSP algorithm constructions for a fixed degree-$d$. Alternatively, if we are given a fixed approximate block-encoding, solving the above for $d$
\begin{align}
    d \sim d_c(\epsilon_b, f(\cdot))
    \label{eq:polyd}
\end{align}
where $d_c(\epsilon_b, f(\cdot))$ is a critical degree as a function of the block-encoding error $\epsilon_b$ and the target function $f(\cdot)$. This suggests that any QSP algorithms longer than $O(d_c)$ the query depth may not be useful anymore.

For a QSP-based real-time evolution \cite{10.1145/3313276.3316366} (see sec.\ref{ss:RTE} and Eq. \eqref{FCQSP}), \eqref{eq:eps_be} becomes
\begin{align}
    \epsilon_b = \frac{1}{d}e^{\alpha|t|e}e^{(d-\alpha|t|)W(-\frac{\alpha|t|}{d-\alpha|t|}e^{-e\frac{\alpha|t|}{d-\alpha|t|}})}
\end{align}
and \eqref{eq:polyd} becomes 
\begin{align}
    d\sim d_c(\epsilon_b(d,\alpha,|t|);f(H,t)=e^{iHt})
\end{align}
For $\epsilon_b\to0$, $d$ approaches infinity which means the QSP polynomial degree can be arbitrarily large since no error from block-encoding will be accumulated.

For a GQSP-based real-time evolution \cite{motlagh2023generalized} (see sec.\ref{ss:RTE} and Eq. \eqref{GQSP}), \eqref{eq:eps_be} becomes
\begin{align}
    \epsilon_b=\frac{1}{d}e^{(d-t)W(-\frac{1}{d-t})}
\end{align}

and \eqref{eq:polyd} becomes
\begin{align}
    d_c=t+\frac{W(1/\epsilon_b)}{\log(W(1/\epsilon_b))}\Big[1+O\big(\frac{\log\log W(1/\epsilon_b)}{ W(1/\epsilon_b)}\big)\Big]
\end{align}

For $\epsilon_b\to0$, this becomes $d_c=t+\frac{\log(1/\epsilon_b)}{\log\log(1/\epsilon_b)}[1+o(1)],$ which diverges to infinity as expected at a rate slower than $\log(1/\epsilon_b)$ but faster than $\log\log(1/\epsilon_b)$.

\textbf{Stochastic case.} While the above error tradeoff discussion assumes deterministic errors on both the block-encoding and the polynomial transform, when running QSP-type of algorithms on early fault-tolerant quantum computers, stochastic errors will be inevitable. The structure of the error tradeoff will be richer in this case and can be analyzed in the following way.

Suppose $\tilde{U}$ is an approximate block encoding of $\tilde{H}$ labeled by projector $\Pi$
            \begin{align}
                \tilde{H} = \Pi \tilde{U} \Pi
            \end{align}
where $\tilde{H}$ is close to a target Hamiltonian $H$. Assume we would like to realize a target non-linear transform $f(x)$ with a degree-$d$ polynomial transform $P(\Phi_d; x)$ using QSP for $\Phi_d$ being the QSP phase angles. The noise and errors on the QSP phase angles render the final polynomial transformation to be $P(\tilde{\Phi}_d; x)$. For a general setting where the error on the phase angles and the Hamiltonians is stochastic, described by two probability distributions, $p(\tilde{\Phi}_d)$ and $q(H)$, we can quantify the distance between the final erroneous polynomial transform and the ideal non-linear transformations over all realizations of the error channel as
    \begin{align}
                \mathcal{L} = \int \delta H ~d\tilde{\Phi}_d~  p(\tilde{\Phi}_d) q(H) ~| P(\tilde{\Phi}_d, \tilde{H}) - f(H) |
                \label{eq:poly-distance}
    \end{align}
where the norm $|\cdot|$ represents the $L_1$ norm. We remark that \eqref{eq:poly-distance} has some remarkable consequences. It suggests that the two errors from approximate block-encoding and QSP phase rotation angles can cancel, which gives an overall smaller error, even though each individual error may be large. 

The goal of designing robust QSP algorithms therefore reduces to exploring the properties of the distance $\mathcal{L}$ and trying to find ways to minimize it. In particular, the following questions are worth considering: (i) Under what conditions $\mathcal{L} \rightarrow 0$ is first-order in small changes for $\delta H$ and $d\tilde{\Phi}_d$? (ii) Are there any lower or upper bounds for $\mathcal{L}$? (iii) For cases where $\mathcal{L}$ does not vanish in 1st-order change for $H$ and $\tilde{\Phi}_d$, are there conditions to guarantee $\mathcal{L}$ is a convex function of $H$ and $\tilde{\Phi}_d$? A solution to (iii) will guarantee the existence of efficient classical algorithms to find the optimal distribution to minimize $\mathcal{L}$.

\subsection{Software to find phase angles}
\label{ssec:soft-poly}

The early work on QSP determines the qubit rotation phases through algorithms similar to Remez exchange algorithms in digital signal processing \cite{pachon2009barycentric} and numerical optimizations \cite{PRXQuantum.2.040203}. However, this approach becomes unstable as the polynomial degree increases, often failing to find valid phase angles for high-degree transformations. A later work \cite{PhysRevA.103.042419} proposed improved optimization strategies capable of accurately approximating polynomials of degree greater than $10^4$ with error below $10^{-12}$ using symmetric phase factors. An implementation of this symmetric phase factor optimization can be found in the \textit{qsppack} codebase~\cite{qsppack}. \textit{pyqsp} package \cite{pyqsp} has many implementation of state-of-the-art phase finding algorithms as well.Later improvements based on gradient-free fixed-point iteration techniques \cite{referee_1, referee_2} and on methods derived from nonlinear Fourier analysis (NLFA) \cite{referee_3, referee_4} have been developed to enhance stability. Ref.
~\cite{laneve2025generalizedquantumsignalprocessing} has python implementation of the newer phase finding algorithms.

To overcome numerical instability and avoid explicit polynomial root finding, several analytical and algebraic approaches have been developed. One method \cite{Haah2019product} relies on decomposing trigonometric polynomial factors and updating the phases one by one through algebraic relations, thereby achieving stable phase extraction. Another work\cite{chao2020findinganglesquantumsignal} achieves machine-precision phase recovery by recursively decomposing the target unitary $U$ into a product of lower-degree unitaries $U_1 U_2$, halving the effective polynomial degree at each step. In addition, a small perturbation was introduced to stabilize near-vanishing highest-order terms, referred to as \textit{capitalization}. Once further decomposition is not feasible, a system of nonlinear equations derived from recursive polynomial relations is solved, eliminating the need for direct root finding. More recently, Ref. \cite{Ying2022stablefactorization} formulates QSP angle finding using Prony's method, also avoiding root finding with stable factorization.

The angle finding algorithm based on Prony's method for QSP has been generalized to GQSP \cite{yamamoto2024robustanglefindinggeneralized} recently. The GQSP angle-finding algorithm includes truncating the target polynomial of degree $d$ using methods such as Remez exchange. Following truncation, the function is partitioned as a linear combination of a finite set of simpler functions $\{f_j\}$ that satisfy certain constraints for $j$ being an integer. Phase factors for each $\{f_j\}$ can be calculated using several direct methods, like root finding, or indirect methods, like optimization using L-BFGS. Other quantum software providers like Pennylane \cite{pennylane_qsvt} also have python implementations of phase angle finding methods for QSVT.

\section{Scalable to Parallel and Distributed Architectures}
\label{sec:scale}

Scaling up the number of qubits on a single quantum chip has its own challenge in control and fabrication. Multiple quantum chips with interconnects are emerging quantum computing architecture that can overcome these challenges \cite{ang2024arquin,li2024heterogeneous,Main_2025,beals2013efficient}. Future distributed quantum hardware requires parallel and distributed quantum computational science methods. We review recent progress on parallel (Sec. \ref{ssec:para-qsp}) and distributed (Sec. \ref{ssec:dist-qsp}) quantum signal processing algorithms in this section.

\subsection{From Serial to Parallel QSP}
\label{ssec:para-qsp}

To split a long QSP algorithm into multiple shorter ones, one idea is to break a high-degree polynomial into a collection of many low-degree ones. There are two fundamentally different ideas for achieving this. One is to factorize the polynomial into a product of low-degree ones. The other is to slice the polynomial into a collection of many piecewise smooth functions, where the polynomial for each input segment is naturally well-approximated by a low-degree one (for example, cubic splines).

Ref. \cite{martyn2025parallel} provides the first construction of a parallel QSP algorithm via the first route -- polynomial factorization (Fig. \ref{fig2}c). The parallel QSP algorithm factorizes a large-degree polynomial of a special form, where each small-degree polynomial is then executed on a separate quantum computer. Finally, a generalized SWAP test is applied to the output of each quantum computer to ``glue'' the short polynomial together into a larger one. The algorithm can reduce a depth $O(d)$ QSP algorithm into $k$ parallel threads of depth $O(d/k)$ QSP algorithm with a sampling overhead of $O(\mathrm{poly}(d) 2^{O(k)})$, thus achieving a space-time resource tradeoff. This parallel QSP algorithm has important applications to entropy estimation and partition function evaluation -- two useful subroutines for simulating physical sciences. We also note several other recent advances in parallel Hamiltonian simulation \cite{Zhang2024parallelquantum}.

\subsection{From Parallel to Distributed QSP}
\label{ssec:dist-qsp}

Parallel QSP achieves a space-time tradeoff by distributing the polynomial transformation into different QPUs. However, the generalized SWAP test stage used to stitch all the individual polynomials together still requires controlled SWAP operations between two QPUs, rendering it not fully distributed. 

As is known from quantum information theory, classical communication combined with quantum entanglement can establish a quantum communication channel to teleport either a quantum state or a quantum gate from one QPU to the other \cite{}. This means that the controlled SWAP operation, and more generally, the entire generalized SWAP test, can be realized by consuming pre-shared entanglement with classical communication between all the QPUs, achieving a fully distributed QSP algorithm. Indeed, this has been achieved in a recent work \cite{qracd2025}. It was shown that there is a constant-depth realization of the generalized SWAP test stage in parallel QSP by consuming GHZ state with a width of $O(k)$ for $k$ parallel threads (independent of the number of qubits in each local QPU). The key to achieving this is the construction of a multi-party SWAP test subroutine with parallel Toffoli gates and constant-depth fanout gates.

\section{Applications}
\label{sec:app}

Using the previous techniques in block-encoding and polynomial transformation, we present some applications in this section. Secs. \ref{ss:RTE} and \ref{ss:ITE} start with a pedagogical overview of existing methods for two common computational tasks for physical science, i.e., the real- and imaginary-time evolution. Sec. \ref{subsec: parameter estimation} describes the use of the QSP algorithm in expectation value and parameter estimation from a Bayesian perspective, highlighting its adaptability to both NISQ and fault-tolerant eras. Secs. \ref{ss:chem}, \ref{ss:phys}, and \ref{ss:opt} then provide examples of how QSP can be used to tackle problems in chemistry, physics, and optimization problems. We note that there are many other important application problems \cite{xlpd-fb1g,PRXQuantum.2.030342,PhysRevA.102.052411,Kane2025blockencodingbosons,PRXQuantum.5.020332} that are not explicitly discussed in this section, but the philosophy of BE and polynomial transform should generally apply.

\subsection{Real-Time Evolution}
\label{ss:RTE}

Simulating real-time dynamics governed by quantum Hamiltonians is foundational for applications using the time-dependent Schrödinger's equation (TDSE). If the Hamiltonian $H$ is time-independent, then TDSE gives the solution:$\ket{\psi(t)}=e^{-iHt}\ket{\psi(0)}$.
Therefore, we want efficient representation and implementation of the unitary time evolution operator $U=e^{-iHt}$ for practical large-scale computations on quantum devices. Beyond Trotter, one well-known way to do this is using linear combination of unitaries over a truncated Taylor series expansion of $e^{-i Ht}$ \cite{2015Berry}. In addition, the original QSP algorithm \cite{Low_2017} can realize Hamiltonian simulation by combined with LCU due to the parity constraint on QSP polynomials. This makes the algorithm \emph{incoherent}, meaning that measurement on some ancilla qubits are required to post-select the simulation results. GQSP can directly simulate the complex exponential $e^{-i H t}$ because the parity constraint on the polynomial transform it can realize is lifted. 
Between these two works, there are some effort to addressed the post-selection problem by using amplitude amplification (AA) techniques on top of the LCU QSP Hamiltonian simulation. Becaue of the importance of AA technique, we give a brief overview of it and highlight one \emph{fully coherent} Hamiltonian simulation technique \cite{martyn2023efficient}.

By applying Euler's theorem, we get $e^{-iHt}=\cos(Ht)-i\sin(Ht)$.
Both of the summands have definite parity, even and odd, respectively. By using QSP, we can find polynomial approximations to each one of them separately.
Using Jacobi-Anger expansion, each summand can be expressed as an infinite sum of well-known family of polynomials such as Chebyshev polynomials and Bessel functions. We can obtain $\epsilon$-approximations to $\cos(Ht)$ and $i\sin(Ht)$ by truncating these infinite series at a sufficiently large index $K$, which can be determined by a function $r(\tau, \epsilon)$, defined as \(r(\tau, \epsilon)=\tau \exp\big[{W(\log(1/\epsilon)/|\tau|)}\big]\in (\tau, \infty)\) where $W(x)$ is the Lambert-$W$ function. Putting these all together, 
\begin{equation}
    \Theta\bigg(\alpha |t|+\log(1/\epsilon)/\log\Big(e+\log(1/\epsilon)/(\alpha|t|)\Big)
    \label{FCQSP}
\end{equation} degrees of polynomial are needed to simulate $H$ for time $t$, where $\alpha>\lVert H\rVert$ is a normalization factor.\cite{10.1145/3313276.3316366}

For a fully coherent Hamiltonian simulation, if we are provided a single copy of an initial state $\ket{\psi_0}$, we must be able to prepare a time-evolved state $\ket{\psi}$ such that $\lVert\ket{\psi} - e^{-iHt}\ket{\psi_0}\rVert\leq\epsilon$ with success probability of at least $1 - \delta$. However, for QSP-LCU, we cannot do further coherent computations, which limits our success probability. By employing conventional amplitude amplification (AA), our query complexity becomes \cite{10.1063/5.0124385}
\begin{equation}
    \Theta\Bigg(\log\Big(\frac{1}{\delta}\Big)\Big(\alpha |t|+\frac{\log(1/\epsilon)}{\log(e+\log(1/\epsilon))/(\alpha|t|)}\Big)\Bigg).
\end{equation}

There is another way of block-encoding that does not require AA. For this approach from Ref. \cite{10.1063/5.0124385}, we aim to approximate $e^{-ix\tau}$ as a polynomial, where $x$ is the input variable and $\tau$ is a real parameter representing the effective time of simulation.  We may design such a polynomial by
estimating the even extension of the complex exponential (EECE), 
\(EECE(x;\tau):=\cos(\tau x)-i\sin(\tau x)\text{sign}(x)\),
which is always an even function.
For $x>0$, the function becomes \(EECE(x>0;\tau)=\cos(\tau x)-i\sin(\tau x)=e^{-i\tau x}\).
But this requires all the eigenvalues of $H$ to be positive. Fortunately, this can be done for arbitrary $H$, where we can block-encode a normalized Hamiltonian $H/\alpha$ with eigenvalues in the range $[-1, 1]$. With our encoding of $H/\alpha$, we can use a linear pre-transformation to block-encode an operator whose spectrum is proportional to that of $H/\alpha$ but shrunken to be in the range $[(1-\beta)/2,(1+\beta)/2] \subset [0, 1]$ for some chosen $\beta < 1$. This leads us to block-encode the final rescaled Hamiltonian $1/2[I+\beta H/\alpha]$, which is fairly easy. Here, $I$ is the identity operator of size $2^n \times2^n.$ Let $U_{H/\alpha}$ be the block-encoding of the $n$ qubit operator $H/\alpha$ and $U_{\beta I}$ be a block encoding of $\beta I$. Then we can construct a block encoding of $H/\alpha \cdot \beta I = \beta H/\alpha$. We may easily construct $U_{\beta I}$ with an $x$-rotation applied to an ancilla qubit, 
\begin{equation}
R_x(2 \cos^{-1}(\beta)) \otimes I = \begin{bmatrix}
    \beta I & *\\
    * & *
\end{bmatrix}
= U_{\beta I}
\end{equation}
where $R_x(\theta)$ is the $x$-rotation through an angle $\theta$. Equivalently, the block encoding of $\beta H/\alpha$ is $U_{\beta H/\alpha} = R_x(2 \cos^{-1}
(\beta)) \otimes U_{H/\alpha}$. This additional x-rotation is not costly and can be cheaply prepared from QSP.

We may introduce another ancilla qubit to obtain an encoding of $1/2(I + \beta H/\alpha) =: \tilde{H}$, which has eigenvalues in the range $[(1-\beta)/2,(1+\beta)/2]
\subset [0, 1]$. In total, with the addition of two ancilla qubits, we can block encode the rescaled Hamiltonian $\tilde{H}$. If the initial Hamiltonian $H/\alpha$ were encoded in the $\ket{0}\bra{0}$ matrix element of a unitary, then the procedure encodes $\tilde{H}$ in the $\ket{000}\bra{000}$ matrix element of a new unitary. This improved algorithm achieves fully coherent one-shot Hamiltonian simulation \cite{10.1063/5.0124385} with query complexity of \(\Theta\Big(\alpha|t|+\log(1/\epsilon)+\log(1/\delta)\Big)\).

GQSP also provides us with an algorithm for Hamiltonian simulation, where we can implement an $\epsilon$-approximation of $e^{it \sin H}$ and $e^{it \cos H}$ for $t \in \mathbb{R}$ with a polynomial of 
\begin{equation}
    O(t + \log(1/\epsilon)/\log\log(1/\epsilon))
    \label{GQSP}
\end{equation} degree by query a block-encoding of $e^{iH}$. \cite{motlagh2023generalized}

\subsection{Imaginary-Time Evolution}
\label{ss:ITE}

Imaginary time evolution (ITE) is a powerful tool for ground state finding \cite{McArdle_2019} and quantum Gibbs-state sampling \cite{kosugi2022probabilisticimaginarytimeevolutionusing}. In the ground state finding problem, ITE can be considered as ``cooling'' the system down to its ground state \cite{motlagh2024groundstatepreparationdynamical}. We consider a transformation from real time $t$ to imaginary time $\tau=-it$, called Wick rotation. Substituting $t=i\tau$ into the TDSE, we obtain the ITE state
\(
\ket{\psi(\tau)}=\sum_nC_ne^{-E_n\tau}\ket{\psi_n(0)}
\),
where $E_n$ is the eigen energy of $n^{th}$ eigen wavefunction $\ket{\psi_n}$ and $C_n$ are the complex amplitude coefficients. For $\tau\to\infty$, all the terms decay exponentially, but the ground state wavefunction $\ket{\psi_0}$ decays the most slowly because of the lowest energy $E_0$ in its exponential. Therefore, after evolving for a sufficiently long time $\tau\gg 1/(E_1-E_0)$, the higher energy wavefunctions get filtered out, while the ground state remains. 

Another useful application is Gibbs state preparation. For a quantum system with Hamiltonian $H$ and inverse temperature $\beta=1/(k_BT)$, the Gibbs state has the following density matrix: $\rho_\beta=e^{-\beta H}/\text{Tr}(e^{-\beta H})$, where $k_B$ is the Boltzmann constant. When the initial state is maximally mixed $(I/2^n)$ and $\tau$ represents the inverse temperature, then the ITE state becomes the Gibbs state. In quantum computing, Gibbs sampling acts as an efficient subroutine for quantum simulation of thermal systems \cite{Poulin_2009, rouzé2024efficientthermalizationuniversalquantum, rajakumar2024gibbssamplinggivesquantum, Bergamaschi_2024}, quantum machine learning \cite{Coopmans_2024}, optimization and probabilistic inference \cite{Low_2014, FATHALLAH2024109307}, studying open quantum systems \cite{Cohn_2020}, and thermalization processes \cite{rouzé2024efficientthermalizationuniversalquantum}.

ITE requires implementing the nonunitary operator $e^{-x\tau}$, which cannot be directly realized on quantum hardware. A general strategy is to embed this non-unitary map into a larger unitary through block encoding, then recover the desired transformation by postselection. Several modern QITE algorithms can be understood within this unified framework, including series-expansion--based block encoding, polynomial approximations, and hardware-efficient neural-network circuits.

Ref. \cite{Silva2023} develops a fragmented master QITE algorithm that prepares the normalized state $F_{\beta}(H)\ket{\Psi}/{\lVert F_{\beta}(H)\ket{\Psi}\rVert}$, where $F_\beta(H):=e^{-\beta(H-\lambda_{\rm min})}$, and $\lambda_{\rm min}$ is the minimum eigenvalue of $H$. The method relies on primitives that block-encode an $\epsilon$-approximation of $F_{\beta}(H)$ using an ancilla register and a combination of Chebyshev and Fourier expansions. In the Chebyshev-based primitive, one assumes oracle access to a block-encoding $\mathcal{O}_1$ of the Hamiltonian $H$. A truncated Chebyshev series approximates the imaginary-time propagator, requiring \(q_1={O}\Big(e\beta/2+\log\big(1/\epsilon)/\log(e+2\log(1/\epsilon)/(e\beta)\big)\Big)\)
queries to $\mathcal{O}_1$, with classical preprocessing cost ${O}(\mathrm{poly}(q_1))$. For small $\beta$, $q_1$ asymptotically scales as ${O}\Big(\sqrt{\beta\ \log(1/\epsilon)}\Big)$. 

In contrast, the Fourier-based primitive assumes access to a unitary oracle $\mathcal{O}_2$ containing the real-time evolution $e^{-iHt}$ for a fixed time $t=\frac{\pi}{2}(1+{\gamma}/{\beta})^{-1}$. A Fourier expansion of the propagator yields an implementation requiring \(q_2={O}\Big((\beta/\gamma+1)\rm\ log(4/\epsilon)\Big)\)
queries to $\mathcal{O}_2$ and $\mathcal{O}_2^\dagger$ and $g_2+{O}(1)$ gates per query, where the gates are obtained in classical runtime ${O}(\mathrm{poly}(q_2))$, and $g_2$ is the gate complexity of $\mathcal{O}_2$, and $\alpha=e^{-\beta(1+\lambda_{\rm min})-\gamma}$. The master QITE algorithm chains these primitives so that the overall query cost is essentially the sum of the Chebyshev and Fourier contributions, achieving trace-distance error $\epsilon$.

An alternative approach, presented in ref. \cite{zhang2025quantumimaginarytimeevolutionpolynomial}, approximates $e^{\tau(x-\lambda)}$ directly using low-degree polynomials. In this formulation, $x$ represents a rescaled eigenvalue of the Hamiltonian in $[-1,1]$, so approximating $e^{\tau(x-\lambda)}$ reproduces the imaginary-time amplification of the low-energy eigencomponents. By selecting a stabilization shift $\lambda \in(0,1]$, the algorithm maintains a constant success probability lower bound near $\gamma^2/e^2$, where $\gamma$ is the initial ground-state overlap. Assuming $\gamma$ is not exponentially small, the method prepares the normalized ITE state with error $\tilde{O}(\mathrm{poly}(\tau^{-1}))$ using $\tilde{O}(\mathrm{poly}(n\tau))$ controlled-Pauli queries and only one ancilla qubit.

Moving from polynomial approximations to a more hardware-oriented perspective, neural-network quantum circuits offer an alternative mechanism for implementing imaginary-time evolution. Ref. \cite{PhysRevResearch.7.013306} shows that restricted and deep Boltzmann Machine-type circuits can exactly block-encode each Trotter step $e^{-\Delta\tau H}$. These ans\"atze are closed under imaginary-time propagation and therefore yield exact thermal states (and ground states in the limit $\tau\to\infty$) with analytic circuit parameters. For $k$-local Hamiltonians, the qubit count scales linearly with system size and total imaginary time $\tau$, and mid-circuit measurement allows either constant circuit width with depth $O(n\tau)$ or $O(n)$ ancillas with depth $\sim O(\tau)$.

\subsection{Expectation Value and Parameter Estimation}\label{subsec: parameter estimation}

In the context of estimating the expectation values of observables and other parameters (such as the overlap of two states), one can think of QSP polynomials as being likelihood functions that facilitate estimating the values of parameters through Bayesian inference methods. That is, consider Bayes' rule
\begin{equation}\label{eq: Bayes' rule}
    \mathbb{P}(A|B) = \frac{\mathbb{P}(B|A)\mathbb{P}(A)}{\mathbb{P}(B)}
\end{equation}where $\mathbb{P}(A)$ and $\mathbb{P}(B)$ are the probabilities of observing outcomes $A$ and $B$, respectively. $\mathbb{P}(A|B)$ reads as ``the probability of observing $A$ conditioned on $B$ being observed,'' and similarly for $\mathbb{P}(B|A)$ (the \emph{likelihood} function). $\mathbb{P}(A)$ is known as the \emph{prior} distribution. That is, it represents our current best guess for what the distribution of measurement outcomes is \emph{prior} to performing any additional measurements. $\mathbb{P}(A|B)$ is known as the \emph{posterior} distribution. That is, it represents our best \emph{post-measurement} guess as to what the distribution of outcomes is, given that our measurement yielded outcome $B$. Bayesian inference refers to an iterative process of repeated measurements, updating the prior distribution in the next iteration with the posterior distribution of the current iteration.

The versatility of QSP allows us to design likelihood functions $\mathbb{P}(B|A)$ that improve the information gain per measurement beyond what is achievable using independent measurements. Furthermore, because the QSP degree is an integer that we control, we can easily interpolate between the low information gain (shallow circuit) and optimal information gain (deep circuit) regimes. This suggests that adapting near-term algorithms that admit shallow circuits (but have suboptimal runtime scaling) to utilize QSP in some fashion will be a fruitful strategy for bridging the gap between the near-term and the fault-tolerant eras.

\begin{figure}[htb!]
    \centering
    \includegraphics[width=0.35\linewidth]{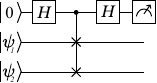}
    \caption{The circuit used for the SWAP test.}
    \label{fig:swap_test_circ}
\end{figure}

Let us illustrate this idea with a simple example: estimating the overlap of two states: $|\braket{\psi_1|\psi_2}|^2$, where each state is prepared as $\ket{\psi_j} = U_j \ket{0}$. The conventional method of estimating this overlap parameter is the SWAP test, which utilizes the circuit given in Fig.~\ref{fig:swap_test_circ}.

The probability of a measurement of the ancilla qubit yielding the outcome $\ket{0}$ is $\frac{1}{2} + \frac{1}{2}|\braket{\psi_1|\psi_2}|^2$. Denote $a = \sqrt{\frac{1}{2} + \frac{1}{2}|\braket{\psi_1|\psi_2}|^2}$. Estimating the parameter $a$ to precision $\epsilon$ using a sequence of independent measurements of the ancilla will incur a runtime that scales as $\mathcal{O}(\frac{1}{\epsilon^2})$. We now illustrate how this can be improved to $\mathcal{O}(\frac{1}{\epsilon}\log_2(\frac{1}{\epsilon}))$ using a variation of the method outlined in the supplementary information of Ref.~\cite{liu_bootstrap_2023}.

Suppose we could prepare a QSP polynomial $P(a)$ that is a step function on the interval $\left[\frac{1}{\sqrt{2}},1\right]$ centered about some value $a_o$. That is,
\begin{equation}
    P(a) = \begin{cases}
        1 & \text{if } a \geq a_0 \\
        0 & \text{if } a < a_0.
    \end{cases}
\end{equation}Sampling from a QSP circuit corresponding to this polynomial will yield a probability $|P(a)|^2$ of measuring the ancilla to be $\ket{0}$. Thus, if we prepare this QSP circuit and the measurement outcome is $\ket{0}$, then we know that the value of $a$ must be in the range $\left[a_0,1\right]$. Similarly, if the measurement outcome is $\ket{1}$ we know $a$ is in the interval $\left[\frac{1}{\sqrt{2}},a_0\right)$.

If we perform a sequence of measurements wherein we iteratively choose $a_0$ to subdivide the interval in which we know it must be, then we can estimate the value of $a$ to precision $\epsilon = \frac{1}{2^n}$ with $n = \log_2(\frac{1}{\epsilon})$ measurements. In practice, we cannot prepare an exact step function. We can only prepare an approximate step function with a rising edge centered about $a_0$ with width $w$ using a QSP degree that scales as $\mathcal{O}(\frac{1}{w})$.~\cite{10.1063/5.0124385} We need $w$ to be smaller than $\epsilon$ to some constant multiplicative factor in order for the method to achieve precision $\epsilon$ with high probability. Thus, the total runtime of this method scales as $\mathcal{O}(\frac{1}{\epsilon}\log_2(\frac{1}{\epsilon}))$.

These concepts of using QSP polynomials and likelihood functions to speed up estimation subroutines beyond $\mathcal{O}(\frac{1}{\epsilon^2})$ can be extended to observable expectation value estimation as well. For example, the authors of Ref.~\cite{knill_optimal_2007} showed how to estimate the expectation value of observables with $\mathcal{O}(\frac{1}{\epsilon})$ scaling using either phase estimation or amplitude estimation (both of which can be formulated as QSP sequences~\cite{PRXQuantum.2.040203, rall_amplitude_2023}) as subroutines.

The tunability of the QSP degree $d$ allows one to use QSP techniques to accelerate estimation subroutines beyond $\mathcal{O}(\frac{1}{\epsilon^2})$ within a given circuit depth budget. This is highly desirable in contexts such as quantum chemistry, where the $\mathcal{O}(\frac{1}{\epsilon^2})$ scaling is known to lead to prohibitively long runtimes~\cite{gonthier_measurements_2022}. For example, Ref.~\cite{wang_accelerated_2019} developed a method called $\alpha$-VQE, which incorporates quantum phase estimation as a subroutine of the variational quantum eigensolver (VQE)~\cite{peruzzo_variational_2014} in such a way that the number of samples scales as $\mathcal{O}(\frac{1}{\epsilon^{2(1-\alpha)}})$. Here, $\alpha \in [0,1]$ is a user-tunable parameter that controls whether the algorithm is more ``QPE-like'' or ``VQE-like.'' When $\alpha=0$, one retains the shallow circuit depth and $\mathcal{O}(\frac{1}{\epsilon^2})$ measurement counts of VQE. Similarly, when $\alpha=1$, one obtains the deep circuit depths and $\mathcal{O}(\frac{1}{\epsilon})$ measurement counts of QPE. Similarly, Ref.~\cite{wang_minimizing_2021} showed how one could design likelihood functions that increase the information gain per measurement for the purpose of reducing the number of measurements needed for VQE. The circuits used in this work bear a strong resemblance to QSP sequences, suggesting it is likely that this method could be formulated in terms of QSP. Furthermore, other classes of likelihood functions could be designed using QSP as a framework. 

For example, the use of QSP to accelerate the SWAP test can be easily extended to accelerate the estimation of Pauli string expectation values in VQE by replacing the SWAP test circuit with a Hadamard test circuit. Additionally, one could consider using polynomials that approximate smoother functions such as Gaussian curves. In this modified scheme, one would use Bayesian inference where the likelihood functions and prior distributions are both Gaussian functions. This is similar to how Bayesian inference has been incorporated into quantum phase estimation to improve its noise robustness.~\cite{BayesianQPE, BayesianQPE_experiment}

\subsection{Chemistry}
\label{ss:chem}

Chemistry problems of interest can be broadly divided into two categories: static (e.g., electronic ground-state energy estimation, finite-temperature state preparation) and dynamic (e.g., electron dynamics, reaction dynamics). In sharp contrast to doing quantum chemistry on classical computers, where methods for treating state preparation and dynamics are entirely different, QSP algorithms allow treating both static and dynamic problems on equal footing with minimal methodology change.

In one of such useful algorithms for electron dynamics, an $n$-orbital free-fermionic Hamiltonian with sparse one-electron integrals can be block-encoded (combined with QSP) with a circuit depth of $\mathrm{polylog}(n)$ \cite{stroeks2024solving}. For interacting fermions, multiple techniques are connected together to achieve the desired result. One such technique involves transforming the Hamiltonian into the interaction picture and then block-encoding the Hamiltonians \cite{Babbush_2019}. 

It is also possible to develop similar techniques in the second quantization framework. A low-rank recursive block-encoding strategy can be used to implement a single Trotter step via qubitization and then extended to multiple steps \cite{Low_2023}. The target unitary evolution is obtained only with finite success probability due to ancillary qubits used in block encoding, which can be improved by means of amplitude amplification (AA) referred to in Sec. \ref{ss:RTE}. Similar to Trotter methods, randomized QSP-based algorithms have also been developed that mix different polynomials, effectively halving the overall cost of Hamiltonian simulation \cite{Martyn_2025}.

Since finite-temperature quantum chemistry problems are inherently non-unitary, block-encoding reduces the challenge of operating them on quantum computers. For static problems, a resource-efficient approach uses just one ancillary qubit to block-encode the system Hamiltonian within a unitary operator. This enables a polynomial approximation to the partition function, and for Gibbs state preparation in the canonical ensemble, a polynomial approximation to $e^{-\beta x}$ (see Sec. \ref{ss:ITE}) is applied \cite{powers2022exploringfinitetemperatureproperties}. For a recent review on quantum computing for chemistry beyond the ground state, see Ref. \cite{bidart2025quantumcomputinggroundstate}.

\textbf{Simulating chemical reaction dynamics.} A well-followed approach to solve chemistry problems is to adopt the Born-Oppenheimer (BO) approximation, where the electrons are treated quantum mechanically but the nuclei classically. While this is useful for many applications and can be solved on classical computers approximately, the BO approximation breaks down in important problems involving highly coupled nuclei-electron dynamics, including but not limited to photo-induced vision processes \cite{doi:10.1021/jp400401f}, photovoltaic energy conversion \cite{STIER200333}, and proton-coupled electron transfer (PCET) \cite{doi:10.1021/jp805876e, 10.1063/1.482053, 10.1063/1.1814635}. These are challenging to solve on classical computers accurately, as they involve dealing with an added exponential scaling of nuclear quantum degrees of freedom that couples to the electrons. 

On quantum computers, simulating the full electron-nuclei dynamics is polynomial scaling as shown in Ref. \cite{kassal2008polynomial} for Trotter algorithms. Just as for spin systems that product formulas and QSP can have different performance in various parameter regimes \cite{childs2018toward}, QSP algorithms may provide advantage over Trotter algorithms for chemical reactions in some regime. Here, we outline a QSP-based quantum computational framework for molecular reaction dynamics. While not all initial quantum states can be prepared efficiently, physically motivated states such as molecular ground states can often be created and used as inputs for subsequent dynamical simulation. 

We begin by constructing a block-encoding $U_{\rm sys}$ of all total Hamiltonian $H_{\rm sys}$ of the entire reactant system in second- or first-quantization. For first-quantized simulations, wavelet bases are an attractive option because they offer orthogonality, tunable locality, and structured sparsity, which help reduce the cost of block-encoding \cite{Georges_2025, white2023nestedgaussletbasissets}. After encoding the Hamiltonian, an initial state needs to be prepared using QSP or other algorithms. Note that this initial state does not have to be ground state as chemical reactions may often happen from excited or a finite-temperature state. Locality in the initial reactant state can be exploited to improve the efficiency of this state preparation as the molecules may actually be far apart thus not interacting that much.

For non–Born–Oppenheimer (non-BO) simulations, we encode the full system Hamiltonian $\hat{H}_{\rm sys}=\hat{T}_e+\hat{T}_{N}+\hat{V}_{eN}+\hat{V}_{ee}+\hat{V}_{NN}$, where $\hat{T}_{e},\hat{T}_{N},\hat{V}_{eN},\hat{V}_{ee},\hat{V}_{NN}$ are the electronic kinetic energy, nuclear kinetic energy, electron-nuclear interaction, electron-electron interaction, and nuclear-nuclear interaction, respectively. This contrasts with the Born–Oppenheimer Hamiltonian $\hat{H}_{\rm BO}=\hat{T}_{e}+\hat{V}_{eN} + \hat{V}_{ee}$ where nuclear motion is treated parametrically. 
Since the kinetic and Coulomb operators possess distinct algebraic structure and sparsity, this should be leveraged to improve the efficiency of the block-encoding circuit.

Following the state preparation, we can then evolve the molecular system under the total Hamiltonian $H_{\rm sys}$ using QSP algorithms. Note that in realistic chemical reactions, there will be energy dissipation from the system to solvent or environment, which may need additional dissipative algorithms to be used to capture that \cite{Lin_2025}. After evolving the molecular system, relevant observables (dipole moment, bond order, and population on products) can be extracted. One way to measure an observable $\braket{\psi|\hat{O}|\psi}$ is to find the Pauli representation of $\hat{O}$ and measure each Pauli strings. Another less obvious way is to first construct a block-encoding of $\sqrt{\hat{O}}$ and then using variants of the Hadamard test on the block-encoding to obtain the expectation value by measure the single ancilla qubit. The adaptive measurement scheme in Sec. \ref{subsec: parameter estimation} can be naturally incorporated.
If we select observables that distinguish reactants from products (e.g., spatial distributions, occupation numbers, vibrational signatures). One example is that for population transfer to products, we measure the projector onto the product subspace.

\begin{figure}[htbp!]
    \centering
    \includegraphics[width=1.0\linewidth]{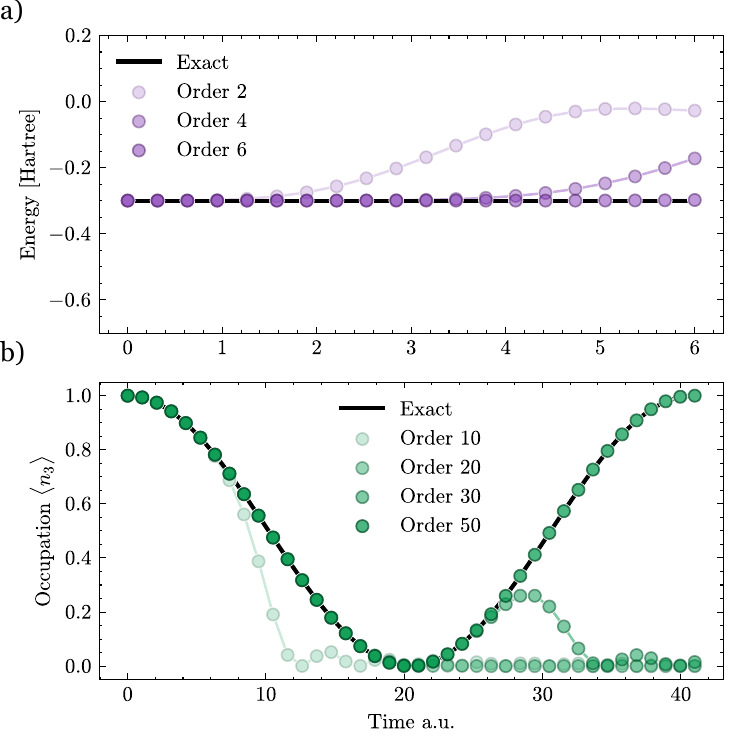}
    \caption{a) Comparison between the exact and QSP-simulated time evolution of the H$_2$ molecular Hamiltonian (Jordan–Wigner mapping of molecular orbitals) using the STO-3G basis. The initial state corresponds to the computational basis state $\lvert 0110 \rangle$. Increasing the polynomial order improves the agreement between the QSP-based simulation and the exact energy evolution. The system is in excited state with an energy of roughly -0.3 Ha. b) Occupation number $\langle \hat{n}_3 \rangle$ on qubit 3 (spin-up of the $\sigma_u$ molecular orbital) as a function of time for the same initial state. The time scale is in atomic units (a.u.) equal to $\hbar/E_h$, for $E_h$ the Hartree energy unit.}
    \label{fig:energy}
\end{figure}

\textbf{A minimal example.} In order to illustrate the convergence between the exact evolution and QSP-simulated energies for different polynomial orders, we compared QSP-based simulations with the exact time evolution of the H$_2$ molecular Hamiltonian (Jordan–Wigner mapping) for the initial state which corresponds to the computational basis state $\lvert 0110 \rangle$ (Fig.~\ref{fig:energy}a). Additionally, we calculate the occupation number for the same system and compare for different polynomial degrees in Fig.~\ref{fig:energy}b.

\subsection{Physics}
\label{ss:phys}

\begin{figure}[htb!]
    \centering
    \includegraphics[width=0.8\linewidth]{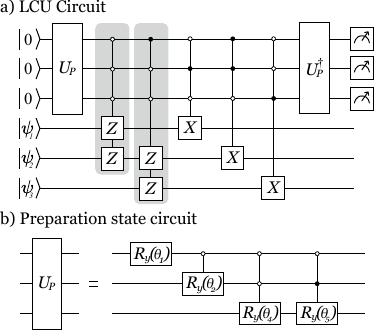}
    \caption{a) Complete circuit for block-encoding the Ising system with $n=3$ using the LCU method. The Hamiltonian consists of five distinct terms, requiring $\log_2(5) \approx 3$ ancilla qubits for the encoding. The multi-controlled fanout gates (shaded) need to be compiled to available gate sets on given hardware. b) State preparation circuit. Some quantum gates do not appear in the circuit because their corresponding rotation angles are equal to zero.}
    \label{fig:lcu_example}
\end{figure}

While the line between what problems fall under ``chemistry'' and ``physics'' is blurred, one can think of the dividing line as follows. In \textit{ab initio} chemistry, one is (in general) attempting to find properties of molecules modeled by lattices with arbitrary all-to-all connectivity (\emph{i.e.} long-range interactions) for the purposes of studying chemical reaction mechanisms. In physics, it is often the case that one is often interested in studying condensed matter systems, which can be modeled by lattice model Hamiltonians containing nearest-neighbor interaction terms for purposes such as studying phase transitions, magnetic properties, and superconductivity. For example, the transverse Ising model
\begin{equation}
    \hat{H} = -J\sum_{\braket{i,j}}Z_iZ_j + g\sum_{i}X_i
    \label{eq:ising_transversal}
\end{equation}
describes a lattice of spin-$\frac{1}{2}$ particles with nearest-neighbor coupling strength $J$ subject to a magnetic field with strength $g$. The particles only interact with each other in the $\hat{z}$-direction. The direction of the magnetic field runs perpendicular to this interaction in the $\hat{x}$-direction. One can also consider the Heisenberg model Hamiltonian,
\begin{equation}
    \hat{H} = -\sum_{\braket{i,j}}\left(J_{ij}^x X_i X_j~+~J_{ij}^yY_iY_j~+~J_{ij}^zZ_iZ_j\right) + g\sum_{i}Z_i
\end{equation}which models a lattice of spin$\frac{1}{2}$ systems that can interact in all directions and are subject to an external $\hat{z}$-direction magnetic field. In order to illustrate this we consider a 3-spin system from Eq. \eqref{eq:ising_transversal} to simulate the magnetization.
\begin{align}
    H = J(Z_1 Z_2 + Z_2 Z_3) +g (X_1 + X_2 + X_3)
    \label{eq:ham-3spin}
\end{align}
For this system we will consider $J=1.0$ and $g=0.25$, but the circuit structure does not depend on these values. Based on the structure of the LCU given in Fig.\ref{fig:LCU}, a concrete realization of block-encoding for this system is given by Fig.\ref{fig:lcu_example}a, where the prepare operation based on Fig. \ref{fig:state_preparation} is given in Fig.\ref{fig:lcu_example}b. The angles that represent the state preparation are $\theta_i \in (0.613, 0.927, 0.0, \pi/2, \pi/2, 0.0, 0.0)$. The values that are $0.0$ do not represent any quantum gate. While narrow, as the number of spin systems increases, the number of required qubits also grows as $\log(2^n)$, where $n$ is the number of spins considered.

\begin{figure}[htb!]
    \centering
    \includegraphics[width=0.8\linewidth]{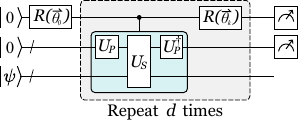}
    \caption{Circuit representation of time evolution simulation of the Ising model for $n=3$, using Generalized Quantum Signal Processing. The post-measurement will apply the polynomial transformation of the quantum state $|\psi\rangle$.}
    \label{fig:gqsp_example}
\end{figure}

Once Eq. \eqref{eq:ham-3spin} is block-encoded, we can implement time-evolution of the Hamiltonian $\exp(-i H t)$. This  block-encoding can also be implemented with previous methods discussed in this work in Sec. \ref{ssec:circ-be}. The circuit implementation is shown in Fig. \ref{fig:gqsp_example} using the GQSP from Sec. \ref{sec:single-poly-single-var}.
The set of angles $\vec{\theta}$ can be found using different frameworks for polynomial transformation explained in Sec. \ref{ssec:soft-poly}. 

\begin{figure}[htb!]
    \centering
    \includegraphics[width=1.0\linewidth]{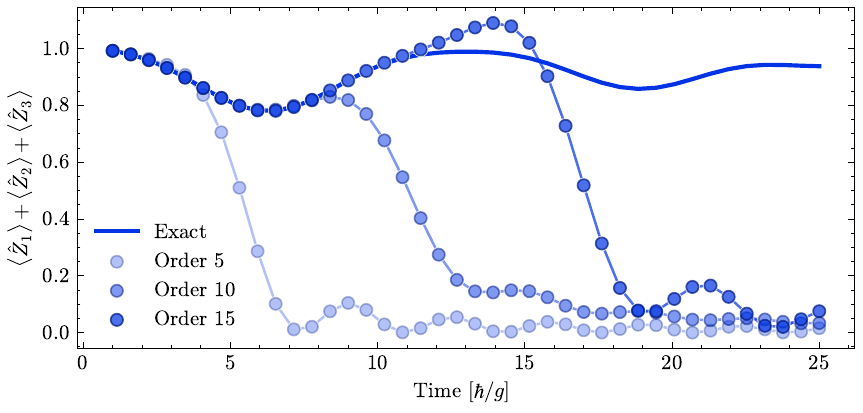}
    \caption{Magnetization for an Ising model of 3-spins in Eq. \eqref{eq:ham-3spin} for different GQSP degrees $d= 5,10,15$ (circle) as compared with the exact result (solid).}
    \label{fig:magnetization}
\end{figure}

Fig. \ref{fig:magnetization} shows the simulation results for the magnetization, $\sum\langle Z_i \rangle$ for an initial state $|\psi_{init}\rangle = |010\rangle$, obtained using GQSP for different orders of magnitude. Higher degrees allow for more accurate simulations at later times.

Platforms such as transmon qubits are themselves lattices of two-level systems with nearest-neighbor connectivity; therefore, it should not be surprising that problems involving such lattice Hamiltonians are key targets for achieving early quantum advantage~\cite{kim_evidence_2023} as the qubit connectivity of the machine closely matches that of the interactions in these Hamiltonians. Quantum signal processing can be readily applied to both static and dynamic problems for such lattice Hamiltonians in much the same way as for chemistry problems. Quantum Phase Estimation, for example, is known to be a particular case of a QSP polynomial~\cite{GrandUnificationAlgos} and is used for solving for the ground state energy of such Hamiltonians at zero temperature. Similarly, eigenstate filtering can be used as a preparation method for the ground state. Block-encoding and QSP can be readily used to perform real-time evolution of the Hamiltonian with respect to some initial state, followed by the estimation of an observable such as the magnetization or spin correlation functions using one of the methods described in Sec.~\ref{subsec: parameter estimation}. Real-time evolution of a Hamiltonian is also an important primitive in some algorithms designed to solve for static properties in methods such as Krylov subspace diagonalization.~\cite{cortes_quantum_2022, yoshioka_krylov_2025} For finite temperature state preparation, one can use QSP to prepare Gibbs distribution states by preparing polynomials that are exponentials of the inverse temperature $\beta$~\cite{GrandUnificationAlgos}.

\subsection{Optimization}
\label{ss:opt}

While one can observe that estimating the ground state of a Hamiltonian drives many applications in chemistry and physics (Secs. \ref{ss:chem}–\ref{ss:phys}), it also extends to optimization problems.
Quantum computers are promising platforms for solving discrete optimization problems, such as MaxCut, minimum vertex cover, and graph coloring, which can all be formulated by quadratic unconstrained binary optimization (QUBO) problems \cite{mazumder2025starterproblemssolvingquadratic}. These problems share a common structure: minimizing a polynomial function of binary variables. The cost function can be mapped to a Hamiltonian whose ground state encodes the optimal solution. However, finding the exact ground state of such Hamiltonians is generally NP-hard, and classical approaches become computationally infeasible for large system sizes due to the exponential scaling of the Hilbert space. Consequently, classical algorithms rely on heuristic or probabilistic methods that yield only approximate solutions. 

In the NISQ era, several non-QSP quantum algorithms have been proposed to approximate ground states of optimization Hamiltonians. Examples include the quantum approximate optimization algorithm (QAOA) \cite{farhi2014quantumapproximateoptimizationalgorithm}, ITE (Sec.\ref{ss:ITE}), and adiabatic quantum evolution \cite{Born1928}. Each method faces specific limitations: QAOA often suffers from barren plateaus that impede gradient-based optimization \cite{McClean2018}; adiabatic evolution typically requires deep circuits and long coherence times; and ITE block-encoding (ITE-BE) schemes demand a large number of shots to achieve reliable post-selection (Sec. \ref{ss:ITE}). Nonetheless, the ITE-BE method demonstrates competitive performance for problems such as MaxCut, requiring only $N$ qubits and $O(|E|)$ circuit depth, where $|E|$ and $N$ denote the number of edges and vertices in the graph, respectively \cite{gtq3-j37b}. 
The algorithm converges to the ground state when the imaginary time $\tau$ is sufficiently large 
as compared to the inverse gap of the Hamiltonian. This algorithm can also be hybridized with QAOA, yielding a tradeoff between the QAOA circuit depth $p$ and the imaginary time $\tau$: a longer $\tau$ allows convergence with a shallow circuit, while a higher $p$ yields faster convergence with increased post-selection probability.

Beyond these non-QSP ground-state-finding strategies, QSP provides a polynomial-transformation–based approach to ground-state preparation. When an initial state with a sufficient overlap with the ground state can be efficiently prepared, and the spectral gap is bounded from below, QSP-based ground state filtering algorithms \cite{lin2020near} achieve near-optimal query complexity for ground state preparation. Even when the spectral gap is unknown, such algorithms can still prepare the ground state with finite success probability. Conceptually, ITE-BE and QSP-based filtering can both be viewed as ground-state–finding frameworks: the former suppresses excited-state amplitudes through imaginary-time dynamics, while the latter constructs a spectral filter polynomial that amplifies the ground-state component directly within the QSP formalism.

\section{Conclusion}
\label{sec:conclusion}

We present a forward-looking Perspective on scalable quantum computational science. Several properties that a quantum computational science method should possess is defined. We highlight the role of block-encoding and polynomial transforms as potential candidates for developing a unified framework for computational science problems. In addition to explain the theory and methods, we also survey current software tools available, and provide illustrative example applications in chemistry and physics, as well as the connection to optimization. Given the rapid development on quantum error correction and hardware, our Perspective provides a timely contribution that helps to bridge the gap between theoretical quantum algorithm community and practical computational scientists. 

Many open challenges remain in order to fully unleash the power of quantum computers for computational science problems. 
For one, as the most important building block, optimal explicit circuit constructions of exact and approximate block-encodings for practical applications are far from established. We think there are ample rooms for constructing approximate block-encodings without violating the design principles outlined. Moreover, the utility of $U(N)$-QSP and M-QSP are much less explored. Discover new applications that can fully leverage these advanced algorithmic structures will be significant for the community. In addition, understanding the performance of QSP algorithms on practical early fault-tolerant serial and distributed quantum computers will be important to unveil a new era of quantum computing method development as quantum machines evolve and mature.

One the software side, results on QSP are mostly scattered. An open-source quantum computing software platform dedicated to QSP method development that has convenient interface with domain application software packages should be established. Phase-angle finding algorithm and automatic block-encoding compilation and circuit construction should be integrated. To enable QSP algorithms executed on near-term and early fault-tolerant hardware, abstractions of multi-qubit gate should be broken and hardware-level details such as pulse-engineering and quantum control techniques should be integrated together with QSP algorithm. This co-design perspective help to minimize the overall resource cost and bring quantum advantage closer. We look forward to engage domain computational scientists with expertise on practical applications to develop QSP algorithms, and broadly, to embrace scalable quantum computational science.

\begin{acknowledgments}
This work is supported by the U.S. Department of Energy, Office of Science, Advanced Scientific Computing Research, under contract number DE-SC0025384. This work is also supported in part by NSF OSI and MPS division under award number 2531350 via a subcontract from Duke University.
\end{acknowledgments}

% \stoptoc

% \section*{Author Declarations}
% \textbf{Conflict of Interest}

% The authors has no conflicts to disclose.

% \section*{Author Contributions}

% KJJ: Software (lead); conceptualization (equal); writing – review and editing (lead). ERD: Writing – review and editing (equal). JB: Writing – review and editing (equal). AM: Writing – review and editing (equal). MHH: Software (supporting); writing – review and editing (supporting). YL: Methodology (lead); conceptualization (lead); writing – review and editing (equal).

% \section*{Data Availability Statement}
% The data that support the findings of this study are available from the corresponding author upon reasonable request.

% \resumetoc

\bibliographystyle{unsrt}
\bibliography{ref,ref1}

@article{referee_1,
  doi = {10.22331/q-2024-12-10-1558},
  url = {https://doi.org/10.22331/q-2024-12-10-1558},
  title = {Infinite quantum signal processing},
  author = {Dong, Yulong and Lin, Lin and Ni, Hongkang and Wang, Jiasu},
  journal = {{Quantum}},
  issn = {2521-327X},
  publisher = {{Verein zur F{\"{o}}rderung des Open Access Publizierens in den Quantenwissenschaften}},
  volume = {8},
  pages = {1558},
  month = dec,
  year = {2024}
}

@misc{referee_2,
      title={Fast Phase Factor Finding for Quantum Signal Processing}, 
      author={Hongkang Ni and Lexing Ying},
      year={2024},
      eprint={2410.06409},
      archivePrefix={arXiv},
      primaryClass={quant-ph},
      url={https://arxiv.org/abs/2410.06409}, 
}

@article{referee_3,
  author    = {Alexis, Maxime and Mnatsakanyan, Grigor and Thiele, Christoph},
  title     = {Quantum signal processing and nonlinear Fourier analysis},
  journal   = {Revista Matem{\'a}tica Complutense},
  volume    = {37},
  pages     = {655--694},
  year      = {2024},
  doi       = {10.1007/s13163-024-00494-5},
  url       = {https://doi.org/10.1007/s13163-024-00494-5}
}

@article{referee_4,
author = {Alexis, Michel and Lin, Lin and Mnatsakanyan, Gevorg and Thiele, Christoph and Wang, Jiasu},
title = {Infinite quantum signal processing for arbitrary Szegő functions},
journal = {Communications on Pure and Applied Mathematics},
volume = {79},
number = {1},
pages = {123-174},
doi = {https://doi.org/10.1002/cpa.70007},
url = {https://onlinelibrary.wiley.com/doi/abs/10.1002/cpa.70007},
eprint = {https://onlinelibrary.wiley.com/doi/pdf/10.1002/cpa.70007},
year = {2026}
}

@article{PRXQuantum.2.040203,
  title = {Grand Unification of Quantum Algorithms},
  author = {Martyn, John M. and Rossi, Zane M. and Tan, Andrew K. and Chuang, Isaac L.},
  journal = {PRX Quantum},
  volume = {2},
  issue = {4},
  pages = {040203},
  numpages = {40},
  year = {2021},
  month = {Dec},
  publisher = {American Physical Society},
  doi = {10.1103/PRXQuantum.2.040203},
  url = {https://link.aps.org/doi/10.1103/PRXQuantum.2.040203}
}

@inproceedings{10.1145/3313276.3316366,
author = {Gily\'{e}n, Andr\'{a}s and Su, Yuan and Low, Guang Hao and Wiebe, Nathan},
title = {Quantum singular value transformation and beyond: exponential improvements for quantum matrix arithmetics},
year = {2019},
isbn = {9781450367059},
publisher = {Association for Computing Machinery},
address = {New York, NY, USA},
url = {https://doi.org/10.1145/3313276.3316366},
doi = {10.1145/3313276.3316366},
booktitle = {Proceedings of the 51st Annual ACM SIGACT Symposium on Theory of Computing},
pages = {193–204},
numpages = {12},
location = {Phoenix, AZ, USA},
series = {STOC 2019}
}

@article{Lin_2025,
   title={Dissipative preparation of many-body quantum states: Toward practical quantum advantage},
   volume={1},
   ISSN={3066-0017},
   url={http://dx.doi.org/10.1063/5.0283315},
   DOI={10.1063/5.0283315},
   number={1},
   journal={APL Computational Physics},
   publisher={AIP Publishing},
   author={Lin, Lin},
   year={2025},
   month=sep }

@misc{lu2024quantumsignalprocessingquantum,
      title={Quantum Signal Processing and Quantum Singular Value Transformation on $U(N)$}, 
      author={Xi Lu and Yuan Liu and Hongwei Lin},
      year={2024},
      eprint={2408.01439},
      archivePrefix={arXiv},
      primaryClass={quant-ph},
      url={https://arxiv.org/abs/2408.01439}, 
}

@misc{laneve2024quantumsignalprocessingsun,
      title={Quantum signal processing over SU(N)}, 
      author={Lorenzo Laneve},
      year={2024},
      eprint={2311.03949},
      archivePrefix={arXiv},
      primaryClass={quant-ph},
      url={https://arxiv.org/abs/2311.03949}, 
}

@article{Poulin_2009,
   title={Sampling from the Thermal Quantum Gibbs State and Evaluating Partition Functions with a Quantum Computer},
   volume={103},
   ISSN={1079-7114},
   url={http://dx.doi.org/10.1103/PhysRevLett.103.220502},
   DOI={10.1103/physrevlett.103.220502},
   number={22},
   journal={Physical Review Letters},
   publisher={American Physical Society (APS)},
   author={Poulin, David and Wocjan, Pawel},
   year={2009},
   month=nov }

@misc{rouzé2024efficientthermalizationuniversalquantum,
      title={Efficient thermalization and universal quantum computing with quantum Gibbs samplers}, 
      author={Cambyse Rouzé and Daniel Stilck França and Álvaro M. Alhambra},
      year={2024},
      eprint={2403.12691},
      archivePrefix={arXiv},
      primaryClass={quant-ph},
      url={https://arxiv.org/abs/2403.12691}, 
}

@misc{rajakumar2024gibbssamplinggivesquantum,
      title={Gibbs Sampling gives Quantum Advantage at Constant Temperatures with O(1)-Local Hamiltonians}, 
      author={Joel Rajakumar and James D. Watson},
      year={2024},
      eprint={2408.01516},
      archivePrefix={arXiv},
      primaryClass={quant-ph},
      url={https://arxiv.org/abs/2408.01516}, 
}

@inproceedings{Bergamaschi_2024,
   title={Quantum Computational Advantage with Constant-Temperature Gibbs Sampling},
   url={http://dx.doi.org/10.1109/FOCS61266.2024.00071},
   DOI={10.1109/focs61266.2024.00071},
   booktitle={2024 IEEE 65th Annual Symposium on Foundations of Computer Science (FOCS)},
   publisher={IEEE},
   author={Bergamaschi, Thiago and Chen, Chi-Fang and Liu, Yunchao},
   year={2024},
   month=oct, pages={1063–1085} }

@article{Coopmans_2024,
   title={On the sample complexity of quantum Boltzmann machine learning},
   volume={7},
   ISSN={2399-3650},
   url={http://dx.doi.org/10.1038/s42005-024-01763-x},
   DOI={10.1038/s42005-024-01763-x},
   number={1},
   journal={Communications Physics},
   publisher={Springer Science and Business Media LLC},
   author={Coopmans, Luuk and Benedetti, Marcello},
   year={2024},
   month=aug }

@article{Low_2014,
   title={Quantum inference on Bayesian networks},
   volume={89},
   ISSN={1094-1622},
   url={http://dx.doi.org/10.1103/PhysRevA.89.062315},
   DOI={10.1103/physreva.89.062315},
   number={6},
   journal={Physical Review A},
   publisher={American Physical Society (APS)},
   author={Low, Guang Hao and Yoder, Theodore J. and Chuang, Isaac L.},
   year={2014},
   month=jun }

@article{FATHALLAH2024109307,
title = {Approximate inference on optimized quantum Bayesian networks},
journal = {International Journal of Approximate Reasoning},
volume = {175},
pages = {109307},
year = {2024},
issn = {0888-613X},
doi = {https://doi.org/10.1016/j.ijar.2024.109307},
url = {https://www.sciencedirect.com/science/article/pii/S0888613X24001944},
author = {Walid Fathallah and Nahla Ben Amor and Philippe Leray}
}

@article{Cohn_2020,
   title={Minimal effective Gibbs ansatz: A simple protocol for extracting an accurate thermal representation for quantum simulation},
   volume={102},
   ISSN={2469-9934},
   url={http://dx.doi.org/10.1103/PhysRevA.102.022622},
   DOI={10.1103/physreva.102.022622},
   number={2},
   journal={Physical Review A},
   publisher={American Physical Society (APS)},
   author={Cohn, J. and Yang, F. and Najafi, K. and Jones, B. and Freericks, J. K.},
   year={2020},
   month=aug }

@article{Rossi2022multivariable,
  doi = {10.22331/q-2022-09-20-811},
  url = {https://doi.org/10.22331/q-2022-09-20-811},
  title = {Multivariable quantum signal processing ({M}-{QSP}): prophecies of the two-headed oracle},
  author = {Rossi, Zane M. and Chuang, Isaac L.},
  journal = {{Quantum}},
  issn = {2521-327X},
  publisher = {{Verein zur F{\"{o}}rderung des Open Access Publizierens in den Quantenwissenschaften}},
  volume = {6},
  pages = {811},
  month = sep,
  year = {2022}
}

@article{pachon2009barycentric,
  title={Barycentric-Remez algorithms for best polynomial approximation in the chebfun system},
  author={Pach{\'o}n, Ricardo and Trefethen, Lloyd N},
  journal={BIT Numerical Mathematics},
  volume={49},
  number={4},
  pages={721--741},
  year={2009},
  publisher={Springer}
}

@article{tan2023errorcorrectionquantumalgorithms,
      title={Error Correction of Quantum Algorithms: Arbitrarily Accurate Recovery Of Noisy Quantum Signal Processing}, 
      author={Andrew K. Tan and Yuan Liu and Minh C. Tran and Isaac L. Chuang},
      year={2023},
      journal={arXiv:2301.08542},
      url={https://arxiv.org/abs/2301.08542}
}

@article{Kane2025blockencodingbosons,
  doi = {10.22331/q-2025-05-15-1747},
  url = {https://doi.org/10.22331/q-2025-05-15-1747},
  title = {Block encoding bosons by signal processing},
  author = {Kane, Christopher F. and Hariprakash, Siddharth and Modi, Neel S. and Kreshchuk, Michael and Bauer, Christian W},
  journal = {{Quantum}},
  issn = {2521-327X},
  publisher = {{Verein zur F{\"{o}}rderung des Open Access Publizierens in den Quantenwissenschaften}},
  volume = {9},
  pages = {1747},
  month = may,
  year = {2025}
}

@misc{motlagh2024groundstatepreparationdynamical,
      title={Ground State Preparation via Dynamical Cooling}, 
      author={Danial Motlagh and Modjtaba Shokrian Zini and Juan Miguel Arrazola and Nathan Wiebe},
      year={2024},
      eprint={2404.05810},
      archivePrefix={arXiv},
      primaryClass={quant-ph},
      url={https://arxiv.org/abs/2404.05810}, 
}

@inproceedings{10.1145/2591796.2591854,
author = {Berry, Dominic W. and Childs, Andrew M. and Cleve, Richard and Kothari, Robin and Somma, Rolando D.},
title = {Exponential improvement in precision for simulating sparse Hamiltonians},
year = {2014},
isbn = {9781450327107},
publisher = {Association for Computing Machinery},
address = {New York, NY, USA},
url = {https://doi.org/10.1145/2591796.2591854},
doi = {10.1145/2591796.2591854},
booktitle = {Proceedings of the Forty-Sixth Annual ACM Symposium on Theory of Computing},
pages = {283–292},
numpages = {10},
location = {New York, New York},
series = {STOC '14}
}

@article{PRXQuantum.5.020368,
  title = {Generalized Quantum Signal Processing},
  author = {Motlagh, Danial and Wiebe, Nathan},
  journal = {PRX Quantum},
  volume = {5},
  issue = {2},
  pages = {020368},
  numpages = {16},
  year = {2024},
  month = {Jun},
  publisher = {American Physical Society},
  doi = {10.1103/PRXQuantum.5.020368},
  url = {https://link.aps.org/doi/10.1103/PRXQuantum.5.020368}
}

@article{PhysRevA.110.012612,
  title = {Doubling the efficiency of Hamiltonian simulation via generalized quantum signal processing},
  author = {Berry, Dominic W. and Motlagh, Danial and Pantaleoni, Giacomo and Wiebe, Nathan},
  journal = {Phys. Rev. A},
  volume = {110},
  issue = {1},
  pages = {012612},
  numpages = {8},
  year = {2024},
  month = {Jul},
  publisher = {American Physical Society},
  doi = {10.1103/PhysRevA.110.012612},
  url = {https://link.aps.org/doi/10.1103/PhysRevA.110.012612}
}

@misc{laneve2025generalizedquantumsignalprocessing,
      title={Generalized Quantum Signal Processing and Non-Linear Fourier Transform are equivalent}, 
      author={Lorenzo Laneve},
      year={2025},
      eprint={2503.03026},
      archivePrefix={arXiv},
      primaryClass={quant-ph},
      url={https://arxiv.org/abs/2503.03026}, 
}

@article{PhysRevLett.118.010501,
  title = {Optimal Hamiltonian Simulation by Quantum Signal Processing},
  author = {Low, Guang Hao and Chuang, Isaac L.},
  journal = {Phys. Rev. Lett.},
  volume = {118},
  issue = {1},
  pages = {010501},
  numpages = {5},
  year = {2017},
  month = {Jan},
  publisher = {American Physical Society},
  doi = {10.1103/PhysRevLett.118.010501},
  url = {https://link.aps.org/doi/10.1103/PhysRevLett.118.010501}
}

@misc{bastidas2024complexification,
      title={Complexification of Quantum Signal Processing and its Ramifications}, 
      author={V. M. Bastidas and K. J. Joven},
      year={2024},
      eprint={2407.04780},
      archivePrefix={arXiv},
      primaryClass={quant-ph},
      url={https://arxiv.org/abs/2407.04780}, 
}

@article{dong2024feedforwardquantumsingularvalue,
      title={Feedforward Quantum Singular Value Transformation}, 
      author={Yulong Dong and Dong An and Murphy Yuezhen Niu},
      year={2024},
      journal={arXiv:2408.07803},
      url={https://arxiv.org/abs/2408.07803} 
}

@ARTICLE{11129874,
  author={Liu, Yuan and Martyn, John M. and Sinanan-Singh, Jasmine and Smith, Kevin C. and Girvin, Steven M. and Chuang, Isaac L.},
  journal={IEEE Transactions on Signal Processing}, 
  title={Toward Mixed Analog-Digital Quantum Signal Processing: Quantum AD/DA Conversion and the Fourier Transform}, 
  year={2025},
  volume={73},
  number={},
  pages={3641-3655},
  doi={10.1109/TSP.2025.3599462}}

@article{PhysRevA.103.042419,
  title = {Efficient phase-factor evaluation in quantum signal processing},
  author = {Dong, Yulong and Meng, Xiang and Whaley, K. Birgitta and Lin, Lin},
  journal = {Phys. Rev. A},
  volume = {103},
  issue = {4},
  pages = {042419},
  numpages = {22},
  year = {2021},
  month = {Apr},
  publisher = {American Physical Society},
  doi = {10.1103/PhysRevA.103.042419},
  url = {https://link.aps.org/doi/10.1103/PhysRevA.103.042419}
}

@article{singh2025nonabelianquantumsignalprocessing,
      title={Towards Non-Abelian Quantum Signal Processing: Efficient Control of Hybrid Continuous- and Discrete-Variable Architectures}, 
      author={Shraddha Singh and Baptiste Royer and Steven M. Girvin},
      year={2025},
      journal={arXiv:2504.19992},
      url={https://arxiv.org/abs/2504.19992}
}

@article{SinananSingh2024singleshotquantum,
  doi = {10.22331/q-2024-07-30-1427},
  url = {https://doi.org/10.22331/q-2024-07-30-1427},
  title = {Single-shot {Q}uantum {S}ignal {P}rocessing {I}nterferometry},
  author = {Sinanan-Singh, Jasmine and Mintzer, Gabriel L. and Chuang, Isaac L. and Liu, Yuan},
  journal = {{Quantum}},
  issn = {2521-327X},
  publisher = {{Verein zur F{\"{o}}rderung des Open Access Publizierens in den Quantenwissenschaften}},
  volume = {8},
  pages = {1427},
  month = jul,
  year = {2024}
}

@misc{chao2020findinganglesquantumsignal,
      title={Finding Angles for Quantum Signal Processing with Machine Precision}, 
      author={Rui Chao and Dawei Ding and Andras Gilyen and Cupjin Huang and Mario Szegedy},
      year={2020},
      eprint={2003.02831},
      archivePrefix={arXiv},
      primaryClass={quant-ph},
      url={https://arxiv.org/abs/2003.02831}, 
}

@article{Haah2019product,
  doi = {10.22331/q-2019-10-07-190},
  url = {https://doi.org/10.22331/q-2019-10-07-190},
  title = {Product {D}ecomposition of {P}eriodic {F}unctions in {Q}uantum {S}ignal {P}rocessing},
  author = {Haah, Jeongwan},
  journal = {{Quantum}},
  issn = {2521-327X},
  publisher = {{Verein zur F{\"{o}}rderung des Open Access Publizierens in den Quantenwissenschaften}},
  volume = {3},
  pages = {190},
  month = oct,
  year = {2019}
}

@article{Ying2022stablefactorization,
  doi = {10.22331/q-2022-10-20-842},
  url = {https://doi.org/10.22331/q-2022-10-20-842},
  title = {Stable factorization for phase factors of quantum signal processing},
  author = {Ying, Lexing},
  journal = {{Quantum}},
  issn = {2521-327X},
  publisher = {{Verein zur F{\"{o}}rderung des Open Access Publizierens in den Quantenwissenschaften}},
  volume = {6},
  pages = {842},
  month = oct,
  year = {2022}
}

@article{alexeev2024quantum,
  title={Quantum-centric supercomputing for materials science: A perspective on challenges and future directions},
  author={Alexeev, Yuri and Amsler, Maximilian and Barroca, Marco Antonio and Bassini, Sanzio and Battelle, Torey and Camps, Daan and Casanova, David and Choi, Young Jay and Chong, Frederic T and Chung, Charles and others},
  journal={Future Generation Computer Systems},
  volume={160},
  pages={666--710},
  year={2024},
  publisher={Elsevier}
}

@book{waterman2018introduction,
  title={Introduction to computational biology: maps, sequences and genomes},
  author={Waterman, Michael S},
  year={2018},
  publisher={Chapman and Hall/CRC}
}

@article{mcweeny1989method,
  title={Method of molecular quantum mechanics},
  author={McWeeny, Roy},
  journal={2nd edition},
  year={1989}
}

@book{thijssen2007computational,
  title={Computational physics},
  author={Thijssen, Jos},
  year={2007},
  publisher={Cambridge university press}
}

@book{Djordjevic2021,
  author    = {Ivan B. Djordjevic},
  title     = {Quantum Information Processing, Quantum Computing, and Quantum Error Correction: An Engineering Approach},
  edition   = {2},
  year      = {2021},
  publisher = {Academic Press},
  address   = {Boston, MA},
  isbn      = {978-0-12-821982-9},
  isbn13    = {9780128219829}
}

@article{gottesman2022ft,
      title={Opportunities and Challenges in Fault-Tolerant Quantum Computation}, 
      author={Daniel Gottesman},
      year={2022},
      journal={arXiv:2210.15844},
      url={https://arxiv.org/abs/2210.15844}
}

@article{altman2021quantum,
  title={Quantum simulators: Architectures and opportunities},
  author={Altman, Ehud and Brown, Kenneth R and Carleo, Giuseppe and Carr, Lincoln D and Demler, Eugene and Chin, Cheng and DeMarco, Brian and Economou, Sophia E and Eriksson, Mark A and Fu, Kai-Mei C and others},
  journal={PRX quantum},
  volume={2},
  number={1},
  pages={017003},
  year={2021},
  publisher={APS}
}

@ARTICLE{10293178,
  author={Vale, Rafaella and Azevedo, Thiago Melo D. and Araújo, Ismael C. S. and Araujo, Israel F. and da Silva, Adenilton J.},
  journal={IEEE Transactions on Computer-Aided Design of Integrated Circuits and Systems}, 
  title={Circuit Decomposition of Multicontrolled Special Unitary Single-Qubit Gates}, 
  year={2024},
  volume={43},
  number={3},
  pages={802-811},
  doi={10.1109/TCAD.2023.3327102}}

@misc{shor1997faulttolerantquantumcomputation,
      title={Fault-tolerant quantum computation}, 
      author={Peter W. Shor},
      year={1997},
      eprint={quant-ph/9605011},
      archivePrefix={arXiv},
      primaryClass={quant-ph},
      url={https://arxiv.org/abs/quant-ph/9605011}, 
}

@misc{yamamoto2024robustanglefindinggeneralized,
      title={Robust Angle Finding for Generalized Quantum Signal Processing}, 
      author={Shuntaro Yamamoto and Nobuyuki Yoshioka},
      year={2024},
      eprint={2402.03016},
      archivePrefix={arXiv},
      primaryClass={quant-ph},
      url={https://arxiv.org/abs/2402.03016}, 
}

@article{vandersypen2004nmr,
  title={NMR techniques for quantum control and computation},
  author={Vandersypen, Lieven MK and Chuang, Isaac L},
  journal={Reviews of modern physics},
  volume={76},
  number={4},
  pages={1037--1069},
  year={2004},
  publisher={APS}
}

@article{10.1063/5.0124385,
    author = {Martyn, John M. and Liu, Yuan and Chin, Zachary E. and Chuang, Isaac L.},
    title = {Efficient fully-coherent quantum signal processing algorithms for real-time dynamics simulation},
    journal = {The Journal of Chemical Physics},
    volume = {158},
    number = {2},
    pages = {024106},
    year = {2023},
    month = {01},
    issn = {0021-9606},
    doi = {10.1063/5.0124385},
    url = {https://doi.org/10.1063/5.0124385}
}

@misc{zhang2025quantumimaginarytimeevolutionpolynomial,
      title={Quantum Imaginary-Time Evolution with Polynomial Resources in Time}, 
      author={Lei Zhang and Jizhe Lai and Xian Wu and Xin Wang},
      year={2025},
      eprint={2507.00908},
      archivePrefix={arXiv},
      primaryClass={quant-ph},
      url={https://arxiv.org/abs/2507.00908}, 
}

@misc{dawson2005solovaykitaevalgorithm,
      title={The Solovay-Kitaev algorithm}, 
      author={Christopher M. Dawson and Michael A. Nielsen},
      year={2005},
      eprint={quant-ph/0505030},
      archivePrefix={arXiv},
      primaryClass={quant-ph},
      url={https://arxiv.org/abs/quant-ph/0505030}, 
}

@article{xlpd-fb1g,
  title = {Gate construction of block-encoding for Hamiltonians needed for simulating partial differential equations},
  author = {Guseynov, Nikita and Huang, Xiajie and Liu, Nana},
  journal = {Phys. Rev. Res.},
  volume = {7},
  issue = {3},
  pages = {033100},
  numpages = {23},
  year = {2025},
  month = {Jul},
  publisher = {American Physical Society},
  doi = {10.1103/xlpd-fb1g},
  url = {https://link.aps.org/doi/10.1103/xlpd-fb1g}
}

@article{ang2024arquin,
  title={ARQUIN: architectures for multinode superconducting quantum computers},
  author={Ang, James and Carini, Gabriella and Chen, Yanzhu and Chuang, Isaac and Demarco, Michael and Economou, Sophia and Eickbusch, Alec and Faraon, Andrei and Fu, Kai-Mei and Girvin, Steven and others},
  journal={ACM Transactions on Quantum Computing},
  volume={5},
  number={3},
  pages={1--59},
  year={2024},
  publisher={ACM New York, NY}
}

@article{Main_2025,
   title={Distributed quantum computing across an optical network link},
   volume={638},
   ISSN={1476-4687},
   url={http://dx.doi.org/10.1038/s41586-024-08404-x},
   DOI={10.1038/s41586-024-08404-x},
   number={8050},
   journal={Nature},
   publisher={Springer Science and Business Media LLC},
   author={Main, D. and Drmota, P. and Nadlinger, D. P. and Ainley, E. M. and Agrawal, A. and Nichol, B. C. and Srinivas, R. and Araneda, G. and Lucas, D. M.},
   year={2025},
   month=feb, pages={383–388} }

@article{li2024heterogeneous,
  title={Heterogeneous integration of spin--photon interfaces with a CMOS platform},
  author={Li, Linsen and Santis, Lorenzo De and Harris, Isaac BW and Chen, Kevin C and Gao, Yihuai and Christen, Ian and Choi, Hyeongrak and Trusheim, Matthew and Song, Yixuan and Errando-Herranz, Carlos and others},
  journal={Nature},
  volume={630},
  number={8015},
  pages={70--76},
  year={2024},
  publisher={Nature Publishing Group UK London}
}

@article{unitary-dilation,
 ISSN = {03794024, 18417744},
 URL = {http://www.jstor.org/stable/24718829},
 author = {PEI YUAN WU},
 journal = {Journal of Operator Theory},
 number = {1},
 pages = {25--42},
 publisher = {Theta Foundation},
 title = {UNITARY DILATIONS AND NUMERICAL RANGES},
 urldate = {2025-10-19},
 volume = {38},
 year = {1997}
}

@article{beals2013efficient,
  title={Efficient distributed quantum computing},
  author={Beals, Robert and Brierley, Stephen and Gray, Oliver and Harrow, Aram W and Kutin, Samuel and Linden, Noah and Shepherd, Dan and Stather, Mark},
  journal={Proceedings of the Royal Society A: Mathematical, Physical and Engineering Sciences},
  volume={469},
  number={2153},
  pages={20120686},
  year={2013},
  publisher={The Royal Society Publishing}
}

@article{martyn2025parallel,
  title={Parallel quantum signal processing via polynomial factorization},
  author={Martyn, John M and Rossi, Zane M and Cheng, Kevin Z and Liu, Yuan and Chuang, Isaac L},
  journal={Quantum},
  volume={9},
  pages={1834},
  year={2025},
  publisher={Verein zur F{\"o}rderung des Open Access Publizierens in den Quantenwissenschaften}
}

@article{Zhang2024parallelquantum,
  doi = {10.22331/q-2024-01-15-1228},
  url = {https://doi.org/10.22331/q-2024-01-15-1228},
  title = {Parallel {Q}uantum {A}lgorithm for {H}amiltonian {S}imulation},
  author = {Zhang, Zhicheng and Wang, Qisheng and Ying, Mingsheng},
  journal = {{Quantum}},
  issn = {2521-327X},
  publisher = {{Verein zur F{\"{o}}rderung des Open Access Publizierens in den Quantenwissenschaften}},
  volume = {8},
  pages = {1228},
  month = jan,
  year = {2024}
}

@article{qracd2025,
      title={COMPAS: A Distributed Multi-Party SWAP Test for Parallel Quantum Algorithms}, 
      author={Goldstein-Gelb, Brayden and Liu, Kun and Martyn, John M. and Zhou, Hengyun (Harry)  and Ding, Yongshan and Liu, Yuan},
      year={2025},
      journal={Under review.}
}

@article{PRXQuantum.5.020332,
  title = {Simulating Open Quantum Systems Using Hamiltonian Simulations},
  author = {Ding, Zhiyan and Li, Xiantao and Lin, Lin},
  journal = {PRX Quantum},
  volume = {5},
  issue = {2},
  pages = {020332},
  numpages = {29},
  year = {2024},
  month = {May},
  publisher = {American Physical Society},
  doi = {10.1103/PRXQuantum.5.020332},
  url = {https://link.aps.org/doi/10.1103/PRXQuantum.5.020332}
}

@article{PRXQuantum.2.030342, 
  title = {Quantum Algorithm for Time-Dependent Hamiltonian Simulation by Permutation Expansion},
  author = {Chen, Yi-Hsiang and Kalev, Amir and Hen, Itay},
  journal = {PRX Quantum},
  volume = {2},
  issue = {3},
  pages = {030342},
  numpages = {17},
  year = {2021},
  month = {Sep},
  publisher = {American Physical Society},
  doi = {10.1103/PRXQuantum.2.030342},
  url = {https://link.aps.org/doi/10.1103/PRXQuantum.2.030342}
}

@article{araujo2021divide,
  title={A divide-and-conquer algorithm for quantum state preparation},
  author={Araujo, Ivan F. and Park, Daniel K. and Petruccione, Francesco and others},
  journal={Scientific Reports},
  volume={11},
  number={1},
  pages={6329},
  year={2021},
  publisher={Nature Publishing Group},
  doi={10.1038/s41598-021-85474-1},
  url={https://doi.org/10.1038/s41598-021-85474-1}
}

@misc{bidart2025quantumcomputinggroundstate,
      title={Quantum Computing Beyond Ground State Electronic Structure: A Review of Progress Toward Quantum Chemistry Out of the Ground State}, 
      author={Alan Bidart and Prateek Vaish and Tilas Kabengele and Yaoqi Pang and Yuan Liu and Brenda M. Rubenstein},
      year={2025},
      eprint={2509.19709},
      archivePrefix={arXiv},
      primaryClass={physics.chem-ph},
      url={https://arxiv.org/abs/2509.19709}, 
}

@article{Babbush_2019,
   title={Quantum simulation of chemistry with sublinear scaling in basis size},
   volume={5},
   ISSN={2056-6387},
   url={http://dx.doi.org/10.1038/s41534-019-0199-y},
   DOI={10.1038/s41534-019-0199-y},
   number={1},
   journal={npj Quantum Information},
   publisher={Springer Science and Business Media LLC},
   author={Babbush, Ryan and Berry, Dominic W. and McClean, Jarrod R. and Neven, Hartmut},
   year={2019},
   month=nov }

@article{Low_2023,
   title={Complexity of Implementing Trotter Steps},
   volume={4},
   ISSN={2691-3399},
   url={http://dx.doi.org/10.1103/PRXQuantum.4.020323},
   DOI={10.1103/prxquantum.4.020323},
   number={2},
   journal={PRX Quantum},
   publisher={American Physical Society (APS)},
   author={Low, Guang Hao and Su, Yuan and Tong, Yu and Tran, Minh C.},
   year={2023},
   month=may }

@misc{powers2022exploringfinitetemperatureproperties,
      title={Exploring Finite Temperature Properties of Materials with Quantum Computers}, 
      author={Connor Powers and Lindsay Bassman Oftelie and Daan Camps and Wibe A. de Jong},
      year={2022},
      eprint={2109.01619},
      archivePrefix={arXiv},
      primaryClass={quant-ph},
      url={https://arxiv.org/abs/2109.01619}, 
}

@article{McArdle_2019,
   title={Variational ansatz-based quantum simulation of imaginary time evolution},
   volume={5},
   ISSN={2056-6387},
   url={http://dx.doi.org/10.1038/s41534-019-0187-2},
   DOI={10.1038/s41534-019-0187-2},
   number={1},
   journal={npj Quantum Information},
   publisher={Springer Science and Business Media LLC},
   author={McArdle, Sam and Jones, Tyson and Endo, Suguru and Li, Ying and Benjamin, Simon C. and Yuan, Xiao},
   year={2019},
   month=sep }

@misc{kosugi2022probabilisticimaginarytimeevolutionusing,
      title={Probabilistic imaginary-time evolution by using forward and backward real-time evolution with a single ancilla: first-quantized eigensolver of quantum chemistry for ground states}, 
      author={Taichi Kosugi and Yusuke Nishiya and Hirofumi Nishi and Yu-ichiro Matsushita},
      year={2022},
      eprint={2111.12471},
      archivePrefix={arXiv},
      primaryClass={quant-ph},
      url={https://arxiv.org/abs/2111.12471}, 
}

@article{PhysRevA.102.052411,
  title = {Approximate quantum circuit synthesis using block encodings},
  author = {Camps, Daan and Van Beeumen, Roel},
  journal = {Phys. Rev. A},
  volume = {102},
  issue = {5},
  pages = {052411},
  numpages = {7},
  year = {2020},
  month = {Nov},
  publisher = {American Physical Society},
  doi = {10.1103/PhysRevA.102.052411},
  url = {https://link.aps.org/doi/10.1103/PhysRevA.102.052411}
}

@misc{li2025variationalquantumalgorithmunitary,
      title={Variational Quantum Algorithm for Unitary Dilation}, 
      author={S. X. Li and Keren Li and J. B. You and Y. -H. Chen and Clemens Gneiting and Franco Nori and X. Q. Shao},
      year={2025},
      eprint={2510.19157},
      archivePrefix={arXiv},
      primaryClass={quant-ph},
      url={https://arxiv.org/abs/2510.19157}, 
}

@article{doi:10.1137/22M1484298,
author = {Camps, Daan and Lin, Lin and Van Beeumen, Roel and Yang, Chao},
title = {Explicit Quantum Circuits for Block Encodings of Certain Sparse Matrices},
journal = {SIAM Journal on Matrix Analysis and Applications},
volume = {45},
number = {1},
pages = {801-827},
year = {2024},
doi = {10.1137/22M1484298},
URL = {https://doi.org/10.1137/22M1484298},
eprint = {https://doi.org/10.1137/22M1484298}
}

@misc{eisert2025mindgapsfraughtroad,
      title={Mind the gaps: The fraught road to quantum advantage}, 
      author={Jens Eisert and John Preskill},
      year={2025},
      eprint={2510.19928},
      archivePrefix={arXiv},
      primaryClass={quant-ph},
      url={https://arxiv.org/abs/2510.19928}, 
}

@INPROCEEDINGS{9951292,
  author={Camps, Daan and Van Beeumen, Roel},
  booktitle={2022 IEEE International Conference on Quantum Computing and Engineering (QCE)}, 
  title={FABLE: Fast Approximate Quantum Circuits for Block-Encodings}, 
  year={2022},
  volume={},
  number={},
  pages={104-113},
  doi={10.1109/QCE53715.2022.00029}}

@misc{sfablelsfable,
      title={S-FABLE and LS-FABLE: Fast approximate block-encoding algorithms for unstructured sparse matrices}, 
      author={Parker Kuklinski and Benjamin Rempfer},
      year={2024},
      eprint={2401.04234},
      archivePrefix={arXiv},
      primaryClass={quant-ph},
      url={https://arxiv.org/abs/2401.04234}, 
}

@article{PhysRevA.52.3457,
  title = {Elementary gates for quantum computation},
  author = {Barenco, Adriano and Bennett, Charles H. and Cleve, Richard and DiVincenzo, David P. and Margolus, Norman and Shor, Peter and Sleator, Tycho and Smolin, John A. and Weinfurter, Harald},
  journal = {Phys. Rev. A},
  volume = {52},
  issue = {5},
  pages = {3457--3467},
  numpages = {0},
  year = {1995},
  month = {Nov},
  publisher = {American Physical Society},
  doi = {10.1103/PhysRevA.52.3457},
  url = {https://link.aps.org/doi/10.1103/PhysRevA.52.3457}
}

@article{PhysRevB.103.014111,
  title = {Chiral symmetry in non-Hermitian systems: Product rule and Clifford algebra},
  author = {Rivero, Jose D. H. and Ge, Li},
  journal = {Phys. Rev. B},
  volume = {103},
  issue = {1},
  pages = {014111},
  numpages = {11},
  year = {2021},
  month = {Jan},
  publisher = {American Physical Society},
  doi = {10.1103/PhysRevB.103.014111},
  url = {https://link.aps.org/doi/10.1103/PhysRevB.103.014111}
}

@article{liu_bootstrap_2023,
	title = {Bootstrap Embedding on a Quantum Computer},
	volume = {19},
	issn = {1549-9618},
	url = {https://doi.org/10.1021/acs.jctc.3c00012},
	doi = {10.1021/acs.jctc.3c00012},
	pages = {2230--2247},
	number = {8},
	journaltitle = {Journal of Chemical Theory and Computation},
	journal = {J. Chem. Theory Comput.},
	author = {Liu, Yuan and Meitei, Oinam R. and Chin, Zachary E. and Dutt, Arkopal and Tao, Max and Chuang, Isaac L. and Van Voorhis, Troy},
	urldate = {2024-05-02},
	date = {2023-04-25},
    year = {2023},
	note = {Publisher: American Chemical Society},
}

@article{knill_optimal_2007,
	title = {Optimal quantum measurements of expectation values of observables},
	volume = {75},
	url = {https://link.aps.org/doi/10.1103/PhysRevA.75.012328},
	doi = {10.1103/PhysRevA.75.012328},
	pages = {012328},
	number = {1},
	journaltitle = {Physical Review A},
	journal = {Phys. Rev. A},
	author = {Knill, Emanuel and Ortiz, Gerardo and Somma, Rolando D.},
	urldate = {2023-05-03},
    year = {2023},
	date = {2007-01-24},
	note = {Publisher: American Physical Society},
}

@article{rall_amplitude_2023,
	title = {Amplitude Estimation from Quantum Signal Processing},
	volume = {7},
	url = {https://quantum-journal.org/papers/q-2023-03-02-937/},
	doi = {10.22331/q-2023-03-02-937},
	pages = {937},
	journal = {Quantum},
	author = {Rall, Patrick and Fuller, Bryce},
	urldate = {2024-07-05},
	date = {2023-03-02},
    year = {2023},
	langid = {british},
	note = {Publisher: Verein zur Förderung des Open Access Publizierens in den Quantenwissenschaften},
}

@article{gonthier_measurements_2022,
	title = {Measurements as a roadblock to near-term practical quantum advantage in chemistry: Resource analysis},
	volume = {4},
	url = {https://link.aps.org/doi/10.1103/PhysRevResearch.4.033154},
	doi = {10.1103/PhysRevResearch.4.033154},
	shorttitle = {Measurements as a roadblock to near-term practical quantum advantage in chemistry},
	pages = {033154},
	number = {3},
	journaltitle = {Physical Review Research},
	journal = {Phys. Rev. Res.},
	author = {Gonthier, Jérôme F. and Radin, Maxwell D. and Buda, Corneliu and Doskocil, Eric J. and Abuan, Clena M. and Romero, Jhonathan},
	urldate = {2023-05-03},
	date = {2022-08-26},
    year = {2022},
	note = {Publisher: American Physical Society},
}

@article{wang_accelerated_2019,
	title = {Accelerated Variational Quantum Eigensolver},
	volume = {122},
	url = {https://link.aps.org/doi/10.1103/PhysRevLett.122.140504},
	doi = {10.1103/PhysRevLett.122.140504},
	pages = {140504},
	number = {14},
	journaltitle = {Physical Review Letters},
	journal = {Phys. Rev. Lett.},
	author = {Wang, Daochen and Higgott, Oscar and Brierley, Stephen},
	urldate = {2023-05-02},
	date = {2019-04-12},
    year = {2019},
	note = {Publisher: American Physical Society},
}

@misc{farhi2014quantumapproximateoptimizationalgorithm,
      title={A Quantum Approximate Optimization Algorithm}, 
      author={Edward Farhi and Jeffrey Goldstone and Sam Gutmann},
      year={2014},
      eprint={1411.4028},
      archivePrefix={arXiv},
      primaryClass={quant-ph},
      url={https://arxiv.org/abs/1411.4028}, 
}

@misc{mazumder2025starterproblemssolvingquadratic,
      title={Five Starter Problems: Solving Quadratic Unconstrained Binary Optimization Models on Quantum Computers}, 
      author={Arul Mazumder and Sridhar Tayur},
      year={2025},
      eprint={2401.08989},
      archivePrefix={arXiv},
      primaryClass={quant-ph},
      url={https://arxiv.org/abs/2401.08989}, 
}

@article{Born1928,
  author = {Born, M. and Fock, V.},
  title = {Beweis des Adiabatensatzes},
  journal = {Zeitschrift für Physik},
  year = {1928},
  volume = {51},
  number = {3},
  pages = {165--180},
  issn = {0044-3328},
  doi = {10.1007/BF01343193},
  url = {https://doi.org/10.1007/BF01343193}
}

@article{McClean2018,
  author = {McClean, Jarrod R. and Boixo, Sergio and Smelyanskiy, Vadim N. and Babbush, Ryan and Neven, Hartmut},
  title = {Barren plateaus in quantum neural network training landscapes},
  journal = {Nature Communications},
  year = {2018},
  volume = {9},
  number = {1},
  pages = {4812},
  issn = {2041-1723},
  doi = {10.1038/s41467-018-07090-4},
  url = {https://doi.org/10.1038/s41467-018-07090-4}
}

@article{peruzzo_variational_2014,
	title = {A variational eigenvalue solver on a photonic quantum processor},
	volume = {5},
	rights = {2014 The Author(s)},
	issn = {2041-1723},
	url = {https://www.nature.com/articles/ncomms5213},
	doi = {10.1038/ncomms5213},
	pages = {4213},
	number = {1},
	journaltitle = {Nature Communications},
	journal = {Nat Commun},
	author = {Peruzzo, Alberto and {McClean}, Jarrod and Shadbolt, Peter and Yung, Man-Hong and Zhou, Xiao-Qi and Love, Peter J. and Aspuru-Guzik, Alán and O’Brien, Jeremy L.},
	urldate = {2023-04-29},
	date = {2014-07-23},
    year = {2014},
	langid = {english},
	note = {Number: 1
Publisher: Nature Publishing Group},
}

@article{wang_minimizing_2021,
	title = {Minimizing Estimation Runtime on Noisy Quantum Computers},
	volume = {2},
	url = {https://link.aps.org/doi/10.1103/PRXQuantum.2.010346},
	doi = {10.1103/PRXQuantum.2.010346},
	pages = {010346},
	number = {1},
	journaltitle = {{PRX} Quantum},
	journal = {{PRX} Quantum},
	author = {Wang, Guoming and Koh, Dax Enshan and Johnson, Peter D. and Cao, Yudong},
	urldate = {2023-05-03},
	date = {2021-03-19},
    year = {2021},
	note = {Publisher: American Physical Society},
}

@article{kim_evidence_2023,
	title = {Evidence for the utility of quantum computing before fault tolerance},
	volume = {618},
	copyright = {2023 The Author(s)},
	issn = {1476-4687},
	url = {https://www.nature.com/articles/s41586-023-06096-3},
	doi = {10.1038/s41586-023-06096-3},
	language = {en},
	number = {7965},
	urldate = {2023-06-15},
	journal = {Nature},
	author = {Kim, Youngseok and Eddins, Andrew and Anand, Sajant and Wei, Ken Xuan and van den Berg, Ewout and Rosenblatt, Sami and Nayfeh, Hasan and Wu, Yantao and Zaletel, Michael and Temme, Kristan and Kandala, Abhinav},
	month = jun,
	year = {2023},
	note = {Number: 7965
Publisher: Nature Publishing Group},
	pages = {500--505},
}

@article{cortes_quantum_2022,
	title = {Quantum {Krylov} subspace algorithms for ground- and excited-state energy estimation},
	volume = {105},
	url = {https://link.aps.org/doi/10.1103/PhysRevA.105.022417},
	doi = {10.1103/PhysRevA.105.022417},
	number = {2},
	urldate = {2025-11-03},
	journal = {Physical Review A},
	author = {Cortes, Cristian L. and Gray, Stephen K.},
	month = feb,
	year = {2022},
	note = {Publisher: American Physical Society},
	pages = {022417},
}

@article{yoshioka_krylov_2025,
	title = {Krylov diagonalization of large many-body {Hamiltonians} on a quantum processor},
	volume = {16},
	copyright = {2025 The Author(s)},
	issn = {2041-1723},
	url = {https://www.nature.com/articles/s41467-025-59716-z},
	doi = {10.1038/s41467-025-59716-z},
	language = {en},
	number = {1},
	urldate = {2025-09-22},
	journal = {Nature Communications},
	author = {Yoshioka, Nobuyuki and Amico, Mirko and Kirby, William and Jurcevic, Petar and Dutt, Arkopal and Fuller, Bryce and Garion, Shelly and Haas, Holger and Hamamura, Ikko and Ivrii, Alexander and Majumdar, Ritajit and Minev, Zlatko and Motta, Mario and Pokharel, Bibek and Rivero, Pedro and Sharma, Kunal and Wood, Christopher J. and Javadi-Abhari, Ali and Mezzacapo, Antonio},
	month = jun,
	year = {2025},
	note = {Publisher: Nature Publishing Group},
	pages = {5014},
}

@article{BayesianQPE,
  title = {Efficient Bayesian Phase Estimation},
  author = {Wiebe, Nathan and Granade, Chris},
  journal = {Phys. Rev. Lett.},
  volume = {117},
  issue = {1},
  pages = {010503},
  numpages = {6},
  year = {2016},
  month = {Jun},
  publisher = {American Physical Society},
  doi = {10.1103/PhysRevLett.117.010503},
  url = {https://link.aps.org/doi/10.1103/PhysRevLett.117.010503}
}

@article{BayesianQPE_experiment,
  title = {Experimental Bayesian Quantum Phase Estimation on a Silicon Photonic Chip},
  author = {Paesani, S. and Gentile, A. A. and Santagati, R. and Wang, J. and Wiebe, N. and Tew, D. P. and O'Brien, J. L. and Thompson, M. G.},
  journal = {Phys. Rev. Lett.},
  volume = {118},
  issue = {10},
  pages = {100503},
  numpages = {6},
  year = {2017},
  month = {Mar},
  publisher = {American Physical Society},
  doi = {10.1103/PhysRevLett.118.100503},
  url = {https://link.aps.org/doi/10.1103/PhysRevLett.118.100503}
}

@article{10.1063/1.482053,
    author = {Soudackov, Alexander and Hammes-Schiffer, Sharon},
    title = {Derivation of rate expressions for nonadiabatic proton-coupled electron transfer reactions in solution},
    journal = {The Journal of Chemical Physics},
    volume = {113},
    number = {6},
    pages = {2385-2396},
    year = {2000},
    month = {08},
    issn = {0021-9606},
    doi = {10.1063/1.482053},
    url = {https://doi.org/10.1063/1.482053},
    eprint = {https://pubs.aip.org/aip/jcp/article-pdf/113/6/2385/19175240/2385_1_online.pdf},
}

@article{10.1063/1.1814635,
    author = {Soudackov, Alexander and Hatcher, Elizabeth and Hammes-Schiffer, Sharon},
    title = {Quantum and dynamical effects of proton donor-acceptor vibrational motion in nonadiabatic proton-coupled electron transfer reactions},
    journal = {The Journal of Chemical Physics},
    volume = {122},
    number = {1},
    pages = {014505},
    year = {2004},
    month = {12},
    issn = {0021-9606},
    doi = {10.1063/1.1814635},
    url = {https://doi.org/10.1063/1.1814635},
    eprint = {https://pubs.aip.org/aip/jcp/article-pdf/doi/10.1063/1.1814635/10864939/014505_1_online.pdf},
}

@article{doi:10.1021/jp400401f,
author = {Ruckenbauer, Matthias and Barbatti, Mario and M{\"u}ller, Thomas and Lischka, Hans},
title = {Nonadiabatic Photodynamics of a Retinal Model in Polar and Nonpolar Environment},
journal = {The Journal of Physical Chemistry A},
volume = {117},
number = {13},
pages = {2790-2799},
year = {2013},
doi = {10.1021/jp400401f},
    note ={PMID: 23470211},

URL = { 
    
        https://doi.org/10.1021/jp400401f
    
    

},
eprint = { 
    
        https://doi.org/10.1021/jp400401f
    
    

}

}

@article{STIER200333,
title = {Non-adiabatic molecular dynamics simulation of ultrafast solar cell electron transfer},
journal = {Journal of Molecular Structure: THEOCHEM},
volume = {630},
number = {1},
pages = {33-43},
year = {2003},
note = {WATOC '02 Special Issue},
issn = {0166-1280},
doi = {https://doi.org/10.1016/S0166-1280(03)00167-2},
url = {https://www.sciencedirect.com/science/article/pii/S0166128003001672},
author = {William Stier and Oleg V. Prezhdo}
}

@misc{white2023nestedgaussletbasissets,
      title={Nested Gausslet Basis Sets}, 
      author={Steven R. White and Michael J. Lindsey},
      year={2023},
      eprint={2309.10704},
      archivePrefix={arXiv},
      primaryClass={physics.chem-ph},
      url={https://arxiv.org/abs/2309.10704}, 
}

@article{Georges_2025,
   title={Quantum simulations of chemistry in first quantization with any basis set},
   volume={11},
   ISSN={2056-6387},
   url={http://dx.doi.org/10.1038/s41534-025-00987-1},
   DOI={10.1038/s41534-025-00987-1},
   number={1},
   journal={npj Quantum Information},
   publisher={Springer Science and Business Media LLC},
   author={Georges, Timothy N. and Bothe, Marius and Sünderhauf, Christoph and Berntson, Bjorn K. and Izsák, Róbert and Ivanov, Aleksei V.},
   year={2025},
   month=apr }

@article{doi:10.1021/jp805876e,
author = {Hammes-Schiffer, Sharon and Soudackov, Alexander V.},
title = {Proton-Coupled Electron Transfer in Solution, Proteins, and Electrochemistry},
journal = {The Journal of Physical Chemistry B},
volume = {112},
number = {45},
pages = {14108-14123},
year = {2008},
doi = {10.1021/jp805876e},
    note ={PMID: 18842015},

URL = { 
    
        https://doi.org/10.1021/jp805876e
    
    

},
eprint = { 
    
        https://doi.org/10.1021/jp805876e
    
    

}

}

@article{aleksandrowicz2019qiskit,
  title        = {Qiskit: An open-source framework for quantum computing},
  author       = {Aleksandrowicz, Gadi and Alexander, Thomas and Barkoutsos, Panagiotis and Bello, Lucia and Ben-Haim, Yael and Bucher, Dominik and Cabrera, Diego and Carballo-Franquis, Antonio and Chen, Adrian and Chen, Chun-Fu and others},
  journal      = {Zenodo},
  year         = {2019},
  doi          = {10.5281/zenodo.2562111}
}

@article{cirq_developers2023cirq,
  title        = {Cirq: A python framework for creating, editing, and invoking Noisy Intermediate Scale Quantum (NISQ) circuits},
  author       = {Cirq Developers},
  journal      = {Zenodo},
  year         = {2023},
  doi          = {10.5281/zenodo.4062499}
}

@article{Silva2023,
  author    = {Thais L. Silva and M{\'a}rcio M. Taddei and Stefano Carrazza and Leandro Aolita},
  title     = {Fragmented imaginary-time evolution for early-stage quantum signal processors},
  journal   = {Scientific Reports},
  year      = {2023},
  volume    = {13},
  number    = {1},
  pages     = {18258},
  doi       = {10.1038/s41598-023-45540-2},
  url       = {https://doi.org/10.1038/s41598-023-45540-2}
}

@article{sivarajah2020tket,
  title        = {t|ket⟩: A retargetable compiler for NISQ devices},
  author       = {Sivarajah, Seyon and Dilkes, Siân and Cowtan, Alexander and Simmons, Will and Edgington, Andrew and Duncan, Ross},
  journal      = {Quantum Science and Technology},
  volume       = {6},
  number       = {1},
  pages        = {014003},
  year         = {2020},
  publisher    = {IOP Publishing}
}

@article{PhysRevResearch.7.013306,
  title = {Exact block encoding of imaginary time evolution with universal quantum neural networks},
  author = {Rrapaj, Ermal and Rule, Evan},
  journal = {Phys. Rev. Res.},
  volume = {7},
  issue = {1},
  pages = {013306},
  numpages = {14},
  year = {2025},
  month = {Mar},
  publisher = {American Physical Society},
  doi = {10.1103/PhysRevResearch.7.013306},
  url = {https://link.aps.org/doi/10.1103/PhysRevResearch.7.013306}
}

@article{gtq3-j37b,
  title = {Classical optimization with imaginary-time block encoding on quantum computers: The MaxCut problem},
  author = {Zhong, Dawei and Francis, Akhil and Rrapaj, Ermal},
  journal = {Phys. Rev. A},
  volume = {112},
  issue = {4},
  pages = {042420},
  numpages = {10},
  year = {2025},
  month = {Oct},
  publisher = {American Physical Society},
  doi = {10.1103/gtq3-j37b},
  url = {https://link.aps.org/doi/10.1103/gtq3-j37b}
}

@misc{fable_repo,
  title        = {FABLE: Fast Approximate Block-Encoding Library},
  howpublished ={\url{https://github.com/QuantumComputingLab/fable}},
  author       = {{Quantum Computing Lab}},
  year         = {2024},
  note         = {GitHub repository, accessed 2025-11-01}
}

@misc{pennylane_fable,
  title        = {PennyLane FABLE Operator and Block-Encoding Tutorials},
  howpublished ={\url{https://pennylane.ai/qml/demos/tutorial_block_encoding}},
  author       = {{PennyLane Team}},
  year         = {2024},
  publisher    = {Xanadu},
  note         = {PennyLane documentation, accessed 2025-11-01}
}

@misc{classiq_vqls,
  title        = {Variational Quantum Linear Solver (VQLS) with LCU},
  howpublished ={\url{https://docs.classiq.io/latest/explore/algorithms/vqls/lcu_vqls/vqls_with_lcu/}},
  author       = {{Classiq Technologies}},
  year         = {2025},
  note         = {Online documentation, accessed 2025-11-01}
}

@misc{openfermion_lcu,
  title        = {OpenFermion: LCU Utilities for Hamiltonian Synthesis},
  howpublished ={\url{https://quantumai.google/reference/python/openfermion/circuits/lcu_util}},
  author       = {{Google Quantum AI}},
  year         = {2024},
  note         = {Software library, accessed 2025-11-01}
}

@misc{riverlane_lcu,
  title        = {Pauli-LCU Repository},
  howpublished ={\url{https://github.com/riverlane/pauli_lcu}},
  author       = {{Riverlane Ltd.}},
  year         = {2024},
  note         = {GitHub repository, accessed 2025-11-01}
}

@misc{qualtran_docs,
  title        = {Qualtran: Modular Quantum Resource Estimation Framework},
  howpublished ={\url{https://qualtran.readthedocs.io}},
  author       = {{Google Quantum AI}},
  year         = {2024},
  note         = {Documentation and SDK, accessed 2025-11-01}
}

@misc{pennylane_lcu,
  title        = {LCU Block-Encoding Tutorial},
  howpublished = {\url{https://pennylane.ai/qml/demos/tutorial_lcu_blockencoding}},
  author       = {{PennyLane Team}},
  year         = {2024},
  publisher    = {Xanadu},
  note         = {Tutorial and example circuits, accessed 2025-11-01}
}

@misc{qrisp_lcu,
  title        = {Qrisp LCU Primitive Reference},
  howpublished = {\url{https://qrisp.eu/reference/Primitives/LCU.html}},
  author       = {{Qrisp Developers}},
  year         = {2024},
  note         = {Library documentation, accessed 2025-11-01}
}

@misc{explicit_be,
  title        = {Explicit Block Encodings for Sparse Matrices},
  howpublished = {\url{https://github.com/QuantumComputingLab/explicit-block-encodings}},
  author       = {{Quantum Computing Lab}},
  year         = {2024},
  note         = {GitHub repository, accessed 2025-11-01}
}

@misc{sqgd_repo,
  title        = {Sequential Quantum Gate Decomposer},
  howpublished = {\url{https://github.com/rakytap/sequential-quantum-gate-decomposer}},
  author       = {Rakyta, P{\'e}ter},
  year         = {2024},
  note         = {GitHub repository, accessed 2025-11-01}
}

@misc{qgopt_docs,
  title        = {QGOpt: Optimization Toolkit for Quantum Gates},
  howpublished = {\url{https://qgopt.readthedocs.io}},
  author       = {{QGOpt Developers}},
  year         = {2024},
  note         = {Python library documentation, accessed 2025-11-01}
}

@misc{gate_decomp_jl,
  title        = {GateDecompositions.jl},
  howpublished = {\url{https://github.com/BBN-Q/GateDecompositions.jl}},
  author       = {{BBN-Q Team}},
  year         = {2024},
  note         = {Julia package for gate decomposition, accessed 2025-11-01}
}

@misc{dcsp_repo,
  title        = {Divide and Conquer State Preparation (DCSP)},
  howpublished = {\url{https://github.com/adjs/dcsp}},
  author       = {ADJS},
  year         = {2023},
  note         = {GitHub repository, accessed 2025-11-01}
}

@misc{unary_iteration,
  title        = {Unary Iteration Compilation},
  howpublished = {\url{https://pennylane.ai/compilation/unary-iteration}},
  author       = {{PennyLane Team}},
  year         = {2024},
  note         = {Compilation primitive reference, accessed 2025-11-01}
}

@misc{pennylane_amp_amp_demo,
  title        = {Introduction to Amplitude Amplification},
  howpublished = {\url{https://pennylane.ai/qml/demos/tutorial_intro_amplitude_amplification}},
  author       = {{PennyLane Team}},
  year         = {2024},
  note         = {Online tutorial, accessed 2025-11-01}
}

@misc{qrisp_amp_amp_docs,
  title        = {Amplitude Amplification Primitive},
  howpublished = {\url{https://qrisp.eu/reference/Primitives/amplitude_amplification.html}},
  author       = {{QRISP Developers}},
  year         = {2024},
  note         = {Documentation, accessed 2025-11-01}
}

@misc{oaa_docs,
  title        = {Oblivious Amplitude Amplification Algorithms},
  howpublished = {\url{https://docs.classiq.io/latest/explore/algorithms/oblivious_amplitude_amplification/oblivious_amplitude_amplification/}},
  author       = {{Classiq Technologies}},
  year         = {2024},
  note         = {Documentation, accessed 2025-11-01}
}

@misc{qtnm_tts,
  title        = {qtnm-tts: Taylor-Series Time Evolution for Quantum Simulation},
  howpublished = {\url{https://github.com/CQCL/qtnm-tts}},
  author       = {{Cambridge Quantum / Quantinuum}},
  year         = {2024},
  note         = {GitHub repository, accessed 2025-11-01}
}

@misc{qite_repo,
  title        = {QITE: Quantum Imaginary Time Evolution},
  howpublished = {\url{https://github.com/mariomotta/QITE}},
  author       = {Motta, Mario},
  year         = {2024},
  note         = {GitHub repository, accessed 2025-11-01}
}

@misc{ncsuQCSrepo,
  author       = {Liu, Yuan},
  title        = {{Quantum Computational Science Repository}},
  howpublished = {\url{https://github.com/ncstate-ece/quantum-computational-science}},
  year         = {2025},
  note         = {North Carolina State University, accessed 2025-11-01}
}

@Misc{qsppack,
author =   {Dong, Yulong and Meng, Xiang and Whaley, K. Birgitta and Lin, Lin},
title =    {QSPPACK},
howpublished = {\url{https://github.com/qsppack/QSPPACK}},
year = {2021}
}

@misc{pennylane_qsvt,
  title        = {Pennylane QSVT Phase Angle},
  url          = {https://pennylane.ai/datasets/collection/phase-angles},
  author       = {{PennyLane Team}},
  publisher    = {Xanadu}
}

@article{acharya2024quantum,
	title = {Quantum error correction below the surface code threshold},
	volume = {638},
	issn = {1476-4687},
	url = {https://doi.org/10.1038/s41586-024-08449-y},
	doi = {10.1038/s41586-024-08449-y},
	number = {8052},
	journal = {Nature},
	author = {Acharya, Rajeev and Abanin, Dmitry A. and Aghababaie-Beni, Laleh and Aleiner, Igor and others},
	month = feb,
	year = {2025},
	pages = {920--926}
}

@article{gidney2024magic,
  title={Magic state cultivation: growing T states as cheap as CNOT gates},
  author={Gidney, Craig and Shutty, Noah and Jones, Cody},
  journal={arXiv preprint arXiv:2409.17595},
  year={2024}
}

@article{koch2019quantum,
  title={Quantum control of molecular rotation},
  author={Koch, Christiane P and Lemeshko, Mikhail and Sugny, Dominique},
  journal={Reviews of Modern Physics},
  volume={91},
  number={3},
  pages={035005},
  year={2019},
  publisher={APS}
}

@article{egan2021fault,
	title = {Fault-tolerant control of an error-corrected qubit},
	volume = {598},
	issn = {1476-4687},
	url = {https://doi.org/10.1038/s41586-021-03928-y},
	doi = {10.1038/s41586-021-03928-y},
	number = {7880},
	journal = {Nature},
	author = {Egan, Laird and Debroy, Dripto M. and Noel, Crystal and Risinger, Andrew and Zhu, Daiwei and Biswas, Debopriyo and Newman, Michael and Li, Muyuan and Brown, Kenneth R. and Cetina, Marko and Monroe, Christopher},
	month = oct,
	year = {2021},
	pages = {281--286},
}

@article{putterman2025hardware,
	title = {Hardware-efficient quantum error correction via concatenated bosonic qubits},
	volume = {638},
	issn = {1476-4687},
	url = {https://doi.org/10.1038/s41586-025-08642-7},
	doi = {10.1038/s41586-025-08642-7},
	number = {8052},
	journal = {Nature},
	author = {Putterman, Harald and Noh, Kyungjoo and Hann, Connor T. and MacCabe, Gregory S. and Aghaeimeibodi, et.al},
	month = feb,
	year = {2025},
	pages = {927--934},
}

@article{bluvstein2024logical,
	title = {Logical quantum processor based on reconfigurable atom arrays},
	volume = {626},
	issn = {1476-4687},
	url = {https://doi.org/10.1038/s41586-023-06927-3},
	doi = {10.1038/s41586-023-06927-3},
	number = {7997},
	journal = {Nature},
	author = {Bluvstein, Dolev and Evered, Simon J. and Geim, Alexandra A. and Li, Sophie H. and Zhou, Hengyun and Manovitz, Tom and Ebadi, Sepehr and Cain, Madelyn and Kalinowski, Marcin and Hangleiter, Dominik and Bonilla Ataides, J. Pablo and Maskara, Nishad and Cong, Iris and Gao, Xun and Sales Rodriguez, Pedro and Karolyshyn, Thomas and Semeghini, Giulia and Gullans, Michael J. and Greiner, Markus and Vuletić, Vladan and Lukin, Mikhail D.},
	month = feb,
	year = {2024},
	pages = {58--65},
}

@article{brock2025quantum,
  title={Quantum error correction of qudits beyond break-even},
  author={Brock, Benjamin L and Singh, Shraddha and Eickbusch, Alec and Sivak, Volodymyr V and Ding, Andy Z and Frunzio, Luigi and Girvin, Steven M and Devoret, Michel H},
  journal={Nature},
  volume={641},
  number={8063},
  pages={612--618},
  year={2025},
  publisher={Nature Publishing Group}
}

@article{chan2024quantum,
  title={Quantum chemistry, classical heuristics, and quantum advantage},
  author={Chan, Garnet Kin-Lic},
  journal={Faraday Discussions},
  year={2024},
  publisher={Royal Society of Chemistry}
}

@article{stroeks2024solving,
  title={Solving Free Fermion Problems on a Quantum Computer},
  author={Stroeks, Maarten and Lenterman, Daan and Terhal, Barbara and Herasymenko, Yaroslav},
  journal={arXiv preprint arXiv:2409.04550},
  year={2024}
}

@article{liu2024hybrid,
  title={Hybrid oscillator-qubit quantum processors: Instruction set architectures, abstract machine models, and applications},
  author={Liu, Yuan and Singh, Shraddha and Smith, Kevin C and Crane, Eleanor and Martyn, John M and Eickbusch, Alec and Schuckert, Alexander and Li, Richard D and Sinanan-Singh, Jasmine and Soley, Micheline B and  Tsunoda, Takahiro and Chuang, Isaac L. and Wiebe, Nathan and Girvin, Steven M.},
  journal={arXiv preprint arXiv:2407.10381},
  year={2024}
}

@article{GrandUnificationAlgos,
  title = {Grand Unification of Quantum Algorithms},
  author = {Martyn, John M. and Rossi, Zane M. and Tan, Andrew K. and Chuang, Isaac L.},
  journal = {PRX Quantum},
  volume = {2},
  issue = {4},
  pages = {040203},
  numpages = {40},
  year = {2021},
  month = {Dec},
  publisher = {American Physical Society},
  doi = {10.1103/PRXQuantum.2.040203},
  url = {https://link.aps.org/doi/10.1103/PRXQuantum.2.040203}
}

@article{rossi2021mqsp,
  title={Multivariable quantum signal processing (M-QSP): prophecies of the two-headed oracle},
  author={Rossi, Zane M. and Chuang, Isaac L.},
  journal={Quantum},
  volume={6},
  pages={811},
  year={2022},
  publisher={Verein zur F{\"o}rderung des Open Access Publizierens in den Quantenwissenschaften},
  url={https://quantum-journal.org/papers/q-2022-09-20-811/}
}

@article{rossi2023quantum,
  title={Quantum signal processing with continuous variables},
  author={Rossi, Zane M and Bastidas, Victor M and Munro, William J and Chuang, Isaac L},
  journal={arXiv preprint arXiv:2304.14383},
  year={2023},
  url={https://arxiv.org/abs/2304.14383}
}

@article{low2014optimal,
  title = {Optimal arbitrarily accurate composite pulse sequences},
  author = {Low, Guang Hao and Yoder, Theodore J. and Chuang, Isaac L.},
  journal = {Phys. Rev. A},
  volume = {89},
  issue = {2},
  pages = {022341},
  numpages = {14},
  year = {2014},
  month = {Feb},
  publisher = {American Physical Society},
  doi = {10.1103/PhysRevA.89.022341},
  url = {https://link.aps.org/doi/10.1103/PhysRevA.89.022341}
}

@article{motlagh2023generalized,
  title={Generalized quantum signal processing},
  author={Motlagh, Danial and Wiebe, Nathan},
  journal={arXiv preprint arXiv:2308.01501},
  year={2023}
}

@article{low2016methodology,
  title = {Methodology of Resonant Equiangular Composite Quantum Gates},
  author = {Low, Guang Hao and Yoder, Theodore J. and Chuang, Isaac L.},
  journal = {Phys. Rev. X},
  volume = {6},
  issue = {4},
  pages = {041067},
  numpages = {13},
  year = {2016},
  month = {Dec},
  publisher = {American Physical Society},
  doi = {10.1103/PhysRevX.6.041067},
  url = {https://link.aps.org/doi/10.1103/PhysRevX.6.041067}
}

@article{rossi2023modular,
  title={Modular quantum signal processing in many variables},
  author={Rossi, Zane M and Ceroni, Jack L and Chuang, Isaac L},
  journal={arXiv preprint arXiv:2309.16665},
  year={2023}
}

@article{kassal2008polynomial,
  title={Polynomial-time quantum algorithm for the simulation of chemical dynamics},
  author={Kassal, Ivan and Jordan, Stephen P and Love, Peter J and Mohseni, Masoud and Aspuru-Guzik, Al{\'a}n},
  journal={Proceedings of the National Academy of Sciences},
  volume={105},
  number={48},
  pages={18681--18686},
  year={2008},
  publisher={National Academy of Sciences},
  url={https://doi.org/10.1073/pnas.0808245105}
}

@article{yuan2025cobblecompilingblockencodings,
      title={Cobble: Compiling Block Encodings for Quantum Computational Linear Algebra}, 
      author={Charles Yuan},
      year={2025},
      journal={arXiv:2511.01736},
      url={https://arxiv.org/abs/2511.01736}
}

@article{kuklinski2025efficientblockencodingsrequirestructure,
      title={Efficient block-encodings require structure}, 
      author={Parker Kuklinski and Benjamin Rempfer and Justin Elenewski and Kevin Obenland},
      year={2025},
      journal={arXiv:2509.19667},
      url={https://arxiv.org/abs/2509.19667}
}

@article{chakraborty2025quantumsingularvaluetransformation,
      title={Quantum singular value transformation without block encodings: Near-optimal complexity with minimal ancilla}, 
      author={Shantanav Chakraborty and Soumyabrata Hazra and Tongyang Li and Changpeng Shao and Xinzhao Wang and Yuxin Zhang},
      year={2025},
      journal={arXiv:2504.02385},
      url={https://arxiv.org/abs/2504.02385}
}

@article{Martyn_2025,
   title={Halving the cost of quantum algorithms with randomization},
   volume={11},
   ISSN={2056-6387},
   url={http://dx.doi.org/10.1038/s41534-025-01003-2},
   DOI={10.1038/s41534-025-01003-2},
   number={1},
   journal={npj Quantum Information},
   publisher={Springer Science and Business Media LLC},
   author={Martyn, John M. and Rall, Patrick},
   year={2025},
   month=mar }

@article{dong2024feedforward,
    title={Feedforward Quantum Singular Value Transformation},
    author={Yulong Dong and Dong An and Murphy Yuezhen Niu},
    year={2024},
    journal={arXiv:2408.07803}
}

@article{Low_2019,
   title={Hamiltonian Simulation by Qubitization},
   volume={3},
   ISSN={2521-327X},
   url={http://dx.doi.org/10.22331/q-2019-07-12-163},
   DOI={10.22331/q-2019-07-12-163},
   journal={Quantum},
   publisher={Verein zur Forderung des Open Access Publizierens in den Quantenwissenschaften},
   author={Low, Guang Hao and Chuang, Isaac L.},
   year={2019},
   month={Jul},
   pages={163}
}

@article{Low_2017,
   title={Optimal Hamiltonian Simulation by Quantum Signal Processing},
   volume={118},
   ISSN={1079-7114},
   url={http://dx.doi.org/10.1103/PhysRevLett.118.010501},
   number={1},
   journal={Physical Review Letters},
   publisher={American Physical Society (APS)},
   author={Low, Guang Hao and Chuang, Isaac L.},
   year={2017},
   month={Jan}
}

@book{nielsen2010quantum,
  title={Quantum Computation and Quantum Information},
  author={Nielsen, Michael A and Chuang, Isaac L},
  year={2010},
  publisher={Cambridge University Press}
}

@article{childs2018toward,
  title={Toward the first quantum simulation with quantum speedup},
  author={Childs, Andrew M and Maslov, Dmitri and Nam, Yunseong and Ross, Neil J and Su, Yuan},
  journal={Proceedings of the National Academy of Sciences},
  volume={115},
  number={38},
  pages={9456--9461},
  year={2018},
  publisher={National Academy Sciences},
  url={https://doi.org/10.1073/pnas.1801723115}
}

@Misc{pyqsp,
author =   {Chuang, Isaac and Tan, Andrew and Martyn, John M},
title =    {Py{QSP}: Python {Q}uantum {S}ignal {P}rocessing},
howpublished = {\url{https://github.com/ichuang/pyqsp}},
year = {2020}
}

@misc{Qiskit,
      title={Quantum computing with {Q}iskit},
      author={Javadi-Abhari, Ali and Treinish, Matthew and Krsulich, Kevin and Wood, Christopher J. and Lishman, Jake and Gacon, Julien and Martiel, Simon and Nation, Paul D. and Bishop, Lev S. and Cross, Andrew W. and Johnson, Blake R. and Gambetta, Jay M.},
      year={2024},
      doi={10.48550/arXiv.2405.08810},
      eprint={2405.08810},
      archivePrefix={arXiv},
      primaryClass={quant-ph}
}

@article{2015Berry,
   title={Simulating Hamiltonian Dynamics with a Truncated Taylor Series},
   volume={114},
   ISSN={1079-7114},
   url={http://dx.doi.org/10.1103/PhysRevLett.114.090502},
   number={9},
   journal={Physical Review Letters},
   publisher={American Physical Society (APS)},
   author={Berry, Dominic W. and Childs, Andrew M. and Cleve, Richard and Kothari, Robin and Somma, Rolando D.},
   year={2015},
   month={Mar}
}

@article{childs2012hamiltonian,
  title={Hamiltonian simulation using linear combinations of unitary operations},
  author={Childs, Andrew M and Wiebe, Nathan},
  journal={Quantum Information \& Computation},
  volume={12},
  number={11-12},
  pages={901--924},
  year={2012},
  publisher={Rinton Press, Incorporated Paramus, NJ}
}

@article{sivak2023real,
  title={Real-time quantum error correction beyond break-even},
  author={Sivak, VV and Eickbusch, Alec and Royer, Baptiste and Singh, Shraddha and Tsioutsios, Ioannis and Ganjam, Suhas and Miano, Alessandro and Brock, BL and Ding, AZ and Frunzio, Luigi and Girvin, SM and Schoelkopf, RJ and Devoret,  MH},
  journal={Nature},
  volume={616},
  number={7955},
  pages={50--55},
  year={2023},
  publisher={Nature Publishing Group UK London}
}

@article{michael2016new,
  title = {New Class of Quantum Error-Correcting Codes for a Bosonic Mode},
  author = {Michael, Marios H. and Silveri, Matti and Brierley, R. T. and Albert, Victor V. and Salmilehto, Juha and Jiang, Liang and Girvin, S. M.},
  journal = {Phys. Rev. X},
  volume = {6},
  issue = {3},
  pages = {031006},
  numpages = {26},
  year = {2016},
  month = {Jul},
  publisher = {American Physical Society},
  doi = {10.1103/PhysRevX.6.031006},
  url = {https://link.aps.org/doi/10.1103/PhysRevX.6.031006}
}

@article{knill1997theory,
  title={Theory of quantum error-correcting codes},
  author={Knill, Emanuel and Laflamme, Raymond},
  journal={Physical Review A},
  volume={55},
  number={2},
  pages={900},
  year={1997},
  publisher={APS}
}

@article{ni2023beating,
  title={Beating the break-even point with a discrete-variable-encoded logical qubit},
  author={Ni, Zhongchu and Li, Sai and Deng, Xiaowei and Cai, Yanyan and Zhang, Libo and Wang, Weiting and Yang, Zhen-Biao and Yu, Haifeng and Yan, Fei and Liu, Song and Zou, Chang-Ling and Sun,  Luyan and  Zheng, Shi-Biao and Xu, Yuan and Yu, Dapeng},
  journal={Nature},
  volume={616},
  number={7955},
  pages={56--60},
  year={2023},
  publisher={Nature Publishing Group UK London}
}

@article{gupta2024encoding,
  title={Encoding a magic state with beyond break-even fidelity},
  author={Gupta, Riddhi S and Sundaresan, Neereja and Alexander, Thomas and Wood, Christopher J and Merkel, Seth T and Healy, Michael B and Hillenbrand, Marius and Jochym-O’Connor, Tomas and Wootton, James R and Yoder, Theodore J and Cross, Andrew W. and Takita, Maika and Brown, Benjamin J.},
  journal={Nature},
  volume={625},
  number={7994},
  pages={259--263},
  year={2024},
  publisher={Nature Publishing Group UK London}
}

@article{martyn2023efficient,
  title={Efficient fully-coherent quantum signal processing algorithms for real-time dynamics simulation},
  author={Martyn, John M and Liu, Yuan and Chin, Zachary E and Chuang, Isaac L},
  journal={The Journal of Chemical Physics},
  volume={158},
  number={2},
  year={2023},
  publisher={AIP Publishing}
}

@article{cerezo2021variational,
  title={Variational quantum algorithms},
  author={Cerezo, Marco and Arrasmith, Andrew and Babbush, Ryan and Benjamin, Simon C and Endo, Suguru and Fujii, Keisuke and McClean, Jarrod R and Mitarai, Kosuke and Yuan, Xiao and Cincio, Lukasz and Coles,  Patrick J.},
  journal={Nature Reviews Physics},
  volume={3},
  number={9},
  pages={625--644},
  year={2021},
  publisher={Nature Publishing Group UK London}
}

@article{bharti2022noisy,
  title = {Noisy intermediate-scale quantum algorithms},
  author = {Bharti, Kishor and Cervera-Lierta, Alba and Kyaw, Thi Ha and Haug, Tobias and Alperin-Lea, Sumner and Anand, Abhinav and Degroote, Matthias and Heimonen, Hermanni and Kottmann, Jakob S. and Menke, Tim and Mok, Wai-Keong and Sim, Sukin and Kwek, Leong-Chuan and Aspuru-Guzik, Al\'an},
  journal = {Rev. Mod. Phys.},
  volume = {94},
  issue = {1},
  pages = {015004},
  numpages = {69},
  year = {2022},
  month = {Feb},
  publisher = {American Physical Society},
  doi = {10.1103/RevModPhys.94.015004},
  url = {https://link.aps.org/doi/10.1103/RevModPhys.94.015004}
}

@article{cao2019quantum,
author = {Cao, Yudong and Romero, Jonathan and Olson, Jonathan P. and Degroote, Matthias and Johnson, Peter D. and Kieferová, Mária and Kivlichan, Ian D. and Menke, Tim and Peropadre, Borja and Sawaya, Nicolas P. D. and Sim, Sukin and Veis, Libor and Aspuru-Guzik, Alán},
title = {Quantum Chemistry in the Age of Quantum Computing},
journal = {Chemical Reviews},
volume = {119},
number = {19},
pages = {10856-10915},
year = {2019},
doi = {10.1021/acs.chemrev.8b00803},
    note ={PMID: 31469277},

URL = { 
    
        https://doi.org/10.1021/acs.chemrev.8b00803
    
    

},
eprint = { 
    
        https://doi.org/10.1021/acs.chemrev.8b00803
    
    

}

}

@article{tan2023perturbative,
  title = {Perturbative model of noisy quantum signal processing},
  author = {Tan, Andrew K. and Liu, Yuan and Tran, Minh C. and Chuang, Isaac L.},
  journal = {Phys. Rev. A},
  volume = {107},
  issue = {4},
  pages = {042429},
  numpages = {20},
  year = {2023},
  month = {Apr},
  publisher = {American Physical Society},
  doi = {10.1103/PhysRevA.107.042429},
  url = {https://link.aps.org/doi/10.1103/PhysRevA.107.042429}
}

@article{lin2020near,
  title={Near-optimal ground state preparation},
  author={Lin, Lin and Tong, Yu},
  journal={Quantum},
  volume={4},
  pages={372},
  year={2020},
  publisher={Verein zur F{\"o}rderung des Open Access Publizierens in den Quantenwissenschaften}
}

@article{pytket,
   title={t|ket⟩: a retargetable compiler for NISQ devices},
   volume={6},
   ISSN={2058-9565},
   url={http://dx.doi.org/10.1088/2058-9565/ab8e92},
   DOI={10.1088/2058-9565/ab8e92},
   number={1},
   journal={Quantum Science and Technology},
   publisher={IOP Publishing},
   author={Sivarajah, Seyon and Dilkes, Silas and Cowtan, Alexander and Simmons, Will and Edgington, Alec and Duncan, Ross},
   year={2020},
   month=nov, pages={014003} }

@article{vasconcelos2025methodsreducingancillaoverheadblock,
      title={Methods for Reducing Ancilla-Overhead in Block Encodings}, 
      author={Francisca Vasconcelos and András Gilyén},
      year={2025},
      journal={arXiv:2507.07900},
      url={https://arxiv.org/abs/2507.07900}
}

\end{document}